\documentclass[12pt]{article}
\usepackage[left=2.5cm, bottom=2.5cm, right=2.5cm, top=2.5cm]{geometry}

\usepackage{sectsty}
\usepackage{rotating}
\usepackage{tikz}
\usepackage{amssymb}
\usepackage{graphicx}
\usepackage{paralist}
\usepackage{mathbbol}
\usepackage{nicefrac}
\usetikzlibrary{patterns}
\usepackage{subcaption}
\makeatletter
\def\tikz@arc@opt[#1]{
  {%
    \tikzset{every arc/.try,#1}%
    \pgfkeysgetvalue{/tikz/start angle}\tikz@s
    \pgfkeysgetvalue{/tikz/end angle}\tikz@e
    \pgfkeysgetvalue{/tikz/delta angle}\tikz@d
    \ifx\tikz@s\pgfutil@empty%
      \pgfmathsetmacro\tikz@s{\tikz@e-\tikz@d}
    \else
      \ifx\tikz@e\pgfutil@empty%
        \pgfmathsetmacro\tikz@e{\tikz@s+\tikz@d}
      \fi%
    \fi
    \tikz@arc@moveto
    \xdef\pgf@marshal{\noexpand%
    \tikz@do@arc{\tikz@s}{\tikz@e}
      {\pgfkeysvalueof{/tikz/x radius}}
      {\pgfkeysvalueof{/tikz/y radius}}}%
  }%
  \pgf@marshal%
  \tikz@arcfinal%
}
\let\tikz@arc@moveto\relax
\def\tikz@arc@movetolineto#1{%
  \def\tikz@arc@moveto{\tikz@@@parse@polar{\tikz@arc@@movetolineto#1}(\tikz@s:\pgfkeysvalueof{/tikz/x radius} and \pgfkeysvalueof{/tikz/y radius})}}
\def\tikz@arc@@movetolineto#1#2{#1{\pgfpointadd{#2}{\tikz@last@position@saved}}}
\tikzset{%
  move to start/.code=\tikz@arc@movetolineto\pgfpathmoveto,%
  line to start/.code=\tikz@arc@movetolineto\pgfpathlineto}
\makeatother
\makeatletter
\newcommand{\labeltext}[3][]{%
    \@bsphack%
    \csname phantomsection\endcsname
    \def\tst{#1}%
    \def\labelmarkup{}
    \def\refmarkup{}%
    \ifx\tst\empty\def\@currentlabel{\refmarkup{#2}}{\label{#3}}%
    \else\def\@currentlabel{\refmarkup{#1}}{\label{#3}}\fi%
    \@esphack%
    \labelmarkup{#2}
}
\makeatother
\usepackage{pgfplots}
\usetikzlibrary{intersections, pgfplots.fillbetween}
\usetikzlibrary{decorations.pathreplacing}
\usepackage{amsmath}
\usepackage{amsthm}
\usepackage{bm}
\usepackage{multirow}
\usepackage[normalem]{ulem}
\usepackage{pdflscape}
\usepackage{pgfplots}
\usepackage{mathtools}
\usepackage{theoremref}
\usepackage{xcolor}
\usepackage{changepage}
\colorlet{linkequation}{blue}
\usepackage[authoryear,longnamesfirst,round]{natbib}
\setcitestyle{authoryear,open={(},close={)},aysep={},citesep={;}}
\colorlet{linkequation}{blue}
\usepackage[colorlinks=true,allcolors=blue]{hyperref}
\newcommand*{\SavedEqref}{}
\let\SavedEqref\eqref
\renewcommand*{\eqref}[1]{%
  \begingroup
    \hypersetup{
      linkcolor=linkequation,
      linkbordercolor=linkequation,
      breaklinks=true,   
    }%
    \SavedEqref{#1}%
  \endgroup
}
\newcommand\bref[1]{{\hypersetup{linkcolor=blue}\autoref{#1}}}
\makeatletter
\renewcommand\@makefnmark{\hbox{\@textsuperscript{\normalfont\color{black}\@thefnmark}}}
\makeatother
\useunder{\uline}{\ul}{}
\definecolor{gray1}{gray}{0.8}
\definecolor{gray2}{gray}{0.6}
\definecolor{gray3}{gray}{0.4}

\newcommand*\diff{\mathop{}\!\mathrm{d}}

\pgfplotsset{compat=1.17}
\theoremstyle{plain}
\newtheorem{thm}{Theorem}
\theoremstyle{plain}
\newtheorem{prop}{Proposition}
\theoremstyle{plain}
\newtheorem{cor}{Corollary}
\theoremstyle{definition}

\theoremstyle{definition}

\theoremstyle{plain}

\theoremstyle{definition}
\theoremstyle{plain}

\newtheorem{taggedtheoremx}{Theorem}
\newenvironment{taggedtheorem}[1]
 {\renewcommand\thetaggedtheoremx{#1}\taggedtheoremx}
 {\endtaggedtheoremx}

\theoremstyle{definition}
\newtheorem{rem}{Remark}
\theoremstyle{remark}

\usetikzlibrary{arrows,shapes,positioning}
\tikzstyle{every lower node part}=[font=5pt]
\sectionfont{\fontsize{14}{16}\selectfont}
\usepackage{setspace}

\renewcommand{\thefigure}{\Roman{figure}}

\usepackage{caption}
\usepackage{subcaption}

\makeatletter
\renewcommand{\p@subfigure}{\thefigure.}
\makeatother

\newcommand*{\lb}{\underline}
\newcommand*{\ub}{\overline}

\renewcommand{\omega}{x}

\begin{document}

\title{\textbf{Monotone Function Intervals: Theory and Applications}\thanks{We thank the coeditors and three anonymous reviewers for their generous comments. We are also grateful to Ian Ball, Dirk Bergemann, Alex Bloedel, Simon Board, Ben Brooks, Peter Caradonna, Roberto Corrao, Xiaohong Chen, Joyee Deb, Piotr Dworczak, Mira Frick, Tan Gan, Marina Halac, Ryota Iijima, Emir Kamenica, Nicolas Lambert, Elliot Lipnowski, Alessandro Lizzeri, Jay Lu, Leslie Marx, Moritz Meyer-ter-Vehn, Stephen Morris, Barry Nalebuff, Pietro Ortoleva, Aniko \"{O}ry, Wolfgang Pesendorfer, Benjamin Polak, Doron Ravid, Anne-Katrin Roesler, Hamid Sabourian, Fedor Sandomirskiy, Christoph Schlom, Ludvig Sinander, Vasiliki Skreta, Nicholas Stephanopoulos, Philipp Strack, Roland Strausz, Tomasz Strzalecki, Omer Tamuz, Can Urgun, Quitz\'{e} Valenzuela-Stookey, Martin Vaeth, Dong Wei, Mark Whitmeyer, Alexander Wolitzky, Wenji Xu, Nathan Yoder, Jidong Zhou, Pavel Zryumov, and the participants of various seminars and conferences for their helpful comments.  We are grateful to Bianca Battaglia for her exceptional copyediting. We also thank Jialun (Neil) He and Nick Wu for their research assistance. All errors are our own.}
}
\author{
    Kai Hao Yang\thanks{Yale School of Management, Email: kaihao.yang@yale.edu} 
    \and 
    Alexander K. Zentefis\thanks{Yale School of Management, Email: alexander.zentefis@yale.edu} }
\date{April 12, 2024}

\maketitle

\begin{abstract} 

A monotone function interval is the set of monotone functions that lie pointwise between two fixed monotone functions. We characterize the set of extreme points of monotone function intervals and apply this to a number of economic settings. First, we leverage the main result to characterize the set of distributions of posterior quantiles that can be induced by a signal, with applications to political economy, Bayesian persuasion, and the psychology of judgment. Second, we combine our characterization with properties of convex optimization problems to unify and generalize seminal results in the literature on security design under adverse selection and moral hazard.

\end{abstract}
\begin{flushleft}
\textbf{\vfill{}
JEL classification:} D72, D82, D83, D86, G23 \textbf{}\linebreak{}
\textbf{Keywords:} Extreme points, monotone functions, pointwise dominance, posterior quantiles, gerrymandering, Bayesian persuasion, misconfidence, security design \\
\par\end{flushleft}

\vfill{}

\thispagestyle{empty} 

\pagebreak{}

\setcounter{footnote}{0}

\setcounter{page}{1} 

\begin{spacing}{1.25}

\section{Introduction}
Monotone functions play a crucial role in many economic settings. In standard equilibrium analysis, demand curves and supply curves are monotone. In moral hazard problems, many contracts are monotone. In information economics, distributions of a one-dimensional unknown state can be summarized by monotone cumulative distribution functions (CDFs). Among all orderings, the pointwise dominance order is one of the most natural ways to compare monotone functions: outward/inward shifts of supply and demand, limited liability in contract theory, and the first-order stochastic dominance order of CDFs are all expressed in terms of pointwise dominance of monotone functions. 

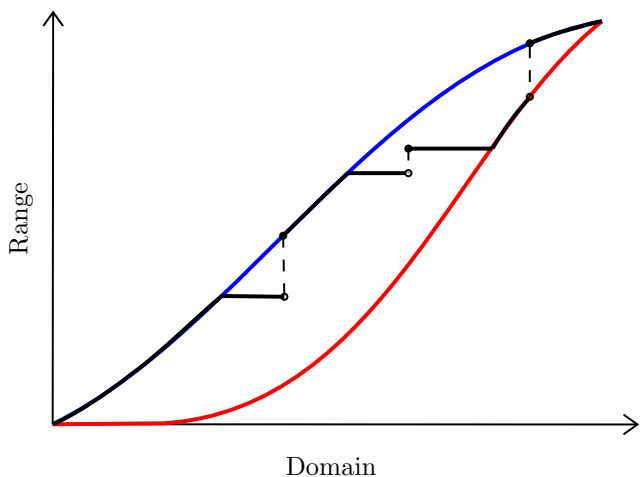
\begin{figure}[h!]
\begin{center}
        \tikzset{every picture/.style={line width=0.75pt}} 

\begin{tikzpicture}[x=0.75pt,y=0.75pt,yscale=-1,xscale=1]

\draw [line width=0.75]  (173.96,284.95) -- (468.87,285.05)(174.21,76.95) -- (173.96,284.95) -- cycle (461.87,280.05) -- (468.87,285.05) -- (461.86,290.05) (169.2,83.95) -- (174.21,76.95) -- (179.2,83.95)  ;
\draw [color={blue}  ,draw opacity=1 ][line width=1.5]    (173.96,284.95) .. controls (270,237.5) and (340,102.5) .. (451,81.5) ;
\draw [color={red}  ,draw opacity=1 ][line width=1.5]    (229.96,284.45) .. controls (341,277.4) and (379,139.5) .. (451,81.5) ;
\draw [color={red}  ,draw opacity=1 ][line width=1.5]    (173.96,284.95) -- (229.96,284.45) ;
\draw [line width=1.5]    (173.96,284.95) .. controls (184.6,280.6) and (201.4,269.4) .. (213.4,259.8) ;
\draw [line width=1.5]    (213.4,259.8) .. controls (227.8,249.8) and (252.2,225.4) .. (259.4,220.2) ;
\draw [line width=1.5]    (259.4,220.2) -- (290.8,220.6) ;
\draw [line width=0.75]  [dash pattern={on 4.5pt off 4.5pt}]  (290.2,189.8) -- (290.8,220.6) ;
\draw   (289.25,220.6) .. controls (289.25,219.74) and (289.94,219.05) .. (290.8,219.05) .. controls (291.66,219.05) and (292.35,219.74) .. (292.35,220.6) .. controls (292.35,221.46) and (291.66,222.15) .. (290.8,222.15) .. controls (289.94,222.15) and (289.25,221.46) .. (289.25,220.6) -- cycle ;
\draw  [fill={rgb, 255:red, 0; green, 0; blue, 0 }  ,fill opacity=1 ] (288.65,189.8) .. controls (288.65,188.94) and (289.34,188.25) .. (290.2,188.25) .. controls (291.06,188.25) and (291.75,188.94) .. (291.75,189.8) .. controls (291.75,190.66) and (291.06,191.35) .. (290.2,191.35) .. controls (289.34,191.35) and (288.65,190.66) .. (288.65,189.8) -- cycle ;
\draw [line width=1.5]    (290.2,189.8) .. controls (296.2,183.8) and (316.2,164.2) .. (323,158.2) ;
\draw [line width=1.5]    (323,158.2) -- (351.8,158.2) ;
\draw   (351.8,158.2) .. controls (351.8,157.34) and (352.49,156.65) .. (353.35,156.65) .. controls (354.21,156.65) and (354.9,157.34) .. (354.9,158.2) .. controls (354.9,159.06) and (354.21,159.75) .. (353.35,159.75) .. controls (352.49,159.75) and (351.8,159.06) .. (351.8,158.2) -- cycle ;
\draw [line width=0.75]  [dash pattern={on 4.5pt off 4.5pt}]  (353.4,145.8) -- (353.35,158.2) ;
\draw  [fill={rgb, 255:red, 0; green, 0; blue, 0 }  ,fill opacity=1 ] (351.85,145.8) .. controls (351.85,144.94) and (352.54,144.25) .. (353.4,144.25) .. controls (354.26,144.25) and (354.95,144.94) .. (354.95,145.8) .. controls (354.95,146.66) and (354.26,147.35) .. (353.4,147.35) .. controls (352.54,147.35) and (351.85,146.66) .. (351.85,145.8) -- cycle ;
\draw [line width=1.5]    (351.85,145.8) -- (395.8,145.8) ;
\draw [line width=1.5]    (395.8,145.8) .. controls (400.6,137.4) and (408.2,126.6) .. (414.6,120.2) ;
\draw   (413.05,119.65) .. controls (413.05,118.79) and (413.74,118.1) .. (414.6,118.1) .. controls (415.46,118.1) and (416.15,118.79) .. (416.15,119.65) .. controls (416.15,120.51) and (415.46,121.2) .. (414.6,121.2) .. controls (413.74,121.2) and (413.05,120.51) .. (413.05,119.65) -- cycle ;
\draw [line width=0.75]  [dash pattern={on 4.5pt off 4.5pt}]  (414.6,94.2) -- (414.6,119.65) ;
\draw  [fill={rgb, 255:red, 0; green, 0; blue, 0 }  ,fill opacity=1 ](413.05,92.65) .. controls (413.05,91.79) and (413.74,91.1) .. (414.6,91.1) .. controls (415.46,91.1) and (416.15,91.79) .. (416.15,92.65) .. controls (416.15,93.51) and (415.46,94.2) .. (414.6,94.2) .. controls (413.74,94.2) and (413.05,93.51) .. (413.05,92.65) -- cycle ;
\draw [line width=1.5]    (414.6,92.65) .. controls (425.8,87.4) and (446.2,82.6) .. (451,81.5) ;

\draw (150.7,200.9) node [anchor=north west][inner sep=0.75pt] [font=\footnotesize] [rotate=-270] [align=left] {{Range}};
\draw (290,300) node [anchor=north west] [inner sep=0.75pt] [font=\footnotesize] [align=left] {Domain};
\end{tikzpicture}
    \end{center}
\caption{An Extreme Point of a Monotone Function Interval}
\label{fig1}
\end{figure}

In this paper, we provide a systematic way to study an arbitrary convex set of monotone functions that are bounded pointwise from above and below by two monotone functions. Without loss, we focus on sets of nondecreasing and right-continuous functions bounded by two nondecreasing functions, such as the blue and red curves in \bref{fig1}. We refer to these sets as \emph{monotone function intervals}. 
Our main result (\bref{thm1}) characterizes the \emph{extreme points} of monotone function intervals. We then show how this abstract characterization has several economic applications. 





Specifically, we show that a nondecreasing, right-continuous function is an extreme point of a monotone function interval if and only if the function either coincides with one of the two bounds or is constant on an interval in its domain. Wherever the function is constant on an interval, it must coincide with one of the two bounds at one of the endpoints of the interval, as illustrated by the black curve in \bref{fig1}. This characterization, together with two well-known properties of extreme points, leads to many economic applications. The first property, formally known as Choquet's theorem, is that any element of a compact and convex set can be represented as a convex combination of the extreme points. The second property is that any (well-defined) convex optimization problem must have an extreme point of the feasible set as its solution. We consider two classes of economic applications, each exploiting one of the two aforementioned properties of extreme points. 



In the first class of applications, we use \bref{thm1} and Choquet's theorem to characterize the set of distributions of posterior quantiles. Consider a one-dimensional state and a signal (i.e., a Blackwell experiment). Each signal realization induces a posterior belief. For every posterior belief, one can compute the posterior mean. Strassen's theorem \citep*{S65} implies that the distribution of these posterior means is a mean-preserving contraction of the prior. Conversely, every mean-preserving contraction of the prior is the distribution of posterior means under some signal. Instead of posterior means, one can derive many other statistics of a posterior. The characterization of the extreme points of monotone function intervals leads to an analog of Strassen's theorem, which characterizes the set of distributions of posterior \emph{quantiles} (\bref{thm2} and \bref{thm3}). The set of distributions of posterior quantiles coincides with an interval of CDFs bounded by a natural upper and lower truncation of the prior. 

We apply \bref{thm2} and \bref{thm3} to three settings: gerrymandering, quantile-based persuasion, and apparent over/underconfidence (misconfidence). These settings all share concerns over ordinal rather than cardinal outcomes. First, gerrymandering is connected to distributions of posterior quantiles, since voters' political ideologies are only ordinal. When the distribution of voters' political ideologies in an election district is interpreted as a posterior, the median voter theorem implies that the ideological position of the elected representative in that district is a posterior median. Since an electoral map corresponds to a distribution of posteriors under this interpretation, \bref{thm2} and \bref{thm3} characterize the compositions of the legislative body that a gerrymandered map can create. Second, in Bayesian persuasion, \bref{thm2} and \bref{thm3} bring tractability to persuasion problems where the sender's indirect payoff is a function of posterior \emph{quantiles}: an ordinal analog of the widely studied environment where the sender's indirect payoff is a function of posterior \emph{means}. This is the case when the sender's payoff is state-independent and the receiver chooses an action to minimize the expected \emph{absolute}---as opposed to \emph{quadratic}---distance to the state, or when the receiver is not an expected utility maximizer, but a quantile maximizer \citep{manski1988ordinal,rostek2010quantile}.  Third, the literature on the psychology of judgment documents that individuals appear to be overconfident or underconfident when evaluating themselves relative to a population. Since individuals are asked to rank themselves relative to the population according to the median of their posterior beliefs in many experiments in this literature, \bref{thm3} implies the seminal result of \citet*{BD11}, who provide a necessary and sufficient condition for \emph{apparent} overconfidence (e.g., more than 50\% of individuals ranking themselves above the population median) to imply \emph{true} overconfidence (i.e., individuals are not Bayesian).


In the second class of applications, we use \bref{thm1}, together with the optimality of extreme points in convex problems, to study security design with limited liability. Consider the canonical security design problem where the security issuer designs a security that specifies payments to the security holder contingent on the realized return of an asset. Two assumptions are commonly adopted in the security design literature. The first assumption is that any security must be nondecreasing in the asset’s return, so that the security holder would not have an incentive to sabotage the asset. 
 The second is limited liability, which places natural upper and lower bounds on the security's payoff given each realized return. Under these two assumptions, the set of securities coincides with a monotone function interval bounded by the identity function and the constant function $0$. 


Two seminal papers adopt these assumptions in their analyses of the security design problem. \citet{innes1990limited} studies the problem under moral hazard, whereas \citet{DD99} study it under adverse selection. Both papers show that a standard debt contract is optimal, which promises either a constant payment or the asset’s realized return, whichever is smaller. Many papers in security design are built upon the \citet{innes1990limited} or \cite{DD99} environment. (See, for example, \citealp{schmidt1997managerial,casamatta2003financing} and \citealp{eisfeldt2004endogenous, biais2005strategic}.)

The optimality of standard debt in \citet{innes1990limited} and \citet{DD99} relies on a crucial assumption: the distribution of the asset return satisfies the monotone likelihood ratio property (MLRP). Therefore, the structure of optimal securities without MLRP remains relatively under-explored. \bref{thm1} provides new insights into this question. Security design in these settings is a convex constrained optimization problem, so there must be an extreme point of the feasible set of securities that is optimal. Because securities are elements of a monotone function interval, \bref{thm1} helps identify the extreme points of the feasible set. These extreme points correspond to \emph{contingent} debt contracts. Contingent debts are a natural generalization of standard debts, in the sense that their face values may depend on the realized return of the asset. Meanwhile, just like standard debts, contingent debts do not create equity shares between the security issuer and the security holder, and marginal returns are (almost) always fully allocated to one of the two parties. In essence, this result separates the effects of limited liability from those of MLRP on the optimal security.

Overall, this paper uncovers the common underlying role of monotone function intervals in many topics in economics, and offers a unifying approach to answering canonical economic questions that have been previously answered by separate, case-specific approaches. 

\paragraph{Related Literature.}

This paper relates to several areas. The main result connects to characterizations of extreme points of convex sets in mathematics. Pioneering in this area, \citet*{HLP29} characterize the extreme points of a set of vectors $x$ majorized by another vector $x_0$ in $\mathbb{R}^n$, and show that the set of extreme points equals the set of permutations of $x_0$.\footnote{A vector $x \in \mathbb{R}^n$ majorizes $y \in \mathbb{R}^n$ if $\sum_{i=1}^k x_{(i)} \geq \sum_{i=1}^k y_{(i)}$ for all $k \in \{1,\ldots,n\}$, with equality at $k=n$, where $x_{(j)}$ and $y_{(j)}$ are the $j$-th smallest component of $x$ and $y$, respectively.} 
\citet*{R67} extends this result to infinite dimensional spaces. \citet{K73} characterizes the extreme points of the set of probability measures over $[0,1]$ that dominate a given probability measure $\mu_0$ in the convex order, and shows that the set of extreme points equals the set of probability measures obtained by applying to $\mu_0$ a Markov transition defined on a closed set $F$ that splits each element in the support of $\mu_0$ into the nearest elements of $F$. Analogously, \citet{L75} characterizes the extreme points of the set of probability measures over $[0,1]$ that are dominated by a given non-atomic probability measure $\mu_0$ in the stochastic order, and shows that the set of extreme points equals the set of probability measures obtained by applying to $\mu_0$ a Markov transition defined on a closed set $F$ that moves each element in the support of $\mu_0$ to the nearest larger element in $F$.

In economics, \citet*{KMS21} characterize the extreme points of monotone functions on $[0,1]$ that majorize (or are majorized by) some given monotone function, which is equivalent to the set of probability measures that dominate (or are dominated by) a given probability measure in the convex order. They then apply this characterization to various economic settings, including mechanism design, two-sided matching, mean-based persuasion, and delegation.\footnote{See also \citet*{ABSY22}. Several recent papers in economics also exploit properties of extreme points to derive economic implications. See, for instance, \citet{BBM15} and \citet{LM17}.} \citet{CS23} and \citet{N23} 
characterize the extreme points of the same sets subject to finitely many additional linear constraints. In comparison, this paper characterizes the extreme points of monotone functions that are in between two given monotone functions on $\mathbb{R}$ in the pointwise order, which is equivalent to the set of probability measures in between two given probability measures in the stochastic order,\footnote{\bref{thma1} in the appendix also characterizes the extreme points of this set, subject to finitely-many additional linear constraints.} and applies the characterization to voting, quantile-based persuasion, self ranking, and security design.



The first application of the extreme point characterization to the distributions of posterior quantiles is related to belief-based characterizations of signals, which date back to the seminal contributions of \citet*{B53} and \citet{S65}. Blackwell's and Strassen's characterizations also lead to the characterization of the set of distributions of posterior means. This paper's characterization of the set of distributions of posterior quantiles (\bref{thm2} and \bref{thm3}) can be regarded as an analog. In a recent paper, \citet{KW24} provide an alternative proof of \bref{thm2}, which does not require the use of extreme points.  


The application to gerrymandering relates to the literature on redistricting, particularly to \citet*{owen1988optimal}, \citet*{friedman2008optimal}, \citet*{gul2010strategic}, and \citet*{kolotilin2023economics}, who also adopt the belief-based approach and model a district map as a way to split the population distribution of voters. Existing work mainly focuses on a political party's optimal gerrymandering when maximizing either its expected number of seats or its probability of winning a majority. In contrast, this paper characterizes the \emph{feasible compositions} of a legislative body that a district map can induce. The application to Bayesian persuasion relates to that large literature (see \citealp{K19} for a comprehensive survey), in particular to communication problems where only posterior means are payoff-relevant (e.g., \citealp*{GK16,RS17,DM19,Aetal22}). This paper complements that literature by providing a foundation for solving communication problems where only the posterior \emph{quantiles} are payoff-relevant. 


Finally, the application to security design connects this paper to that large literature. \citet{allen2022security} provide a recent survey. In this application, we base our economic environments on \citet{innes1990limited}, which involves moral hazard, and \citet{DD99}, which involves adverse selection. This paper generalizes and unifies results in those seminal works under a common structure. 

\paragraph{Outline.} The rest of the paper proceeds as follows. \bref{sec:characterization} presents the paper's central theorem: the characterization of the extreme points of monotone function intervals (\bref{thm1}). \bref{s3} applies \bref{thm1} to characterize the set of distributions of posterior quantiles. Economic applications related to the quantile characterization (gerrymandering, quantile-based persuasion, and apparent misconfidence) follow in \bref{sec:quantile_applications}. \bref{s4} applies \bref{thm1} to security design with limited liability. \bref{sec:conclusion} concludes. 
 
\section{Extreme Points of Monotone Function Intervals}
\label{sec:characterization}

\subsection{Notation}
Let $\mathcal{F}$ be the set of nondecreasing and right-continuous functions on $\mathbb{R}$.\footnote{Whenever needed, $\mathcal{F}$ is endowed with the topology defined by weak convergence (i.e., $\{F_n\} \to F$ if $\lim_{n \to \infty} F_n(x)=F(x)$ for all $x$ at which $F$ is continuous), as well as the Borel $\sigma$-algebra induced by this topology.} For any $\underline{F}, \overline{F} \in \mathcal{F}$ such that $\underline{F}(x) \leq \overline{F}(x)$ for all $x \in \mathbb{R}$ ($\underline{F} \leq \overline{F}$ henceforth), let 
\[
\mathcal{I}(\underline{F},\overline{F}):=\{H \in \mathcal{F}\,|\underline{F}(x) \leq H(x) \leq \overline{F}(x), \, \forall x \in \mathbb{R}\}.
\]
Namely, $\mathcal{I}(\underline{F},\overline{F})$ is the set of nondecreasing, right-continuous functions that dominate $\underline{F}$ and simultaneously are dominated by $\overline{F}$ pointwise. We refer to $\mathcal{I}(\underline{F},\overline{F})$ as the \emph{interval} of monotone functions bounded by $\underline{F}$ and $\overline{F}$. For any $H \in \mathcal{F}$ and for any $x \in \mathbb{R}$, let $H(x^-):=\lim_{y \uparrow x}H(y)$ denote the left-limit of $F$ at $x$.

\subsection{Extreme Points of Monotone Function Intervals}
For any $\underline{F},\overline{F} \in \mathcal{F}$ with $\underline{F} \leq \overline{F}$, the interval $\mathcal{I}(\underline{F},\overline{F})$ is convex. Recall that $H \in \mathcal{I}(\underline{F},\overline{F})$ is an \emph{extreme point} of the convex set $\mathcal{I}(\underline{F},\overline{F})$ if $H$ cannot be written as a convex combination of two distinct elements of $\mathcal{I}(\underline{F},\overline{F})$. Our main result, \bref{thm1}, characterizes the extreme points of $\mathcal{I}(\underline{F},\overline{F})$. 

\begin{figure}
\centering
\tikzset{
solid node/.style={circle,draw,inner sep=1.25,fill=black},
hollow node/.style={circle,draw,inner sep=1.25}
}
\begin{subfigure}[b]{0.4\linewidth}
\begin{tikzpicture}[scale=5]
\draw [<->, very thick] (0,1.05) node (yaxis) [above] {\footnotesize }
        |- (1.05,0) node (xaxis) [right] {\footnotesize$x$};
\draw (0,0) node [left=2pt] {\footnotesize$0$};
\draw [blue, very thick] (0,0)--(0.4,0.8);
\draw [blue, very thick] (0.4,0.8)--(1,1);
\draw [red, very thick] (0,0)--(0.7,0.4);
\draw [red, very thick] (0.7,0.4)--(1,1);

\draw [ultra thick] (0,0)--(0.1,0.2);
\draw [ultra thick] (0.1,0.2)--(0.15,0.2);
\draw (0.15,0.2) node [hollow node] {};
\draw (0.15,0.3) node [solid node] {};
\draw [dashed, thick] (0.15,0.2)--(0.15,0.3);
\draw [ultra thick] (0.15,0.3)--(0.2,0.4);
\draw [ultra thick] (0.2,0.4)--(0.3,0.4);
\draw (0.3,0.4) node [hollow node] {}; 
\draw (0.3,0.6) node [solid node] {};
\draw [dashed, thick] (0.3,0.4)--(0.3,0.6);
\draw [ultra thick] (0.3,0.6)--(0.35,0.7);
\draw [ultra thick] (0.35,0.7)--(0.55,0.7);
\draw [ultra thick] (0.55,0.75)--(0.875,0.75);
\draw [ultra thick] (0.875,0.75)--(0.9,0.8);
\draw (0.55,0.7) node [hollow node] {};
\draw (0.55,0.75) node [solid node] {};
\draw [dashed, thick] (0.55,0.7)--(0.55,0.75);
\draw (0.9,0.8) node [hollow node] {};
\draw (0.9,0.9) node [solid node] {};
\draw [dashed, thick] (0.9,0.8)--(0.9,0.9);
\draw [ultra thick] (0.9,0.9)--(0.95,0.9);
\draw [ultra thick] (0.95,0.9)--(1,1);






\node[align=center] at (0.5,-0.1) {\footnotesize Domain};

\node[rotate=90, align=center] at (-0.1,0.5) {\footnotesize Range};

\matrix [draw=black, thin, anchor=north west, row sep=0pt] at (0.05,1.1) {
    \draw [blue, very thick] (0,0) -- (0.2,0); & \node[right] {\footnotesize $\overline{F}$}; \\
    \draw [black, very thick] (0,0) -- (0.2,0); & \node[right] {\footnotesize $H$}; \\
    \draw [red, very thick] (0,0) -- (0.2,0); & \node[right] {\footnotesize $\underline{F}$}; \\
};

\end{tikzpicture}
\caption{An Extreme Point, $H$}
\label{fig2a}
\end{subfigure}%
\begin{subfigure}[b]{0.4\linewidth}
\begin{tikzpicture}[scale=5]
\draw [<->, very thick] (0,1.05) node (yaxis) [above] {\footnotesize }
        |- (1.05,0) node (xaxis) [right] {\footnotesize $x$};
\draw (0,0) node [left=2pt] {\footnotesize$0$};
\draw [blue, very thick] (0,0)--(0.4,0.8);
\draw [blue, very thick] (0.4,0.8)--(1,1);
\draw [red, very thick] (0,0)--(0.7,0.4);
\draw [red, very thick] (0.7,0.4)--(1,1);


\draw [ultra thick] (0,0)--(0.2335,0.467);
\draw [ultra thick] (0.2335,0.467)--(0.4,0.467);
\draw [ultra thick] (0.6,0.533)--(0.8,0.9);
\draw [ultra thick] (0.4,0.467)--(0.6,0.533);
\draw [ultra thick] (0.8,0.9)--(1,1);

\draw [dashed, thick] (0.4,0.467) rectangle (0.6,0.533);

\draw [gray2, thick] (0.4,0.467)--(0.5,0.533);
\draw [gray2, thick] (0.5,0.467)--(0.6,0.533);
\draw [gray2, thick] (0.5,0.533)--(0.6,0.533);
\draw [gray2, thick] (0.4,0.467)--(0.5,0.467);




\node[align=center] at (0.5,-0.1) {\footnotesize Domain};

\node[rotate=90, align=center] at (-0.1,0.5) {\footnotesize Range};

\matrix [draw=black, thin, anchor=north west, row sep=1pt] at (0.05,1.1) {
    \draw [blue, very thick] (0,0) -- (0.2,0); & \node[right] {\footnotesize $\overline{F}$}; \\
    \draw [black, very thick] (0,0) -- (0.2,0); & \node[right] {\footnotesize $H^{\prime}$}; \\
    \draw [red, very thick] (0,0) -- (0.2,0); & \node[right] {\footnotesize $\underline{F}$}; \\
};

\end{tikzpicture}
\caption{Not an Extreme Point, $H^{\prime}$}
\label{fig2b}
\end{subfigure}
\caption{Extreme Points of $\mathcal{I}(\underline{F},\overline{F})$}
\end{figure}
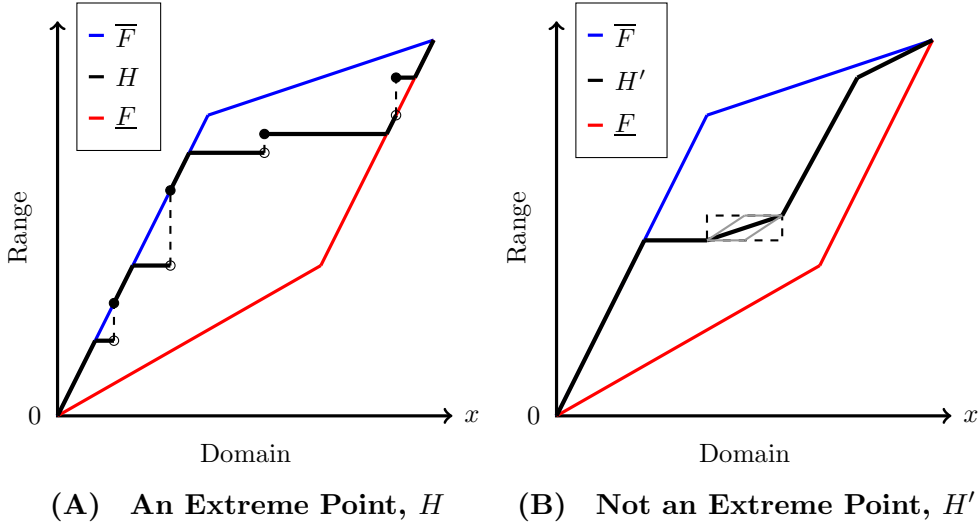

\begin{thm}[Extreme Points of $\mathcal{I}(\underline{F},\overline{F})$]\label{thm1}
For any $\underline{F},\overline{F} \in \mathcal{F}$ such that $\underline{F} \leq \overline{F}$, $H \in \mathcal{F}$ is an extreme point of $\mathcal{I}(\underline{F},\overline{F})$ if and only if there exists a countable collection of intervals $\{[\underline{x}_n,\overline{x}_n)\}_{n=1}^\infty$ such that:
\begin{enumerate}
\item $H(x) \in \{\underline{F}(x),\overline{F}(x)\}$ for all $x \notin \cup_{n=1}^\infty [\underline{x}_n,\overline{x}_n)$.
\item For all $n \in \mathbb{N}$, $H$ is constant on $[\underline{x}_n,\overline{x}_n)$ and either $H(\overline{x}_n^-)=\underline{F}(\overline{x}_n^-)$ or $H(\underline{x}_n)=\overline{F}(\underline{x}_n)$.
\end{enumerate}
\end{thm}
\bref{fig2a} depicts an extreme point $H$ of a monotone function interval $\mathcal{I}(\underline{F},\overline{F})$, where the blue curve is the upper bound $\overline{F}$, and the red curve is the lower bound $\underline{F}$. According to \bref{thm1}, any extreme point $H$ of $\mathcal{I}(\underline{F},\overline{F})$ must either coincide with one of the bounds, or be constant on an interval in its domain, where at least one end of the interval reaches one of the bounds.

Appendix \ref{thm1p} contains the proof of \bref{thm1}. We briefly summarize the argument here. For the sufficiency part, consider any $H$ that satisfies conditions 1 and 2 of \bref{thm1}. Suppose that $H$ can be expressed as a convex combination of two distinct $H_1$ and $H_2$ in $\mathcal{I}(\underline{F},\overline{F})$. Then, for any $x \notin \cup_{n=1}^\infty [\underline{x}_n,\overline{x}_n)$, it must be that $H_1(x)=H_2(x)=H(x)$, since otherwise at least one of $H_1(x)$ and $H_2(x)$ would be either above $\overline{F}(x)$ or below $\underline{F}(x)$. Thus, since $H_1 \neq H_2$, there exists $n \in \mathbb{N}$ such that $H_1(x)\neq H_2(x)$ and $\lambda H_1(x)+(1-\lambda)H_2(x)=H(x)$ for all $x \in [\underline{x}_n,\overline{x}_n)$, for some $\lambda \in (0,1)$. Since $H$ is constant on $[\underline{x}_n,\overline{x}_n)$, and since $H_1$ and $H_2$ are nondecreasing, both $H_1$ and $H_2$ must also be constant on $[\underline{x}_n,\overline{x}_n)$. Suppose that, without loss, $H_1(x)<H(x)<H_2(x)$ for all $x \in [\underline{x}_n,\overline{x}_n)$. If $H(\underline{x}_n)=\overline{F}(\underline{x}_n)$, then $\overline{F}(\underline{x}_n)=H(\underline{x}_n)<H_2(\underline{x}_n)$; whereas if $H(\overline{x}_n^-)=\underline{F}(\overline{x}_n^-)$, then $H_1(\overline{x}_n^-)>H(\overline{x}_n^-)=\underline{F}(\overline{x}_n^-)$. In either case, one of $H_1$ and $H_2$ must not be an element of $\mathcal{I}(\underline{F},\overline{F})$, resulting in a contradiction. 


For the necessity part, consider any $H^{\prime}$ that does not satisfy conditions 1 and 2 of \bref{thm1}. In this case, as depicted in \bref{fig2b}, there exists a rectangle that lies between the graphs of $\underline{F}$ and $\overline{F}$, so that when restricted to this rectangle, the graph of $H^{\prime}$ is not a step function with only one jump. Then, since extreme points of uniformly bounded, nondecreasing functions are step functions with only one jump (see, for example, \citealp{S06,B14}), $H^{\prime}$ can be written as a convex combination of two distinct nondecreasing functions when restricted to this rectangle. Since the rectangle lies in between the graphs of $\underline{F}$ and $\overline{F}$, this, in turn, implies that $H^{\prime}$ can be written as a convex combination of two distinct distributions in $\mathcal{I}(\underline{F},\overline{F})$. 

\begin{rem}
Several assumptions in the setup are for ease of exposition and can be relaxed. First, the domain of $H \in \mathcal{F}$ does not need to be $\mathbb{R}$. \bref{thm1} holds for any monotone function intervals defined on a totally ordered topological space. Second, right-continuity of $H \in \mathcal{F}$ serves as a convention that dictates how a function behaves whenever the function is discontinuous, and is consistent with the natural topology of weak convergence. Lastly, \bref{thm1} can be extended even if the bounds $\lb{F}$ and $\ub{F}$ are nonmonotonic. Indeed, for arbitrary functions $\lb{F},\ub{F}$ such that $\underline{F} \leq \overline{F}$, and for any nondecreasing function $H$, $\lb{F}\leq H \leq \ub{F}$ if and only if $\mathrm{mon}_+(\lb{F})\leq H \leq \mathrm{mon}_-(\ub{F})$, where $\mathrm{mon}_+(\lb{F})$ is the smallest nondecreasing function above $\lb{F}$ and $\mathrm{mon}_-(\ub{F})$ is the largest monotone function below $\ub{F}$.

It is also noteworthy that \bref{thm1} extends to the case where one of the two bounds $\overline{F}$, $\underline{F}$ equals $\pm\infty$, respectively. Consider when $\ub{F}$ is the bound that takes a finite value for all $x$ and $\lb{F}=-\infty$. (The other case follows symmetrically.) Then, $H$ is an extreme point of $\mathcal{I}(\lb{F},\ub{F})$ if and only if there exists a countable collection of intervals $\{[\underline{x}_n,\overline{x}_n)\}_{n=1}^\infty$ such that $H(x)=\ub{F}(x)$ for all $x \notin \cup_{n=1}^\infty [\underline{x}_n,\overline{x}_n)$ and $H(x)=\ub{F}(\underline{x}_n)$ for all $x \in [\underline{x}_n,\overline{x}_n)$ and for all $n$.  
\end{rem}

In the ensuing sections, we demonstrate how the characterization of the extreme points of monotone function intervals can be applied to various economic settings. These applications rely on two crucial properties of extreme points. The first property---formally known as Choquet's theorem---allows one to express any element $H$ of $\mathcal{I}(\underline{F},\overline{F})$ as a convex combination of its extreme points if $\mathcal{I}(\lb{F},\ub{F})$ is compact. As a result, if one wishes to establish some property for every element of $\mathcal{I}(\underline{F},\overline{F})$, and if this property is preserved under convex combinations, then it suffices to establish the property for all extreme points of $\mathcal{I}(\underline{F},\overline{F})$, which is a much smaller set. \bref{s3} uses this first property to characterize the set of distributions of posterior quantiles. The second property of extreme points is that, for any convex optimization problem, one of the solutions must be an extreme point of the feasible set. This property is useful for economic applications because it immediately provides knowledge about the solutions to the underlying economic problem if that problem is convex and if the feasible set is related to a monotone function interval. \bref{s4} uses this second property to analyze security design.

\section{Distributions of Posterior Quantiles}
\label{s3}

\bref{thm1} alongside Choquet's theorem leads to the characterization of the set of distributions of posterior quantiles. This characterization is an analog of the celebrated characterization of the set of distributions of posterior means that follows from Strassen's theorem \citep{S65}. Quantiles are important in settings where only the ordinal values or relative rankings of the relevant variables are meaningful, rather than the cardinal values or numeric differences (e.g., voting, grading or rating schemes, measures of potential losses such as the value-at-risk), or in settings where moments are not well-defined (e.g., finance or insurance). In this regard, the characterization of the set of distributions of posterior quantiles is useful for identifying possible outcomes from a signal (e.g., posterior value-at-risk that arises from a signal), as well as optimal policies (e.g., optimal voter signals in an election) in these settings.

\subsection{Characterization of the Distributions of Posterior Quantiles}
Let $\mathcal{F}_0 \subseteq \mathcal{F}$ be the collection of cumulative distribution functions (CDFs) in $\mathcal{F}$.\footnote{That is, $G \in \mathcal{F}_0$ if and only if $G \in \mathcal{F}$ and $\lim_{x \to \infty}G(x)=1$ and $\lim_{x \to -\infty}G(x)=0$.} Consider a one-dimensional state $\omega \in  \mathbb{R}$ that is drawn from a prior $F$. A \emph{signal} consists of a set of signal realizations $S$ and a joint distribution over $\mathbb{R} \times S$ whose marginal over $\mathbb{R}$ equals $F$. From Blackwell's theorem \citep*{B53,S65}, a signal can be represented by a distribution $\mu \in \Delta(\mathcal{F}_0)$ over posteriors  such that 
\begin{equation}\label{bp}
\int_{\mathcal{F}_0}G(\omega)\mu(\diff G)=F(\omega),    
\end{equation}
for all $\omega \in \mathbb{R}$. Let $\mathcal{M}$ denote the collection of all such distributions. For any $\mu \in \mathcal{M}$, $G \in \mathrm{supp}(\mu)$ can be regarded as the posterior belief after observing a signal realization.\footnote{More precisely, $G$ is a version of the regular conditional distribution of $x$ conditional on a signal realization.} 

For any CDF $G \in \mathcal{F}_0$ and for any $\tau \in (0,1)$, denote the set of $\tau$-quantiles of $G$ by $[G^{-1}(\tau),G^{-1}(\tau^+)]$, where $G^{-1}(\tau):=\inf\{x \in \mathbb{R}|G(x) \geq \tau\}$ is the \emph{quantile function} of $G$ and $G^{-1}(\tau^+):=\lim_{q \downarrow \tau} G^{-1}(q)$ denotes the right-limit of $G^{-1}$ at $\tau$.\footnote{Note that $G^{-1}$ is nondecreasing and left-continuous for all $G \in \mathcal{F}_0$. Moreover, for any $\tau \in (0,1)$ and for any $\omega \in \mathbb{R}$, $G^{-1}(\tau) \leq \omega$ if and only if $G(\omega) \geq \tau$.} Since the $\tau$-quantile for an arbitrary CDF may not be unique, we further introduce a notation for selecting a quantile. We say that a transition probability $r:\mathcal{F}_0 \to \Delta(\mathbb{R})$ is a \emph{$\tau$-quantile selection rule} if, for all $G \in \mathcal{F}_0$, $r(\cdot|G)$ assigns probability 1 to $[G^{-1}(\tau),G^{-1}(\tau^+)]$. In other words, a quantile selection rule $r$ selects (possibly through randomization) a $\tau$-quantile of $G$, for every CDF $G$, whenever it is not unique. Let $\mathcal{R}_\tau$ be the collection of all $\tau$-quantile selection rules. 

For any $\tau \in (0,1)$, for any signal $\mu \in \mathcal{M}$, and for any selection rule $r \in \mathcal{R}_\tau$, let $H^\tau(\cdot|\mu,r)$ denote the distribution of the $\tau$-quantile induced by $\mu$ and $r$. For any $\tau \in (0,1)$, let $\mathcal{H}_\tau$ denote the set of distributions of posterior $\tau$-quantiles that can be induced by some signal $\mu \in \mathcal{M}$ and selection rule $r \in \mathcal{R}_\tau$. 

Using \bref{thm1}, we provide a complete characterization of the set of distributions of posterior quantiles induced by arbitrary signals and selection rules. To this end, define two distributions, $F_L^\tau$ and $F_R^\tau$, as follows:
\[
F_L^\tau(\omega):=\min\left\{\frac{1}{\tau}F(\omega),1\right\}, \quad F_R^\tau(\omega):=\max\left\{\frac{F(\omega)-\tau}{1-\tau},0\right\}.
\]
Note that $F_R^\tau \leq F_L^\tau$ for all $\tau \in (0,1)$. In essence, $F_L^\tau$ is the \emph{left-truncation} of the prior $F$: the conditional distribution of $F$ in the event that $\omega$ is smaller than a $\tau$-quantile of $F$; whereas $F_R^\tau$ is the \emph{right-truncation} of $F$: the conditional distribution of $F$ in the event that $\omega$ is larger than the same $\tau$-quantile. \bref{thm2} below characterizes the set of distributions of posterior quantiles $\mathcal{H}_\tau$.

\begin{thm}[Distributions of Posterior Quantiles]\label{thm2}
For any $\tau \in (0,1)$, 
\[
\mathcal{H}_\tau=\mathcal{I}(F_R^\tau,F_L^\tau).
\]
\end{thm}

\begin{figure}[ht!]
    \begin{center}
        \input{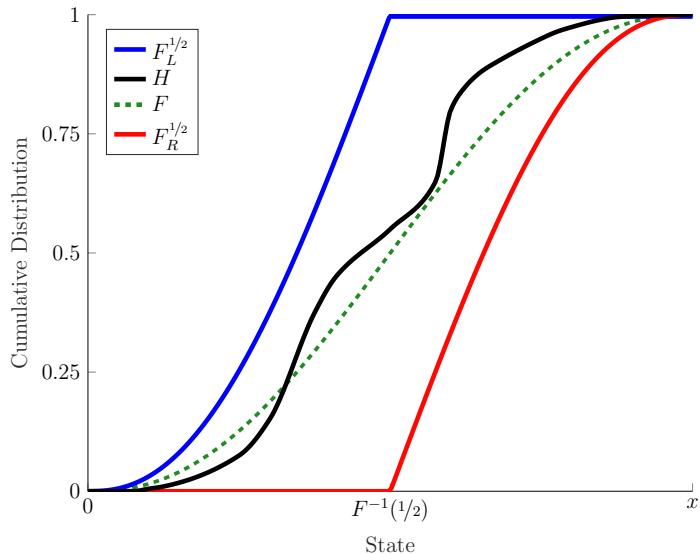}
    \end{center}
    \caption{Distributions of Posterior Medians}
    \label{fig:distributions}
\end{figure}

\bref{thm2} characterizes the set of distributions of posterior $\tau$-quantiles by the monotone function interval $\mathcal{I}(F_R^\tau,F_L^\tau)$. Notice that, because $F_R^\tau$ and $F_L^\tau$ are CDFs, their pointwise dominance relation means that $F_R^\tau$ first-order stochastically dominates $F_L^\tau$. \bref{fig:distributions} illustrates \bref{thm2} for the case when $\tau=\nicefrac{1}{2}$. The distribution $F_L^{\nicefrac{1}{2}}$ is colored blue, whereas the distribution $F_R^{\nicefrac{1}{2}}$ is colored red. The green dotted curve represents the prior, $F$. According to \bref{thm2}, any distribution $H$ bounded by $F_L^{\nicefrac{1}{2}}$ and $F_R^{\nicefrac{1}{2}}$ (for instance, the black curve in the figure) can be induced by a signal $\mu \in \mathcal{M}$ and a selection rule $r \in \mathcal{R}_{\nicefrac{1}{2}}$. Conversely, for any signal and for any selection rule, the graph of the induced distribution of posterior $\tau$-quantiles must fall in the area bounded by the blue and red curves. For example, under the signal that reveals all the information, the distribution of posterior $\nicefrac{1}{2}$-quantiles coincides with the prior, whereas under the signal that does not reveal any information, the distribution of posterior $\nicefrac{1}{2}$-quantiles coincides with the step function that has a jump (of size 1) at $F^{-1}(\nicefrac{1}{2})$.  

\bref{thm2} can be regarded as a natural analog of the well-known characterization of the set of distributions of posterior \emph{means} that follows from \citet{S65}. Strassen's theorem implies that a CDF  $H \in \mathcal{F}_0$ is a distribution of posterior means if and only if $H$ is a mean-preserving contraction of the prior $F$. Instead of posterior means, \bref{thm2} pertains to posterior quantiles. According to \bref{thm2}, $H$ is a distribution of posterior $\tau$-quantiles if and only if $H$ first-order stochastically dominates the left-truncation $F_L^\tau$ and is dominated by the right-truncation $F_R^\tau$. 

\begin{figure}
\centering
\tikzset{
solid node/.style={circle,draw,inner sep=1.25,fill=gray2},
hollow node/.style={circle,draw,inner sep=1.25}
}
\begin{subfigure}[b]{0.4\linewidth}
\begin{tikzpicture}[scale=5]
\draw [<->, very thick] (0,1.05) node (yaxis) [above] {\footnotesize }
        |- (1.05,0) node (xaxis) [right] {\footnotesize$\omega$};
\draw (0,0) node [left=2pt] {\footnotesize$0$};
\draw (0,1) node [left=2pt] {\footnotesize$1$};
\draw [blue, very thick] (0,0)--(0.5,1);
\draw [blue, very thick] (0.5,1)--(1,1);
\draw [red, very thick] (0,0)--(0.5,0);
\draw [red, very thick] (0.5,0)--(1,1);
\draw [ultra thick] (0,0)--(0.25,0.5);
\draw [ultra thick] (0.25,0.5)--(0.75,0.5);
\draw [ultra thick] (0.75,0.5)--(1,1);
\draw (0.25,0) node [below=2pt] {\footnotesize $\underline{x}$};
\draw (0.75,0) node [below=2pt] {\footnotesize $\overline{x}$};
\draw [dotted] (0.25,0.5)--(0.25,0);
\draw [dotted] (0.75,0.5)--(0.75,0);


\node[rotate=90, align=center] at (-0.1,0.5) {\footnotesize Cumulative \footnotesize Distribution};

\matrix [draw=black, thin, anchor=north west, row sep=1pt] at (0.05,1.1) {
    \draw [blue, very thick] (0,0) -- (0.2,0); & \node[right] {\footnotesize $F_L^{\tau}$}; \\
    \draw [black, very thick] (0,0) -- (0.2,0); & \node[right] {\footnotesize $H$}; \\
    \draw [red, very thick] (0,0) -- (0.2,0); & \node[right] {\footnotesize $F_R^{\tau}$}; \\
};

\end{tikzpicture}
\caption{An Extreme Point $H$}
\label{fig4a}
\end{subfigure}%
\begin{subfigure}[b]{0.4\linewidth}
\begin{tikzpicture}[scale=5]
\draw [<->, very thick] (0,1.05) node (yaxis) [above] {\footnotesize }
        |- (1.05,0) node (xaxis) [right] {\footnotesize$\omega$};
\draw (0,0) node [left=2pt] {\footnotesize$0$};
\draw (0,1) node [left=2pt] {\footnotesize$1$};
\draw [gray2, ultra thick] (0,0)--(0.125,0);
\draw [gray2, dashed, thick] (0.125,0)--(0.125,0.5);
\draw [gray2, ultra thick] (0.125,0.5)--(0.25,0.5);
\draw [gray2, ultra thick] (0.25,0.5)--(0.75,1);
\draw [gray2, ultra thick] (0.75,1)--(1,1);
\draw [dotted] (0.25,0.5)--(0.25,0);
\draw [dotted] (0.75,1)--(0.75,0);
\draw (0.25,0) node [below=2pt] {\footnotesize $\underline{x}$};
\draw (0.75,0) node [below=2pt] {\footnotesize $\overline{x}$};
\draw (0.125,0) node [below=2pt] {\footnotesize $\hat{x}$};
\draw (0.125,0) node [hollow node] {};
\draw (0.125,0.5) node [solid node] {};
\draw [dotted] (0.125,0.5)--(0.125,0);
\draw [dotted] (0.125,0.5)--(0,0.5);
\draw (0,0.5) node [left=2pt] {\footnotesize $\tau$};

\node[rotate=90, align=center] at (-0.15,0.5) {\footnotesize Cumulative \footnotesize Distribution};


\matrix [draw=black, thin, anchor=north west, row sep=1pt] at (0.05,1.1) {
    \draw [gray2, very thick] (0,0) -- (0.2,0); & \node[right] {\footnotesize Posterior}; \\
};
\end{tikzpicture}
\caption{Posterior Containing $\hat{x}$}
\label{fig4b}
\end{subfigure}
\caption{Constructing a Signal that Induces $H$}
\end{figure}
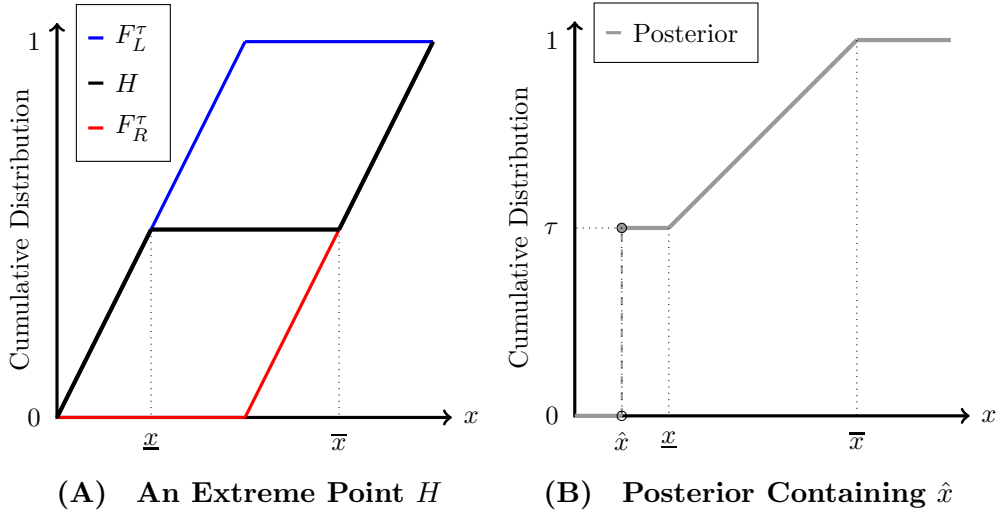

The fact that $\mathcal{H}_\tau \subseteq \mathcal{I}(F_R^\tau,F_L^\tau)$ follows from the martingale property of posterior beliefs. For the converse (i.e., $\mathcal{I}(F_R^\tau,F_L^\tau) \subseteq \mathcal{H}_\tau$), \bref{thm1} and Choquet's theorem imply that it suffices, for each $H$ satisfying conditions 1 and 2, to construct a signal (and a selection rule) that induces $H$  as its distribution of posterior quantile. The proof of \bref{thm2} in Appendix \ref{thm2p} explicitly constructs such signals (and selection rules). To see the intuition, consider an extreme point $H$ of $\mathcal{I}(F_R^\tau, F_L^\tau)$ that takes the following form: 
\[
H(x)=\left\{
\begin{array}{cc}
F_L^\tau(x),&\mbox{if } x< \underline{x}\\
F_L^\tau(\underline{x}),&\mbox{if } x \in [\underline{x},\overline{x})\\
F_R^\tau(x),&\mbox{if } x \geq \overline{x}
\end{array}
\right.,
\]
for some $\underline{x},\overline{x}$ such that $F_L^\tau(\lb{x})=F_R^\tau(\ub{x}^-)$, as depicted in \bref{fig4a}. To construct a signal that has $H$ as its distribution of posterior quantiles, separate all the states $x \notin [\underline{x},\overline{x}]$. Then, take $\alpha$ fraction of the states in $[\underline{x},\overline{x}]$ and pool them uniformly with each separated state below $\underline{x}$, while pooling the remaining $1-\alpha$ fraction uniformly with the separated states above $\overline{x}$.  Since $F_L^\tau(\underline{x})=F_R^\tau(\ub{x}^-)$, when  $\alpha$ is chosen correctly, 
each $x < \underline{x}$, after being pooled with states in $[\underline{x},\overline{x}]$, would become a $\tau$-quantile of the posterior it belongs to, as illustrated in \bref{fig4b}. Similarly, each $x>\ub{x}$ would become a $\tau$-quantile of the posterior it belongs to. Together, by properly selecting the posterior quantiles, the induced distribution of posterior quantiles under this signal would indeed be $H$. 

\paragraph{Distributions of Unique Posterior Quantiles.}
Although the characterization of \bref{thm2} may seem to rely on selection rules $r \in \mathcal{R}_\tau$, the result remains (essentially) the same even when restricted to signals that always induce a unique posterior $\tau$-quantile, provided that the prior $F$ has full support on an interval. \bref{thm3} below formalizes this statement. To this end, let $\widetilde{\mathcal{H}}_\tau\subseteq \mathcal{H}_\tau$ be the set of distributions of posterior $\tau$-quantiles that can be induced by some signal where (almost) all posteriors have a unique $\tau$-quantile. The characterization of $\widetilde{\mathcal{H}}_\tau$ relates to a family of perturbations of the set $\mathcal{I}(F_R^\tau,F_L^\tau)$, denoted by $\{\mathcal{I}(F_R^{\tau,\varepsilon},F_L^{\tau,\varepsilon})\}_{\varepsilon>0}$, where 
\[
F_L^{\tau,\varepsilon}(x):=\left\{
\begin{array}{cc}
\frac{1}{\tau+\varepsilon}F(x),&\mbox{if } x <F^{-1}(\tau)\\
1,&\mbox{if } x \geq F^{-1}(\tau)
\end{array}
\right.;  \mbox{ and } 
F_R^{\tau,\varepsilon}(x):=\left\{
\begin{array}{cc}
0,&\mbox{if } x<F^{-1}(\tau)\\
\frac{F(x)-(\tau-\varepsilon)}{1-(\tau-\varepsilon)},&\mbox{if } x \geq F^{-1}(\tau)
\end{array}
\right.,
\]
for all $\varepsilon\geq 0$ and for all $x \in \mathbb{R}$. Note that $\mathcal{I}(F_R^{\tau,0},F_L^{\tau,0})=\mathcal{I}(F_R^\tau,F_L^\tau)$, and $\{\mathcal{I}(F_R^{\tau,\varepsilon},F_L^{\tau,\varepsilon})\}_{\varepsilon>0}$ is decreasing in $\varepsilon$ under the set-inclusion order.\footnote{As a convention, let $\mathcal{I}(F_R^{\tau,\varepsilon},F_L^{\tau,\varepsilon}) :=\emptyset$ when $\varepsilon \geq \max\{\tau,1-\tau\}$.} 

\begin{thm}[Distributions of Unique Posterior Quantiles]\label{thm3}
For any $\tau \in (0,1)$ and for any $F \in \mathcal{F}_0$ that has a full support on an interval,
\[
\bigcup_{\varepsilon>0} \mathcal{I}(F_R^{\tau,\varepsilon},F_L^{\tau,\varepsilon}) \subseteq \widetilde{\mathcal{H}}_\tau\subseteq \mathcal{I}(F_R^\tau,F_L^\tau).
\]
\end{thm}
According to \bref{thm3}, for any $\varepsilon>0$ and for any $H \in \mathcal{I}(F_R^{\tau,\varepsilon},F_L^{\tau,\varepsilon})$, there exists a signal $\mu$ such that $H$ is the distribution of \emph{unique} posterior $\tau$-quantiles. As a result, the set of distributions of \emph{unique} posterior quantiles is given by the ``interior'' of $\mathcal{I}(F_R^\tau,F_L^\tau)$, and only the ``boundaries'' of $\mathcal{I}(F^\tau_R,F^\tau_L)$ (such as $F_R^\tau$ and $F_L^\tau$ themselves) are lost by requiring uniqueness. In other words, the set of distributions of unique posterior quantiles contains an \emph{open dense} subset of $\mathcal{I}(F_R^\tau,F_L^\tau)$. 

\paragraph{Law of Iterated Quantiles.}As an immediate corollary of \bref{thm2} and \bref{thm3}, an analog of the law of iterated expectations emerges, which we refer to as the \emph{law of iterated quantiles}. 

\begin{cor}[Law of Iterated Quantiles]\label{lit}
Consider any $\tau,q \in (0,1)$. 
\begin{enumerate}
\item For any closed interval $Q \subseteq \mathbb{R}$, $Q=[H^{-1}(\tau),H^{-1}(\tau^+)]$ for some $H \in \mathcal{H}_{q}$ if and only if $Q \subseteq [(F_R^{q})^{-1}(\tau),(F_L^{q})^{-1}(\tau^+)]$.
\item Suppose that the prior $F$ is continuous and has full support on an interval. Then for any $\hat{x} \in \mathbb{R}$, $\hat{x} \in [H^{-1}(\tau), H^{-1}(\tau^+)]$ for some $H \in \widetilde{H}_q$ if and only if $\hat{x} \in [(F_R^q)^{-1}(\tau),(F_L^{q})^{-1}(\tau)]$.
\end{enumerate}
\end{cor}

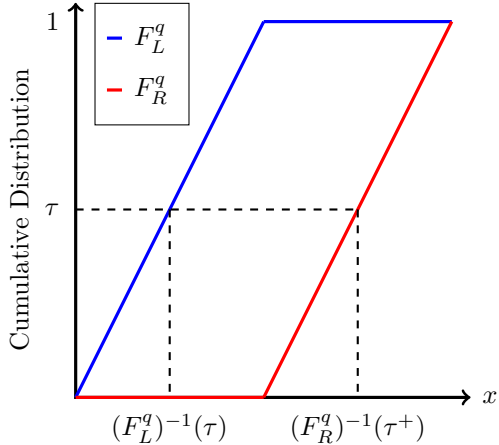
\begin{figure}[h!]
\centering
\tikzset{
Two node styles for game trees: solid and hollow
solid node/.style={circle,draw,inner sep=1,fill=black},
hollow node/.style={circle,draw,inner sep=1}
}
\begin{tikzpicture}[scale=5]
\draw [<->, very thick] (0,1.05) node (yaxis) [above] {\footnotesize }
        |- (1.05,0) node (xaxis) [right] {\footnotesize$\omega$};
        \draw (1,0) node [below=2pt] {\footnotesize$ $};
\draw (0,1) node [left=2pt] {\footnotesize$1$};
\draw [blue, very thick] (0,0)--(0.5,1);
\draw [blue, very thick] (0.5,1)--(1,1); 
\draw [red, very thick] (0,0)--(0.5,0);
\draw [red, very thick] (0.5,0)--(1,1);
\draw [dashed, thick] (0,0.5)--(0.75,0.5);
\draw (0,0.5) node [left=2pt] {\footnotesize$\tau$};
\draw [dashed, thick] (0.25,0.5)--(0.25,0);
\draw [dashed, thick] (0.75,0.5)--(0.75,0); 
\draw (0.25,0) node [below=2pt] {\footnotesize $({F}_L^{q})^{-1}(\tau)$};
\draw (0.75,0) node [below=2pt] {\footnotesize $(F_R^{q})^{-1}(\tau^+)$};


\node[rotate=90, align=center] at (-0.15,0.5) {\footnotesize Cumulative \footnotesize Distribution};

\matrix [draw=black, thin, anchor=north west, row sep=1pt] at (0.05,1.05) {
    \draw [blue, very thick] (0,0) -- (0.2,0); & \node[right] {\footnotesize $F_L^{q}$}; \\
    \draw [red, very thick] (0,0) -- (0.2,0); & \node[right] {\footnotesize $F_R^{q}$}; \\
};

\end{tikzpicture}
\caption{Law of Iterated Quantiles}
\label{fig6}
\end{figure}

The intuition of \bref{lit} is summarized in \bref{fig6}. For any $q,\tau \in (0,1)$, \bref{fig6} plots the interval $\mathcal{I}(F_R^q,F_L^q)$, which, according to \bref{thm2} (and \bref{thm3}), equals all possible distributions of posterior $q$-quantiles. Therefore, the $\tau$-quantiles of posterior $q$-quantiles must coincide with the interval $[(F_L^{q})^{-1}(\tau),(F_R^{q})^{-1}(\tau^+)]$. According to \bref{lit}, while the expectation of posterior means under any signal is always the expectation under the prior, the possible $\tau$-quantiles of posterior $q$-quantiles are exactly $[(F_L^{q})^{-1}(\tau),(F_R^{q})^{-1}(\tau^+)]$. For example, the collection of all possible \emph{medians} of posterior \emph{medians} is the \emph{interquartiles} $[F^{-1}(\nicefrac{1}{4}),F^{-1}(\nicefrac{3}{4}^+)]$ of the prior.

\subsection{Economic Applications}
\label{sec:quantile_applications}

In what follows, we illustrate economic applications of \bref{thm2} and \bref{thm3} through three examples. In these examples, only the ranking---instead of cardinal values---of outcomes are relevant. The first application is to gerrymandering: here, citizens rank candidates' positions relative to their own ideal positions, and the median voter theorem determines who is elected. The second application is to Bayesian persuasion when payoffs depend only on posterior quantiles. The third application is to apparent misconfidence, which explains why people rank themselves better or worse than others.

\paragraph{Limits of Gerrymandering}
\mbox{}\\
We first apply \bref{thm2} and \bref{thm3} to gerrymandering. Existing economic theory on gerrymandering has primarily focused on optimal redistricting or fair redistricting mechanisms (e.g., \citealp{owen1988optimal,friedman2008optimal,gul2010strategic, pegdenetal2017,ely2019cake,friedman2020optimal,kolotilin2023economics}). Another fundamental question is the scope of gerrymandering's impact on a legislature. If \emph{any} electoral map can be drawn, what kinds of legislatures can be created? In other words, what are the ``limits of gerrymandering''? 

\bref{thm2} and \bref{thm3} can be used to answer this question. Consider an environment in which a continuum of citizens vote, and each citizen has single-peaked preferences over positions on political issues. Citizens have different ideal positions $\omega \in \mathbb{R}$, and these positions are distributed according to some $F \in \mathcal{F}_0$. In this setting, a signal $\mu \in \mathcal{M}$ can be thought of as an electoral \emph{map}, which segments citizens into electoral \emph{districts}, such that a district $G \in \mathrm{supp}(\mu)$ is described by the distribution of the ideal positions of citizens who belong to it. Each district elects a \emph{representative}, and election results at the district-level follow the median voter theorem. That is, given any map $\mu \in \mathcal{M}$, the elected representative of each district $G \in \mathrm{supp}(\mu)$ must have an ideal position that is a median of $G$. When there are multiple medians in a district, the representative's ideal position is determined by a selection rule $r \in \mathcal{R}_{\nicefrac{1}{2}}$, which is either flexible or stipulated by election laws.

Given any $\mu \in \mathcal{M}$ and any selection rule $r \in \mathcal{R}_{\nicefrac{1}{2}}$, the induced distribution of posterior medians $H^{\nicefrac{1}{2}}(\cdot|\mu,r)$ can be interpreted as the distribution of the ideal positions of the elected representatives. Meanwhile, the bounds $F_L^{\nicefrac{1}{2}}$ and $F_R^{\nicefrac{1}{2}}$ can be interpreted as distributions of representatives that only reflect one side of voters' political positions relative to the median of the population. Specifically, $F_L^{\nicefrac{1}{2}}$ describes an ``all-left'' legislature, which only reflects citizens' ideal positions that are left of the population median. Likewise, $F_R^{\nicefrac{1}{2}}$ represents an ``all-right'' legislature, which only reflects citizens' ideal positions that are right of the population median. As an immediate implication of \bref{thm2} and \bref{thm3}, \bref{prop1} below characterizes the set of possible compositions of the legislature across all election maps.

\begin{prop}[Limits of Gerrymandering]\label{prop1}
For any $H \in \mathcal{F}_0$, the following are equivalent: 
\begin{enumerate}
\item $H \in \mathcal{I}(F_R^{\nicefrac{1}{2}},F_L^{\nicefrac{1}{2}})$.
\item $H$ is a distribution of the representatives' ideal positions under some map $\mu \in \mathcal{M}$ and some selection rule $r \in \mathcal{R}_{\nicefrac{1}{2}}$. 
\end{enumerate}
If, furthermore, $F \in \mathcal{F}_0$ has full support on an interval, then for any fixed selection rule $\hat{r} \in \mathcal{R}_{\nicefrac{1}{2}}$, every $H \in \cup_{\varepsilon>0} \mathcal{I}(F_R^{\nicefrac{1}{2},\varepsilon},F_L^{\nicefrac{1}{2},\varepsilon})$ is a distribution of the representatives' ideal positions under some map $\mu \in \mathcal{M}$ and selection rule $\hat{r}$.
\end{prop}

\bref{prop1} characterizes the \emph{compositions} of the legislature that gerrymandering can induce. According to \bref{prop1}, \emph{any} composition of the legislative body ranging from the ``all-left'' to the ``all-right'' can be created by some gerrymandered map. In other words, gerrymandering can produce a wide range of legislative bodies that differ from the population distribution, including legislatures that represent only one side of the population median, as well as any legislature ``in between,'' in the sense of first-order stochastic dominance. Meanwhile, the ``all-left'' and ``all-right'' bodies also identify the limits of the scope of gerrymandering: any composition that is more extreme than the ``all-left'' or the ``all-right'' bodies is not possible, regardless of how districts are drawn.\footnote{\citet*{gomberg2021electoral} also study how gerrymandering affects the composition of the legislature. However, the authors assume that each district elects a \emph{mean} candidate as opposed to the median.}

If we further specify the model for the legislature to enact legislation, we may explore the set of possible \textit{legislative outcomes} that can be enacted. One natural assumption for the outcomes, regardless of the details of the legislative model, is that the enacted legislation must be a median of the representatives (i.e., the median voter property holds at the legislative level).\footnote{See \citet{mccarty2001hunt}, \citet{bradbury2005legislative}, and \citet{krehbiel2010pivotal} for evidence that the median legislator is decisive. See also \citet*{cho2009bargaining} for a microfoundation.} Under this assumption, an immediate implication of \bref{lit} is that the set of achievable legislative outcomes  coincides with the interquartile range of the citizenry's ideal positions, as summarized by \bref{gerryoutcome} below.

\begin{cor}[Limits of Legislative Outcomes]\label{gerryoutcome}
Suppose that the median voter property holds both at the district and legislative level. Then an outcome $x \in \mathbb{R}$ can be enacted as legislation under some map if and only if $x \in [F^{-1}(\nicefrac{1}{4}),F^{-1}(\nicefrac{3}{4}^+)]$. 
\end{cor}

According to \bref{gerryoutcome}, while the only Condorcet winners in this setting are the population medians, gerrymandering expands the set of possible legislation to the entire interquartile range of the population's views. Conversely, \bref{gerryoutcome} also suggests it is impossible to enact any legislative outcome \emph{beyond} the interquartile range, regardless of how the districts are drawn. Studying these downstream effects of gerrymandering on enacted legislation is less common in the political economy literature, which tends to stop at the solution of an optimal map. Work that has examined possible legislation under gerrymandering typically focuses on``policy bias,'' which is the gap between majority rule (i.e., the ideal point of the population's median voter) and the ultimate policy that could come out of the legislature under some gerrymandered map \citep{shotts2002gerrymandering,buchler2005competition,gilligan2006public}. \bref{gerryoutcome} unifies existing results on bounding the potential magnitude of policy bias.

Furthermore, as the population becomes more polarized, so that the interquartile range becomes wider, more extreme legislation can pass. For instance, consider two population distributions, $F$ and $\widetilde{F}$, with the same unique median $x^*$, and suppose that $\widetilde{F}$ is more dispersed than $F$ under the rotation order around the common median. That is, $F(x)\geq \widetilde{F}(x)$ for all $x>x^*$ and $F(x)\leq\widetilde{F}(x)$ for all $x<x^*$. Then, it must be that $\widetilde{F}^{-1}(\nicefrac{1}{4}) \leq F^{-1}(\nicefrac{1}{4}) \leq F^{-1}(\nicefrac{3}{4}^+) \leq \widetilde{F}^{-1}(\nicefrac{3}{4}^+)$. By \bref{gerryoutcome}, it then follows that the range of legislation that can be enacted becomes wider as the population distribution becomes more dispersed.

\begin{rem}[Districts on a Geographic Map]
    \label{rem:2d}
    In practice, election districts are drawn on a geographic map. Drawing districts in this manner can be regarded as partitioning a two-dimensional space that is spanned by latitude and longitude. More specifically, let a convex and compact set $\Theta \subseteq [0,1]^2$ denote a geographic map. Suppose that every citizen who resides at the same location $\theta \in \Theta$ shares the same ideal position $\bm{\omega}(\theta)$, where $\bm{\omega}:\Theta \to \mathbb{R}$ is a measurable function. Furthermore, suppose that citizens are distributed on $\Theta$ according to a density function $\phi>0$. Under this setting, Theorem 1 of \citet*{Yang2020generating} ensures that for any $\mu \in \mathcal{M}$ with a countable support, there exists a countable partition of $\Theta$, such that the distributions of citizens' ideal positions within each element coincide with the distributions in the support of $\mu$. If we further assume that $\bm{\omega}$ is non-degenerate, in the sense that each of its indifference curves $\{\theta \in \Theta|\bm{\omega}(\theta)=\omega\}_{\omega \in \mathbb{R}}$ is isomorphic to the unit interval, then Theorem 2 of \citet*{Yang2020generating} ensures that for any $\mu \in \mathcal{M}$, there exists a partition on $\Theta$ that generates the same distributions in each district. Therefore, the splitting of the distribution of citizens' ideal positions has an exact analog to the splitting of geographic areas on a physical map.
\end{rem}


\bref{prop1} can not only characterize the set of feasible maps based on the citizenry's distribution of ideal positions, but can also help identify that distribution itself. Suppose that $H$ is the observed distribution of representatives' ideal positions. \bref{prop1} implies that the population distribution $F$ must have $H$ be dominated by $F_R^{\nicefrac{1}{2}}$ and dominate $F_L^{\nicefrac{1}{2}}$ at the same time. This leads to \bref{gerryidentify} below.  

\begin{cor}[Identification Set of $F$]\label{gerryidentify}
Suppose that $H \in \mathcal{F}_0$ is the distribution of ideal positions of a legislature. Then the distribution of citizens' ideal positions $F$ must satisfy 
\begin{equation}\label{gerryideq}
\frac{1}{2}H(x) \leq F(x) \leq \frac{1+H(x)}{2},
\end{equation}
for all $x \in \mathbb{R}$. Conversely, for any $F \in \mathcal{F}_0$ satisfying \eqref{gerryideq}, there exists a map $\mu \in \mathcal{M}$ and a selection rule $r \in \mathcal{R}_{\nicefrac{1}{2}}$, such that $H$ is the distribution of ideal positions of the legislature. 
\end{cor}
According to \bref{gerryidentify}, the distribution of citizens' ideal positions can be partially identified by \eqref{gerryideq}, even when only the distribution of the representatives' ideal positions can be observed.

\paragraph{Quantile-Based Persuasion}
\mbox{}\\
\noindent Naturally, \bref{thm2} and \bref{thm3} can also be applied to a Bayesian persuasion setting where the sender's indirect utility depends only on posterior quantiles. Consider the Bayesian persuasion problem formulated by \citet{KG11}: A state $\omega \in \mathbb{R}$ is distributed according to a common prior $F$. A sender chooses a signal $\mu \in \mathcal{M}$ to inform a receiver, who then picks an action $a \in A$ after seeing the signal's realization. The ex-post payoffs of the sender and receiver are $u_S(\omega,a)$ and $u_R(\omega,a)$, respectively. \citet*{KG11} show that the sender's optimal signal and the value of persuasion can be characterized by the concave closure of the function $\hat{v}:\mathcal{F}_0 \to \mathbb{R}$, where $\hat{v}(G):=\mathbb{E}_F[u_S(\omega,a^*(G))]$ is the indirect utility of the sender, and $a^*(G) \in A$ is the optimal action of the receiver under posterior $G \in \mathcal{F}_0$ that the sender prefers the most.

When $|\mathrm{supp}(F)| > 2$, this ``concavafication'' method requires finding the concave closure of a multi-variate function, which is known to be computationally challenging, especially when $|\mathrm{supp}(F)| =\infty$.\footnote{A recent elegant contribution by \citet*{KCW22} also provides a tractable method that simplifies these persuasion problems and more using optimal transport.}  For tractability, many papers have restricted attention to preferences where the only payoff-relevant statistic of a posterior is its mean (i.e., $\hat{v}(G)$ is measurable with respect to $\mathbb{E}_G[\omega]$). See, for example, \citet*{GK16}, \citet*{Ketal17}, \citet*{DM19}, \citet*{KMZ22}, and \citet*{ABSY22}. 

A natural analog of this ``mean-based'' setting is for the payoffs to depend only on the posterior quantiles. Just as mean-based persuasion problems are tractable because distributions of posterior means are mean-preserving contractions of the prior, \bref{thm2} and \bref{thm3} provide a tractable formulation of any ``quantile-based'' persuasion problem, as described in \bref{qp} below.  

\begin{prop}[Quantile-Based Persuasion]\label{qp}
Suppose that the prior $F$ has full support on some interval, and suppose that there exists $\tau \in (0,1)$, a selection rule $r \in \mathcal{R}_\tau$, and a measurable function $v_S:\mathbb{R} \to \mathbb{R}$ such that $\hat{v}(G)=\int_\mathbb{R} v_S(\omega)r(\diff x|G)$, for all $G \in \mathcal{F}_0$. Then 
\begin{equation}\label{eq:simplified_bayesian_persuasion}
\mathrm{cav} (\hat{v})[F]= \sup_{H \in \mathcal{I}(F_R^\tau, F_L^\tau)} \int_\mathbb{R}v_S(\omega)H(\diff \omega).
\end{equation}
\end{prop}

By \bref{qp}, any $\tau$-quantile-based persuasion problem can be solved by simply choosing a distribution in $\mathcal{I}(F_R^\tau,F_L^\tau)$ to maximize the expected value of $v_S(\omega)$, rather than concavafying the infinite-dimensional functional $\hat{v}$. Furthermore, since the objective function of \eqref{eq:simplified_bayesian_persuasion} is affine, \bref{thm1} further reduces the search for the solution to only distributions that satisfy its conditions 1 and 2.

As an immediate application, \bref{qp} sheds light on the structure of optimal signals in a class of canonical persuasion problems. Consider the setting where a receiver chooses an action to match the state and a sender has a state-independent payoff (i.e., $u_S(x,a)=v_S(a)$). The typical assumption is that the receiver minimizes a quadratic loss function (i.e., $u_R(x,a):=-(x-a)^2$). 
Under this assumption, the receiver's optimal action $a^*(G)$, given a posterior $G$, equals the posterior expected value $\mathbb{E}_G[x]$, and hence, the sender's problem is mean-measurable. Parameterizing the receiver's loss function in this way makes the sender's persuasion problem tractable, since the distributions of the receiver's actions are equivalent to mean-preserving contractions of the prior. 
However, the shape of the loss function imposes a specific cardinal structure on the receiver's preferences, and it remains unclear how different parameterizations of the receiver's loss could affect the structure of the optimal signal. 

With \bref{qp}, one may now examine the sender's problem when the receiver has a different loss function. When the receiver has an \emph{absolute} loss function (i.e., $u_R(x,a)=-|x-a|$), the optimal action under any posterior must be a posterior median. More generally, when the receiver has a ``pinball'' loss function (i.e., $u_R(x,a)=-\rho_\tau(x-a)$, with $\rho_\tau(y):=y(\tau-\mathbf{1}\{y<0\})$), the optimal action under any posterior must be a posterior $\tau$-quantile. Since the sender's payoff is state-independent, \bref{qp} applies, and the sender's problem can be rewritten via \eqref{eq:simplified_bayesian_persuasion}.\footnote{When applying \bref{qp} to this problem, one may take the selection rule $r$ to be the one that always selects the sender-preferred $\tau$-quantile.} The pinball loss function imposes a different cardinality structure on the receiver's payoff, where the marginal loss remains constant instead of being linear as the action moves further away from the state.

With \bref{qp} and \eqref{eq:simplified_bayesian_persuasion}, one can solve the sender's problem when the receiver has a pinball loss function for some specific sender payoffs. Specifically, for any continuous prior $F$ that has full support on some interval and for any $a \in \mathbb{R}$, let 
\[
H^{L}_{a}(x):=\left\{
\begin{array}{cc}
0,&\mbox{if } x < a\\
F_L^\tau(x),&\mbox{if } x \geq a
\end{array}
\right.; \quad\mbox{ and }
H^{R}_{a}(x):=\left\{
\begin{array}{cc}
F_R^\tau(x),&\mbox{if } x < a\\
1,&\mbox{if } x \geq a
\end{array}
\right.,
\]
for all $x \in \mathbb{R}$. Also, for any $\underline{a},\overline{a} \in \mathbb{R}$ such that $F_L^\tau(\underline{a})=F_R^\tau(\overline{a})=:\eta$, let
\[
H^{C}_{\underline{a},\overline{a}}(x):=\left\{
\begin{array}{cc}
F_L^\tau(x),&\mbox{if } x<\underline{a}\\
\eta,&\mbox{if } x \in [\underline{a},\overline{a})\\
F_R^\tau(x),&\mbox{if } x \geq \overline{a}
\end{array}
\right..
\]
\bref{optqp} summarizes the sender's optimal signal under various sender payoffs $v_S$. 
\begin{cor}\label{optqp}
Suppose that $F$ is continuous and has full support on a compact interval. Suppose that $v_S:\mathbb{R} \to \mathbb{R}$ is upper-semicontinuous. Then:
\begin{compactenum}[(i)]
\item If $v_S$ is quasi-concave and attains its maximum at $a\leq F^{-1}(\tau)$, then $H^{L}_{a}$ solves \eqref{eq:simplified_bayesian_persuasion}. 
\item If $v_S$ is quasi-concave and attains its maximum at $a> F^{-1}(\tau)$, then $H^{R}_{a}$ solves \eqref{eq:simplified_bayesian_persuasion}. 
\item If $v_S$ is strictly quasi-convex, then $H^{C}_{\underline{a},\overline{a}}$ solves \eqref{eq:simplified_bayesian_persuasion} for some $\underline{a},\overline{a}$ such that $F_L^\tau(\underline{a})=\overline{F}^\tau(\overline{a})$. 
\item $F$ is never the unique solution of \eqref{eq:simplified_bayesian_persuasion}. 
\end{compactenum}
\end{cor}

The distribution $H^{L}_{a}$ ($H^{R}_{a}$) can be induced by separating all states below (above) $F^{-1}(\tau)$ and pooling all states above (below) $F^{-1}(\tau)$ with each of these separated states, and then pooling all the posteriors with states below (above) $a$ together. This signal is optimal for the sender if the sender's payoff is quasi-concave and is maximized at $a \leq F^{-1}(\tau)$ ($a>F^{-1}(\tau)$). In particular, for any strictly concave $v_S$ that is maximized at some $a \in \mathbb{R}$, it is optimal for the sender to reveal no information at all if the receiver's loss function is quadratic, but not optimal if the receiver's loss function is an absolute value. Moreover, the nature of the receiver's loss function affects how the optimal signal changes when monotone transformations are applied to $v_S$. Since any monotone transformation of $v_S$ remains quasi-concave and $a$ remains its maximizer, the sender's optimal signal remains unchanged when the receiver's loss function is an absolute value. However, the optimal signal can be very different if the receiver has quadratic loss, since the curvature of $v_S$ may be different.

Likewise, the distribution $H^{C}_{\underline{a},\overline{a}}$ can be induced by separating all states above $\overline{a}$ and below $\underline{a}$, while pooling all the states in $[\underline{a},\overline{a}]$ with each of these separated states. In particular, for any strictly convex $v_S$, it is optimal for the sender to reveal all the information if the receiver's loss function is quadratic, but not optimal if the receiver's loss function is an absolute value. In fact, since $F$ is the distribution of posterior $\tau$-quantiles under the fully revealing signal, it is never the unique optimal signal if the receiver's loss function is an absolute value.


\bref{qp} can also be used to analyze a persuasion problem where the receiver is not an expected utility maximizer but makes decisions according to ordinal models of utility (i.e., quantile maximizers), a class of preferences studied in \citet{manski1988ordinal}, \citet{chambers2007ordinal}, \citet{rostek2010quantile}, and \citet{de2021static}. When selecting among lotteries, a $\tau$-quantile-maximizer chooses the one that gives the highest $\tau$-quantile of the utility distribution.\footnote{As seen in the literature, optimal gerrymandering problems can be studied via a belief-based approach (e.g., \citealp{friedman2008optimal,gul2010strategic,kolotilin2023economics}). As a result, quantile-based persuasion problems are also connected to gerrymandering when finding optimal or equilibrium election maps with only aggregate uncertainty.}

\paragraph{Apparent Misconfidence}
\mbox{}\\
\noindent Another application of the characterization of the set of distributions of posterior quantiles relates to the literature on over/underconfidence (i.e., misconfidence) in the psychology of judgment. The experimental literature has documented that, when individuals are asked to predict their own abilities, a prediction dataset can be very different from the population distribution. 
Instead of attributing this observation to individuals being irrationally overconfident or underconfident, \citet*{BD11} propose an alternative explanation: this difference can be caused by noise in each individual's signal. Individuals can still be fully Bayesian even if the prediction dataset is different from the population distribution. That is, individuals can be \emph{apparently} misconfident due to dispersion of posterior beliefs. Here, we show how \citet{BD11}'s insight can be derived from \bref{thm3}.

Consider the following setting due to \citet{BD11}. There is a unit mass of individuals, and each one of them is attached to a ``type'' $x \in [0,1]$ that is distributed according to a CDF $F \in \mathcal{F}_0$. Common interpretations of $x$ in the literature include skill levels, scores on a standardized test, the probability of being successful at a task, or simply an individual's ranking in the population in percentage terms. Individuals are asked to predict their own type $x$. Given a finite partition $0=z_0<z_1<\ldots<z_K=1$ of $[0,1]$, a \emph{prediction dataset} is a vector $(\theta_k)_{k=1}^K \in [0,1]^K$ with $\sum_{k=1}^K \theta_k = 1$, where $\theta_k$ denotes the share of individuals who predict their own types in $[z_{k-1},z_k)$. 

A prediction dataset $(\theta_k)_{k \in K}$ is said to be \emph{median rationalizable ($\tau$-quantile rationalizable)} if there exists a signal such that the induced posterior has a unique median ($\tau$-quantile) with probability 1, and that for all $k \in \{1,\ldots,K\}$, the probability of the posterior median ($\tau$-quantile) being in the interval $[z_{k-1},z_k)$ is $\theta_k$.\footnote{Namely, $(\theta_k)_{k=1}^K$ is $\tau$-quantile rationalizable if there exists $H \in \widetilde{\mathcal{H}}_\tau$ such that $H(z_k^-)-H(z_{k-1}^-)=\theta_k$. Technically speaking, \citet{BD11} use a less stringent requirement regarding multiple quantiles. However, as shown below, \bref{thm3} generalizes their conclusion even with this stringent requirement.} In other words, a prediction dataset $(\theta_k)_{k=1}^K$ is median ($\tau$-quantile) rationalizable if there exists a Bayesian framework under which the share of individuals who predict $[z_{k-1},z_k)$ equals $\theta_k$ when individuals are asked to predict their types based on the median ($\tau$-quantiles) of their beliefs.\footnote{Experiments in the literature typically ask individuals to make predictions based on their posterior means or medians. When subjects use the posterior mean to predict their types, the set of rationalizable data would be given by mean-preserving contractions of the prior, which follows immediately from Strassen's theorem, as noted by \citet{BD11}.} 
Theorem 1 (Theorem 4) of \citet{BD11}, as stated below, characterizes the median ($\tau$-quantile) rationalizable datasets. As the proof in Appendix \ref{BDp} shows, this characterization can be derived immediately from \bref{thm3}.


\begin{cor}[Rationalizable Apparent Misconfidence]\label{cor2}
For any $\tau \in (0,1)$, for any $F \in \mathcal{F}_0$ with full support on $[0,1]$, and for any partition $0=z_0<z_1<\ldots<z_K=1$ of $[0,1]$, a prediction dataset $(\theta_k)_{k=1}^K$ is $\tau$-quantile rationalizable if and only if for all $k \in \{1,\ldots,K\}$,
\begin{equation}\label{lb}
\sum_{i=1}^k \theta_i < \frac{1}{\tau} F(z_k)
\end{equation}
and 
\begin{equation}\label{ub}
\sum_{i=k}^K \theta_i < \frac{1-F(z_{k-1}^-)}{1-\tau}
\end{equation}
\end{cor}

\begin{rem}
\citet{BD11} further assume that $F(z_k)=\nicefrac{k}{K}$ for all $k$ (i.e., individuals are asked to place themselves into one of the equally populated $K$-ciles of the population). With this assumption, \bref{cor2} specializes to Theorem 4 of \citet*{BD11}, whose proof relies on projection and perturbation arguments and is not constructive. In addition to having a more straightforward proof and yielding a more general result, another benefit of \bref{thm3} is that the signals rationalizing a feasible prediction dataset are semi-constructive: the extreme points of $\mathcal{I}(F_R^{\tau,\varepsilon},F_L^{\tau,\varepsilon})$ are attained by explicitly constructed signals, as shown in the proof of \bref{thm3}.\footnote{It is also noteworthy that, although Theorem 4 of \citet*{BD11} can be used to prove \bref{thm2} indirectly when $F$ admits a density (by taking $K \to \infty$ and establishing proper continuity properties), the same argument cannot be used to prove \bref{thm3}, which is crucial for the proof of \bref{cor2}. This is because of the failure of upper-hemicontinuity when signals that induce multiple quantiles are excluded. Likewise, as shown by \citet{KW24}, \bref{thm2} can be proved by an argument that does not involve the extreme points of $\mathcal{I}(F_R^\tau,F_L^\tau)$. Nonetheless, the same argument, which relies on the flexibility to select a non-unique quantile, cannot be used to prove \bref{thm3}.}  
\end{rem}


\section{Security Design with Limited Liability}
\label{s4}
In this second class of applications, we show how monotone function intervals pertain to security design with limited liability. In security design problems, a security issuer designs a security that specifies how the return of an asset is divided between the issuer and the security holder. Monotone function intervals embed two widely adopted economic assumptions in the security design literature. The first is limited liability, which places natural upper and lower bounds on the security's payoff---a security cannot pay more than the asset's return or less than zero. The second is that the security’s payoff has to be monotone in the asset’s return. These two assumptions imply that the set of feasible securities can be described by a monotone function interval. Recognizing this, we use the second crucial property of extreme points---namely, for any convex optimization problem, one of the solutions must be an extreme point of the feasible set---to generalize and unify several results in security design under a common framework. To do so, we revisit the environments of two seminal papers in the literature: \citet{innes1990limited}, which has moral hazard, and \citet{DD99}, which has adverse selection.


\subsection{Security Design with Moral Hazard}\label{s41}
A risk-neutral entrepreneur issues a security to an investor to fund a project. The project needs an investment $I>0$. If the project is funded, the entrepreneur then exerts costly effort to develop the project. If the effort level is $e \in [0,\bar{e}]$, the project's profit is distributed according to $\Phi(\cdot|e) \in \mathcal{F}_0$, and the (additively separable) effort cost to the entrepreneur is $C(e) \geq 0$.

A security specifies the return to the investor for every realized profit $x \geq 0$ of the project. Both the entrepreneur and the investor have limited liability, and therefore, any security must be a function $H:\mathbb{R}_+ \to \mathbb{R}$ such that $0\leq H(x) \leq x$ for all $x \geq 0$. Moreover, a security is required to be monotone in the project's profit.\footnote{Requiring securities to be monotone is a standard assumption in the security design literature \citep{innes1990limited,nachman1994optimal,DD99}. Monotonicity can be justified without loss of generality if the entrepreneur could contribute additional funds to the project so that only monotone profits would be observed.} Given a security $H$, the entrepreneur chooses an effort level to solve 
\begin{equation}\label{obd}
\sup_{e \in [0,\bar{e}]} \int_0^\infty (x-H(x))\Phi(\diff x|e)-C(e).
\end{equation}
For simplicity, we make the following technical assumptions: 1) The supports of the profit distributions $\{\Phi(\cdot|e)\}_{e \in [0,\bar{e}]}$ are all contained in a compact interval, which is normalized to $[0,1]$. 2) $\Phi(\cdot|e)$ admits a density $\phi(\cdot|e)$ for all $e \in [0,\bar{e}]$. 3) $\phi(x|e)>0$ and is differentiable with respect to $e$ for all $x \in [0,1]$ and for all $e\geq 0$, with its derivative, $\phi_e(x|e)$, dominated by an integrable function in absolute value. 4) $\{\Phi(\cdot|e)\}_{e \in [0,\bar{e}]}$ and $C$ are such that \eqref{obd} admits a solution, and every solution to \eqref{obd} can be characterized by the first-order condition.\footnote{For example, we may assume that $C$ is strictly increasing and strictly convex and that $\int_0^1 x \max\{\phi_{ee}(x|e),0\}\diff x<C''(e)$ for all $e$. Another sufficient condition would be $\phi_{ee}(1|e)<2C''(e)$ and $\phi'_{ee}(x|e) \geq 0$ for all $x$ and for all $e$. Moreover, if there are only finitely many effort levels available to the entrepreneur, the first-order approach would not be necessary to establish the results below, as suggested by \bref{thma1} in the Appendix.} 

The entrepreneur's goal is to design a security to acquire funding from the investor while maximizing the entrepreneur's expected payoff. Specifically, let $\overline{F}(x):=x$ and let $\underline{F}(x):=0$ for all $x \in [0,1]$. The set of securities can be written as $\mathcal{I}(\underline{F},\overline{F})$. The entrepreneur solves 
\begin{align}\label{ent}
\sup_{H \in \mathcal{I}(\underline{F},\overline{F}),\, e \in [0,\bar{e}]} &\left[\int_0^{1} [x-H(x)]\phi( x | e)\diff x-C(e)\right] \notag\\
\mbox{s.t. }& \int_0^{1} [x-H(x)]\phi_e(x|e)\diff x=C'(e)\\
&\int_0^{1} H(x)\phi( x|e)\diff x \geq (1+r)I,\notag
\end{align}
where $r>0$ is the rate of return on a risk-free asset. 

\citet{innes1990limited} characterizes the optimal security in this setting under an additional crucial assumption: the project profit distributions $\{\phi(\cdot|e)\}_{e \in [0,\bar{e}]}$ satisfy the monotone likelihood ratio property \citep{M86}. With this assumption, he shows that every optimal security must be a standard debt contract $H^d(x):=\min\{x,d\}$ for some face value $d>0$. While the simplicity of a standard debt contract is a desirable feature, the monotone likelihood ratio property is arguably a strong condition \citep{hart1995firms}, where higher effort leads to higher probability weights on all higher project profits at any profit level. It remains unclear what the optimal security might be under a more general class of distributions.

Using \bref{thm1}, we can generalize \citet{innes1990limited} and solve the entrepreneur's problem \eqref{ent} without the monotone likelihood ratio property. As we show in \bref{moral} below, \emph{contingent} debt contracts are now optimal. A security $H \in \mathcal{I}(\underline{F},\overline{F})$ is a \textit{contingent debt contract} if there exists a countable collection of intervals $\{[\underline{x}_n,\overline{x}_n)\}_{n=1}^N$, with $N \in \mathbb{N} \cup \{\infty\}$, such that $H$ is constant on $[\underline{x}_n,\overline{x}_n)$, $H(x)=x$ for all $x \notin \cup_{n=1}^\infty [\underline{x}_n,\overline{x}_n)$, and that $H(\underline{x}_n) \neq H(\underline{x}_m)$ for all $n \neq m$. In other words, a contingent debt contract $H$ has $N$ possible face values, $\{d_n=H(\underline{x}_n)\}_{n=1}^N$, where the entrepreneur pays face value $d_n=H(\underline{x}_n)$ if the project's profit falls in $[\underline{x}_n,\overline{x}_n)$, and defaults otherwise. 

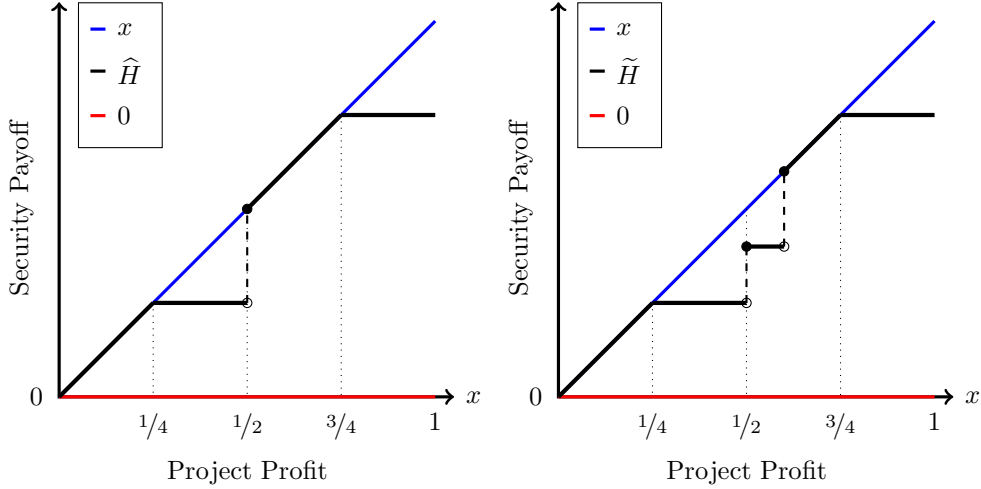
\begin{figure}
\centering
\tikzset{
solid node/.style={circle,draw,inner sep=1.25,fill=black},
hollow node/.style={circle,draw,inner sep=1.25}
}
\begin{subfigure}[b]{0.4\linewidth}%

\begin{tikzpicture}[scale=5]
\draw [<->, very thick] (0,1.05) node (yaxis) [above] {\footnotesize}
        |- (1.05,0) node (xaxis) [right] {\footnotesize$\omega$};
\draw (1,0) node [below=2pt] {\footnotesize$1$};
\draw (0,0) node [left=2pt] {\footnotesize$0$};
\draw [blue, very thick] (0,0)--(1,1);
\draw [red, very thick] (0,0)--(1,0);
\tikzset{
solid node/.style={circle,draw,inner sep=1.25,fill=red},
hollow node/.style={circle,draw,inner sep=1.25}
}
\draw [ultra thick] (0,0)--(0.25,0.25);
\draw [ultra thick] (0.5,0.25)--(0.25,0.25);
\tikzset{
solid node/.style={circle,draw,inner sep=1.25,fill=black},
hollow node/.style={circle,draw,inner sep=1.25}
}
\draw (0.5,0.25) node [hollow node] {};
\draw (0.5,0.5) node [solid node] {};
\draw [dashed, thick] (0.5,0.25)--(0.5,0.5);
\draw [ultra thick] (0.5,0.5)--(0.75,0.75);
\draw [ultra thick] (0.75,0.75)--(1,0.75);
\draw [dotted] (0.25,0.25)--(0.25,0);
\draw [dotted] (0.5,0.5)--(0.5,0);
\draw [dotted] (0.75,0.75)--(0.75,0);
\draw  (0.25,0) node  [below=2pt] {\footnotesize{$\nicefrac{1}{4}$}};
\draw  (0.5,0)node  [below=2pt] {\footnotesize{$\nicefrac{1}{2}$}};
\draw  (0.75,0)node  [below=2pt] {\footnotesize{$\nicefrac{3}{4}$}};

\node[align=center] at (0.5,-0.2) {\footnotesize Project Profit};

\node[rotate=90, align=center] at (-0.1,0.5) {\footnotesize Security Payoff};

\matrix [draw=black, thin, anchor=north west, row sep=1pt] at (0.05,1.05) {
    \draw [blue, very thick] (0,0) -- (0.2,0); & \node[right] {\footnotesize $x$}; \\
    \draw [black, very thick] (0,0) -- (0.2,0); & \node[right] {\footnotesize $\widehat{H}$};\\
    \draw [red, very thick] (0,0)--(0.2,0);&
    \node[right] {\footnotesize $0$};\\
 \\
};
\end{tikzpicture}
\caption{Face Values: $\nicefrac{1}{4}$ and $\nicefrac{3}{4}$}
\label{fig3a}
\end{subfigure}%
\begin{subfigure}[b]{0.5\linewidth}%
\begin{tikzpicture}[scale=5]
\draw [<->, very thick] (0,1.05) node (yaxis) [above] {\footnotesize}
        |- (1.05,0) node (xaxis) [right] {\footnotesize$\omega$};
\draw (1,0) node [below=2pt] {\footnotesize$1$};
\draw (0,0) node [left=2pt] {\footnotesize$0$};
\draw [blue, very thick] (0,0)--(1,1);
\draw [red, very thick] (0,0)--(1,0);
\tikzset{
solid node/.style={circle,draw,inner sep=1.25,fill=red},
hollow node/.style={circle,draw,inner sep=1.25}
}
\draw [ultra thick] (0,0)--(0.25,0.25);
\draw [ultra thick] (0.5,0.25)--(0.25,0.25);
\tikzset{
solid node/.style={circle,draw,inner sep=1.25,fill=black},
hollow node/.style={circle,draw,inner sep=1.25}
}
\draw (0.5,0.25) node [hollow node] {};
\draw (0.5,0.4) node [solid node] {};
\draw [dashed, thick] (0.5,0.25)--(0.5,0.4);
\draw [ultra thick] (0.5,0.4)--(0.6,0.4);
\draw [dashed, thick] (0.6,0.4)--(0.6,0.6);
\draw [ultra thick] (0.6,0.6)--(0.75,0.75);
\draw (0.6,0.4) node [hollow node] {};
\draw (0.6,0.6) node [solid node] {};
\draw [ultra thick] (0.75,0.75)--(1,0.75);
\draw [dotted] (0.25,0.25)--(0.25,0);
\draw [dotted] (0.5,0.5)--(0.5,0);
\draw [dotted] (0.75,0.75)--(0.75,0);
\draw  (0.25,0) node  [below=2pt] {\footnotesize{$\nicefrac{1}{4}$}};
\draw  (0.5,0)node  [below=2pt] {\footnotesize{$\nicefrac{1}{2}$}};
\draw  (0.75,0)node  [below=2pt] {\footnotesize{$\nicefrac{3}{4}$}};

\node[align=center] at (0.5,-0.2) {\footnotesize Project Profit};

\node[rotate=90, align=center] at (-0.1,0.5) {\footnotesize Security Payoff};

\matrix [draw=black, thin, anchor=north west, row sep=1pt] at (0.05,1.05) {
    \draw [blue, very thick] (0,0) -- (0.2,0); & \node[right] {\footnotesize $x$}; \\
    \draw [black, very thick] (0,0) -- (0.2,0); & \node[right] {\footnotesize $\widetilde{H}$};\\
    \draw [red, very thick] (0,0)--(0.2,0);&
    \node[right] {\footnotesize $0$};\\
 \\
};
\end{tikzpicture}
\caption{Defaultable Face Values: $\nicefrac{1}{4}$ and $\nicefrac{3}{4}$}
\label{fig3b}
\end{subfigure}
\caption{Contingent Debt Contracts}
\label{fig3}
\end{figure}

Clearly, every standard debt contract with face value $d$ is a contingent debt contract where $N=1$. Moreover, a contingent debt contract never involves the entrepreneur and investor sharing in the equity of the project (i.e., the derivative of $H$, whenever defined, must be either $0$ or $1$). \bref{fig3a} illustrates a contingent debt contract $\widehat{H}$ with $N=2$, $[\underline{x}_1,\overline{x}_1)=[\nicefrac{1}{4},\nicefrac{1}{2})$, $[\underline{x}_2,\overline{x}_2)=[\nicefrac{3}{4},1)$, $d_1=\nicefrac{1}{4}$, and $d_2=\nicefrac{3}{4}$. Under $\widehat{H}$, if the project's profit $x$ is below $\nicefrac{1}{2}$, the entrepreneur owes debt with face value $\nicefrac{1}{4}$. On the other hand, if the profit is above $\nicefrac{1}{2}$, the entrepreneur owes debt with a higher face value $\nicefrac{3}{4}$. The entrepreneur's required debt payment to the investor is contingent on the realized profit of the project.

Given a contingent debt contract $H$ with face values $\{d_n=H(\underline{x}_n)\}_{n=1}^\infty$, a face value $d_n$ is said to be \emph{non-defaultable} if $d_n < x$, for all $x \in [\underline{x}_n,\overline{x}_n)$. 
That is, a face value $d_n$ is non-defaultable if, conditional on $d_n$ being in effect, the project's profit is always higher than that face value and the entrepreneur always retains some residual surplus after paying off $d_n$. \bref{fig3b} illustrates a contingent debt contract $\widetilde{H}$ with three possible face values, $\nicefrac{1}{4}$, $\nicefrac{1}{2}$, and $\nicefrac{3}{4}$. Here, the face values $\nicefrac{1}{4}$ and $\nicefrac{3}{4}$ are defaultable, whereas the face value $\nicefrac{1}{2}$ is non-defaultable.


\begin{prop}\label{moral}
There is a contingent debt contract with at most two non-defaultable face values that solves the entrepreneur's problem \eqref{ent}. 
\end{prop}


According to \bref{moral}, even without the MLRP assumption, the nature of standard debt contracts, which allocates any additional dollar of the project's profit either fully to the entrepreneur or to the investor, is preserved even without the monotone likelihood ratio assumption. Nonetheless, the entrepreneur may be liable for more when the project earns more. 

The proof of \bref{moral} can be found in Appendix \ref{security}. In essence, since the entrepreneur's objective in \eqref{ent} is affine and the set of feasible contracts is convex, there must exist an extreme point of the feasible set that solves \eqref{ent}. Thus, it suffices to show that any extreme point of the feasible set must correspond to a contingent debt contract with at most two non-defaultable face values. To this end, first note that, by Proposition 2.1 of \citet{W88}, any extreme point $H^*$ of the feasible set of \eqref{ent} can be written as convex combinations of at most three extreme points of $\mathcal{I}(\underline{F},\overline{F})$. Using an argument similar to the proof of \bref{thm1}, the proof in Appendix \ref{security} then shows that any such convex combinations that do not correspond to a contingent debt contract with at most two defaultable face values can be written as a convex combination of two distinct feasible securities in \eqref{ent}, and thus, is not an extreme point.

\paragraph{Optimal Contingent Debts with Finitely Many Face Values.}With additional assumptions on the project's profit distributions $\{\Phi(\cdot|e)\}_{e \in [0,\bar{e}]}$, the structure of the optimal contracts can be further simplified. For any $N \in \mathbb{N}$ and for any $e \in [0,\bar{e}]$,  we say that the function $\phi_e(\cdot|e)/\phi(\cdot|e)$ is \emph{$N$-peaked} if there exists $N$ disjoint intervals $\{I_n\}_{n=1}^N$ in $[0,1]$ such that $\phi_e(x|e)/\phi(x|e)$ is increasing in $x$ on $I_n$ for all $n \in \{1,\ldots,N\}$, and is decreasing in $x$ on $[0,1] \backslash \cup_{n=1}^N I_n$. Note that if $\phi_e(\cdot|e)/\phi(\cdot|e)$ is increasing on $[0,1]$, then it is $N$-peaked with $N=1$. In particular, profit distributions that satisfy MLRP are $N$-peaked with $N=1$.

Furthermore, assume that the risk-free rate of return $r$ and the required investment $I$ are such that $(1+r)I$ is in the interior of the set 
\begin{equation}\label{interior}
\left\{\int_0^1 H(x)\phi(x|e)\diff x\bigg| H \in \mathcal{I}(\underline{F},\overline{F}), \, \int_0^1 (x-H(x))\phi_e(x|e)\diff x=C'(e)\right\},
\end{equation}
for all $e \in [0,\bar{e}]$. \bref{finite} below identifies a sufficient condition for there to be an optimal contingent debt contract with finitely many face values. 


\begin{prop}\label{finite}
Suppose that there exists $N \in \mathbb{N}$ such that for any $e \in [0,\bar{e}]$, the function $\phi_e(\cdot|e)/\phi(\cdot|e)$ is at most $N$-peaked. Then there is a contingent debt contract with at most $N+1$ face values (with at most two of them being non-defaultable) that solves the entrepreneur's problem \eqref{ent}. 
\end{prop}

\begin{figure}
\centering
\tikzset{
solid node/.style={circle,draw,inner sep=1.25,fill=black},
hollow node/.style={circle,draw,inner sep=1.25}
}
\begin{subfigure}[b]{0.4\linewidth}
\begin{tikzpicture}[scale=5]
\draw [<->, very thick] (0,1.05) node (yaxis) [above] {\footnotesize }
        |- (1.05,0) node (xaxis) [right] {\footnotesize$\omega$};
\draw (1,0) node [below=2pt] {\footnotesize$1$};
\draw (0,0) node [below=2pt] {\footnotesize$0$};
\draw [very thick] (0,0.1) to [out=30, in=180] (0.2,0.5);
\draw [very thick] (0.2,0.5) to [out=0, in=180](0.5,0.2);
\draw [very thick] (0.5,0.2) to [out=0, in=180] (0.8,0.45);
\draw [very thick] (0.8,0.45) to [out=0, in=120] (1,0.3);
\draw [dashed, thick] (0,0.3)--(1,0.3);
\draw [dotted, thick] (0.095,0.3)--(0.095,0);
\draw (0.095,0) node [below=2pt] {\footnotesize${d_1}$};
\draw [dotted, thick] (0.37,0.3)--(0.37,0);
\draw (0.37,0) node [below=2pt] {\footnotesize $x^*$};
\draw [dotted, thick] (0.64,0.3)--(0.64,0);
\draw (0.64,0) node [below=2pt] {\footnotesize $d_2$};
\draw (0,0.3) node [left=2pt] {\footnotesize $\lambda^*$};

\node[align=center] at (0.5,-0.2) {\footnotesize Project Profit};


\matrix [draw=black, thin, anchor=north west, row sep=1pt] at (0.05,1.1) {
    \draw [very thick] (0,0) -- (0.2,0); & \node[right] {\footnotesize $\frac{\phi_e(\cdot|e^*)}{\phi(\cdot|e^*)}$}; \\
};

\end{tikzpicture}
\caption{$2$-peaked $\frac{\phi_e(\cdot|e)}{\phi(\cdot|e)}$}
\label{fig7a}
\end{subfigure}%
\begin{subfigure}[b]{0.4\linewidth}
\begin{tikzpicture}[scale=5]
\draw [<->, very thick] (0,1.05) node (yaxis) [above] {\footnotesize }
        |- (1.05,0) node (xaxis) [right] {\footnotesize$\omega$};
\draw (1,0) node [below=2pt] {\footnotesize$1$};
\draw (0,0) node [below=2pt] {\footnotesize$0$};
\draw [blue, very thick] (0,0)--(1,1);
\draw [red, very thick] (0,0)--(1,0);
\draw [very thick] (0,0)--(0.095,0.095);
\draw [very thick] (0.095,0.095)--(0.37,0.095);
\draw [very thick] (0.37,0.37)--(0.64,0.64);
\draw [very thick] (0.64,0.64)--(1,0.64);
\draw (0.095,0) node [below=2pt] {\footnotesize${d_1}$};
\draw [dotted, thick] (0.095,0.095)--(0.095,0);

\draw (0.37,0) node [below=2pt] {\footnotesize $x^*$};
\draw [dotted, thick] (0.64,0.64)--(0.64,0);
\draw (0.64,0) node [below=2pt] {\footnotesize $d_2$};
\draw [dotted, thick] (0.37,0.095)--(0.37,0);

\draw (0.37,0.095) node [hollow node] {};
\draw (0.37,0.37) node [solid node] {};
\draw [dashed, thick] (0.37,0.095)--(0.37,0.37);

\node[align=center] at (0.5,-0.2) {\footnotesize Project Profit};

\node[rotate=90, align=center] at (-0.1,0.5) {\footnotesize Security Payoff};

\matrix [draw=black, thin, anchor=north west, row sep=1pt] at (0.05,1.05) {
    \draw [blue, very thick] (0,0) -- (0.2,0); & \node[right] {\footnotesize $x$}; \\
    \draw [black, very thick] (0,0) -- (0.2,0); & \node[right] {\footnotesize $H^*$}; \\
    \draw [red, very thick] (0,0)--(0.2,0);&
    \node[right] {\footnotesize $0$};\\
};
\end{tikzpicture}
\caption{Optimal Security $H^*$}
\label{fig7b}
\end{subfigure}
\caption{Optimal Contingent Debt with 2 Face Values}
\end{figure}
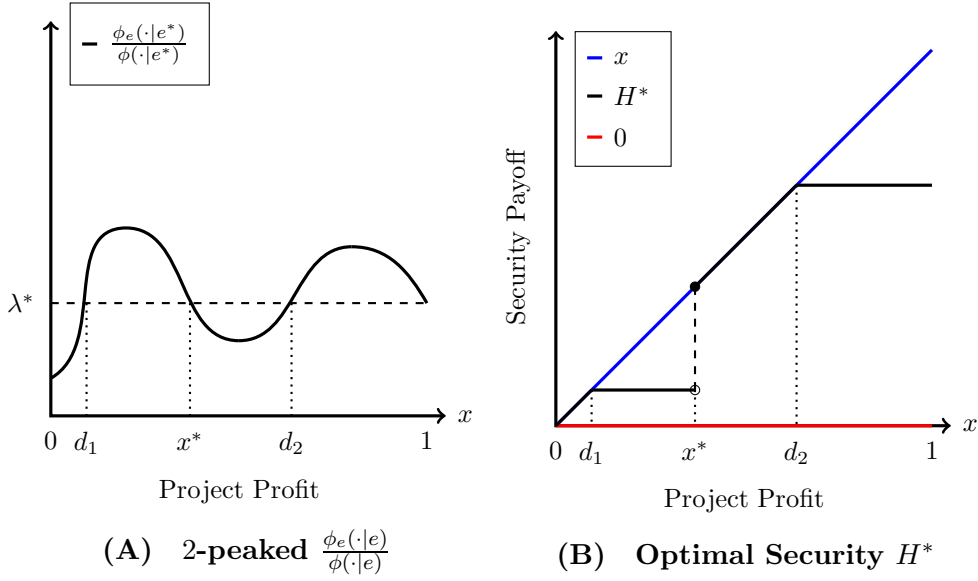

The proof of \bref{finite} can be found in Appendix \ref{afinite}. The essence of the proof is the following observation: under \eqref{interior}, strong duality holds for the entrepreneur's problem \eqref{ent}. Thus, for the optimal effort level $e^* \in [0,\bar{e}]$, an optimal security $H^*$ must also solve
\begin{equation}\label{strong}
\sup_{H \in \mathcal{I}(\lb{F},\ub{F})} \left[\int_0^1 H(x)[(1+\lambda_2^*) \phi(x|e^*)-\lambda^*_1 \phi_e(x|e^*)]\diff x\right],
\end{equation}
where $\lambda_1^* \neq 0, \lambda_2^*\geq 0$ are the Lagrange multipliers for the incentive compatibility and individual rationality constraints, respectively. Since 
\[
(1+\lambda_2^*) \phi(x|e^*) - \lambda_1^* \phi_e(x|e^*) \geq 0 \iff \frac{\phi_e(x|e^*)}{\phi(x|e^*)} \leq \frac{1+\lambda_2^*}{\lambda_1^*}=: \lambda^*,
\]
and since $\phi_e(\cdot|e^*)/\phi(\cdot|e^*)$ is at most $N$-peaked, the set of profits $x$ under which $\phi_e(x|e^*)/\phi(x|e^*)$ is greater than or smaller than $\lambda^*$ must form an interval partition with at most $2N$ elements, as depicted in \bref{fig7a}. It can then be shown that, for a contingent debt contract $H^*$ to be optimal, $H^*$ cannot take two distinct values on any element where $\phi_e(x|e^*)/\phi(x|e^*)>\lambda^*$, and that $H^*(x)=x$ whenever $\phi_e(x|e^*)/\phi(x|e^*)<\lambda^*$, as depicted in \bref{fig7b}. Thus, there must be at most $N+1$ partition elements on which $H^*$ is constant, and hence $H^*$ must be a contingent debt contract with at most $N+1$ face values. In other words, the number of face values in the optimal security is determined by the number of times the function $\phi_e(x|e^*)/\phi(x|e^*)$ crosses the multiplier $\lambda^*$ from below. Using this argument, it also follows that a standard debt contract is optimal whenever $\phi_e(x|e^*)/\phi(x|e^*)$ crosses the multiplier $\lambda^*$  from below only once. In particular, the optimality of standard debt contracts under MLRP follows immediately, as MLRP implies $\phi_e(x|e^*)/\phi(x|e^*)$ crosses $\lambda^*$ from below once.

According to \bref{finite}, if the project's profit distributions $\{\Phi(\cdot|e)\}_{e \in [0,\bar{e}]}$ satisfy the regularity condition so that $\phi_e(\cdot|e)/\phi(\cdot|e)$ is at most $N$-peaked for all $e$, then not only would a contingent debt contract be optimal, but also this optimal contract would be simple, in that it would have at most finitely many face values. 

To better understand the intuition behind \bref{moral} and \bref{finite}, recall that the optimality of standard debt contracts in \citet{innes1990limited} is due to (i) the risk-neutrality and the limited-liability structure of the problem, and (ii) the monotone likelihood ratio property of the profit distributions. Indeed, for any incentive-compatible and individually-rational contract, risk neutrality allows one to construct a standard debt contract with the same expected payment. Meanwhile, MLRP---which implies that higher profit is always a stronger indication of high effort---ensures that this debt contract---which maximally rewards the entrepreneur in the events when the profit is high---incentivizes the entrepreneur to exert higher effort, thus leading to a higher expected project profit. Without the monotone likelihood ratio assumption, simply replicating an individually-rational contract with a standard debt contract may distort incentives and lead to less efficient effort and suboptimal outcomes. This is because when MLRP fails, high profit may sometimes indicate low effort. In this regard, \bref{finite} follows since the optimal contingent debt contract is designed to maximally reward the entrepreneur only when high profit strongly suggests high effort. Moreover, \bref{moral} shows that contingent debt contracts are enough to replicate the expected payments of all other feasible contracts while preserving incentive compatibility. In essence, these propositions separate the effects of risk neutrality and limited liability on security design from the effects of the monotone likelihood ratio property.

\begin{rem}
Contingent debt contracts are similar in spirit to many fixed-income securities observed in practice. The first are state-contingent debt instruments (SCDIs) from the sovereign debt literature, which link a country's principal or interest payments to its nominal GDP \citep{lessard1987capital,shiller1994macro,borensztein2004case}. The second are versions of contingent convertible bonds (CoCos) issued by corporations, which write down the bond's face value after a triggering event like financial distress \citep{albul2015contingent,oster2020contingent}. The third are commodity-linked bonds common to mineral companies and resource-rich developing countries, which tie the amount paid at maturity to the market value of a reference commodity like silver \citep{lessard1977commodity,schwartz1982pricing}.
\end{rem}

\subsection{Security Design with Adverse Selection}
\label{s42}
There is a risk-neutral security issuer with discount rate $\delta \in (0,1)$ and a unit mass of risk-neutral investors. The issuer has an asset that generates a random cash flow $x \geq 0$ in period $t=1$. The cash flow is distributed according to $\Phi_0 \in \mathcal{F}_0$, which is supported on a compact interval normalized to $[0,1]$. Because $\delta<1$, the issuer has demand for liquidity in period $t=0$ and therefore has an incentive to sell a limited-liability security backed by the asset to raise cash. A security is a nondecreasing, right-continuous function $H:[0,1] \to \mathbb{R}_+$ such that $0 \leq H(x) \leq x$ for all $x$. Let $\ub{F}(x):=x$ and $\lb{F}(x):=0$ for all $x \in [0,1]$. The set of securities can again be written as $\mathcal{I}(\underline{F},\overline{F})$.

Given any security $H \in \mathcal{I}(\underline{F},\overline{F})$, the issuer first observes a signal $s \in S$ for the asset's cash flow. Then, taking as given an inverse demand schedule $P:[0,1] \to \mathbb{R}_+$, she chooses a fraction $q \in [0,1]$ of the security to sell. If a fraction $q$ of the security is sold and the signal realization is $s$, the issuer's expected return is 
\[
\underbrace{qP(q)}_{\text{revenue raised in $t=0$}}+\delta\cdot\underbrace{\mathbb{E}[x-qH(x)|s]}_{\text{residual return in $t=1$}}=q(P(q)-\delta \mathbb{E}[H(x)|s])+\delta\mathbb{E}[x|s].
\]
Investors observe the quantity $q$, update their beliefs about $x$, and decide whether to purchase. 

\citet*{DD99} show that, in the unique equilibrium that survives the D1 criterion,\footnote{An equilibrium in this market is a pair $(P,Q)$ of measurable functions such that $Q(\mathbb{E}[H(x)|s])(P\circ Q(\mathbb{E}[H(x)|s])-\delta \mathbb{E}[H(x)|s]) \geq q(P(q)-\delta \mathbb{E}[H(x)|s])$ for all $q \in [0,1]$ with probability 1, and $P\circ Q (\mathbb{E}[H(x)|s])=\mathbb{E}[H(x)|Q(\mathbb{E}[H(x)|s])]$ with probability 1.}  the issuer's profit under a security $H$, when the posterior expected value of the security is $\mathbb{E}[H(x)|s]=z$, is given by 
\[
\Pi(z|H):=(1-\delta)z_0^{\frac{1}{1-\delta}}z^{-\frac{\delta}{1-\delta}},
\]
where $z_0:= \inf_{s \in S}\mathbb{E}[H(x)|s]$. Therefore, let $\Phi(\cdot|s)$ be the conditional distribution of the cash flow $x$ given signal $s$, and let $\Psi:S \to [0,1]$ be the marginal distribution of the signal $s$. The expected value of a security $H$ is then
\[
\Pi(H):=(1-\delta) \left(\inf_{s \in S} \int_0^{1} H(x)\Phi(\diff x|s)\right)^{\frac{1}{1-\delta}}\int_S \left(\int_0^{1}H(x) \Phi(\diff x|s)\right)^{-\frac{\delta}{1-\delta}} \Psi(\diff s).
\]
As a result, the issuer's security design problem can be written as 
\[
\sup_{H \in \mathcal{I}(\underline{F},\overline{F})} \Pi(H).
\]

Using a variational approach, \citet{DD99} characterize several general properties  of the optimal securities without solving for them explicitly.  They then specialize the model by assuming that the signal structure $\{\Phi(\cdot|s)\}_{s \in S}$ has a \emph{uniform worst case}, a condition slightly weaker than the monotone likelihood ratio property that requires the cash flow distribution to be smallest in the sense of first-order stochastic dominance (FOSD) under some $s_0$, conditional on every interval $I$ of $[0,1]$.\footnote{Specifically, they assume that there exists some $s_0 \in S$ such that, for any $s \in S$ and for any interval $I \subset [0,1]$, (i) $\Phi(I|s_0)=0$ implies $\Phi(I|s)=0$, and (ii) the conditional distribution of the asset's cash flow given signal realization $s$ and given that the cash flow falls in an interval $I$---which is denoted $\nicefrac{\Phi|_I(\cdot |s)}{\Phi(I|s)}$---dominates the conditional distribution given signal realization $s_0$, denoted by $\nicefrac{\Phi|_I(\cdot|s_0)}{\Phi(I|s_0)}$, in the sense of FOSD.} With this assumption, \citet*{DD99} show that a standard debt contract $H^d(x):=\min\{x,d\}$ is optimal. 

With \bref{thm1}, we are able to generalize this result and solve for an optimal security while relaxing the uniform-worst-case assumption. Instead of a uniform worst case, we only assume that there is a worst signal $s_0$ such that $\Phi(\cdot|s)$ dominates $\Phi(\cdot|s_0)$ in the sense of FOSD for all $s \in S$. With this assumption, the issuer's security design problem can be written as 
\begin{align}\label{iss}
\sup_{H \in \mathcal{I}(\underline{F},\overline{F}), \underline{z} \in [0,\mathbb{E}[x|s_0]]}& \left[(1-\delta)\underline{z}^{\frac{1}{1-\delta}}\int_S\left(\int_0^{1}H(x)\Phi(\diff x|s)\right)^{-\frac{\delta}{1-\delta}}\Psi(\diff s)\right]\notag\\
\mbox{s.t. }& \int_0^{1} H(x) \Phi(\diff x|s_0)=\underline{z}. 
\end{align}
As shown by \bref{prop2} below, a particular class of contingent debt contracts is always sufficient for the issuer to consider. 

\begin{prop}\label{prop2}
There is a contingent debt contract with at most one non-defaultable face value that solves the issuer's problem \eqref{iss}. Furthermore, if $\Phi(\cdot|s)$ has full support on $[0,1]$ for all $s \in S$, this solution is unique. 
\end{prop}

Overall, this section showcases the unifying role of the extreme points of monotone function intervals in security design. The security design literature has rationalized the existence of different financial securities observed in practice under a variety of economic environments and assumptions. Doing so has strengthened the robustness of these securities as optimal contracts. But that variety also makes it hard to sort the essential modeling ingredients from the inessential ones. And the core features that connect these environments are not readily apparent. 

An advantage of recasting the set of feasible securities as a monotone function interval is that it strips the problem down to its basic elements. Whether the setting has moral hazard or adverse selection, and whether the asset's cash flow distributions exhibit MLRP, are not defining. Limited liability, monotone contracts, and risk neutrality are the core elements that deliver debt as an optimal security. The terms of the debt contract somewhat differ from those of a standard one, as the face value is now contingent on the asset’s cash flow, but the nature of debt contracts, which never has the issuer and investor share in the asset's equity and grants the issuer only residual rights, still prevails.

\begin{rem}
If $x \geq 0$ is interpreted as the loss, instead of the return, of an asset, a security $H \in \mathcal{I}(\lb{F},\ub{F})$ can be regarded as an \emph{insurance contract} that specifies which part of the loss will be covered by the contract (i.e., $x-H(x)$). In this setting, \citet{Getal23} solve for the optimal insurance contract for a monopolist insurer who faces dual-utility-risk-averse agents \citep{Y87} with private information, under the assumption that insurance contracts have to be doubly-monotone: both the coverage $x-H(x)$ and the retention $H(x)$ are monotone. They further note that when the monotonicity assumption on $x-H(x)$ is relaxed, the optimal contracts must be piecewise continuous, with $H'(x) \in \{0,1\}$ almost everywhere. From the perspective of \bref{thm1}, we can see that this is the case because the insurer's objective is affine, and because the extreme points of $\mathcal{I}(\lb{F},\ub{F})$ have the exact same property. 
\end{rem}


\section{Conclusion}
\label{sec:conclusion}
We characterize the extreme points of monotone function intervals and apply this result to several economic problems. We show that any extreme point of a monotone function interval must either coincide with one of the monotone function interval's bounds, or be constant on an interval in its domain, where at least one end of the interval reaches one of the bounds. Using this result, we characterize the set of distributions of posterior quantiles, which coincide with a monotone function interval. We apply this insight to topics in political economy, Bayesian persuasion, and the psychology of judgment. Furthermore, monotone function intervals provide a common structure to security design. We unify and generalize seminal results in that literature when either adverse selection or moral hazard afflicts the environment.

It is worthwhile to acknowledge the paper's  limitations. Regarding the distributions of posterior quantiles, the analysis is restricted to a one-dimensional state space. Moreover, while the characterization parallels the well-known characterization of distributions of posterior means, it provides little intuition for how distributions of other statistics (say, the posterior $k$-th moment) may behave. In particular, while the characterization of the set of distributions of posterior quantiles allows one to compare Bayesian persuasion problems when the receiver has either an absolute loss function or a quadratic loss function, optimal signals under other loss functions remain largely under-explored.   

Regarding security design, the clearest limitation is the absence of risk aversion. This is due to the lack of convexity of the objectives and constraints of security design problems with risk averse agents. The majority of the security design literature features risk neutral agents, and this risk neutrality makes the design problem amenable to being analyzed using extreme points of monotone function intervals. Nevertheless, security design with risk averse agents has gotten less attention among researchers and deserves further study. \citet{allen1988optimal,malamud2010optimal} and \citet{gershkov2023optimal} study the problem and provide many intriguing results thus far.


\bibliography{ref}

@Article{owen1988optimal,
  author    = {Owen, Guillermo and Grofman, Bernard},
  journal   = {Political Geography Quarterly},
  title     = {Optimal Partisan Gerrymandering},
  year      = {1988},
  number    = {1},
  pages     = {5--22},
  volume    = {7},
  publisher = {Elsevier},
}

@Article{friedman2008optimal,
  author  = {Friedman, John N and Holden, Richard T},
  journal = {American Economic Review},
  title   = {Optimal Gerrymandering: Sometimes Pack, but Never Crack},
  year    = {2008},
  number  = {1},
  pages   = {113--44},
  volume  = {98},
}

@Article{BD11,
  author  = {Jean-Pierre Beno{\^{i}}t and Juan Dubra},
  journal = {Econometrica},
  title   = {Apparent Overconfidence},
  year    = {2011},
  number  = {5},
  pages   = {1591--625},
  volume  = {79},
}

@Article{gul2010strategic,
  author  = {Gul, Faruk and Pesendorfer, Wolfgang},
  journal = {American Economic Review},
  title   = {Strategic Redistricting},
  year    = {2010},
  number  = {4},
  pages   = {1616--41},
  volume  = {100},
}

@Unpublished{kolotilin2023economics,
  author = {Kolotilin, Anton and Wolitzky, Alexander},
  note   = {Working Paper},
  title  = {The Economics of Partisan Gerrymandering},
  year   = {2023},
}

@Unpublished{KCW22,
  author = {Kolotilin, Anton and Roberto Corrao and Wolitzky, Alexander},
  note   = {Working Paper},
  title  = {Persuasion and Matching: Optimal Productive Transport},
  year   = {2023},
}

@article{oster2020contingent,
  title={Contingent Convertible Bond Literature Review: Making Everything and Nothing Possible?},
  author={Oster, Philippe},
  journal={Journal of Banking Regulation},
  volume={21},
  pages={343--381},
  year={2020},
  publisher={Springer}
}

@article{CS23,
  title={Optimal Disclosure of Information to Privately
Informed Agents},
  author={Ozan Candogan and Philipp Strack},
  journal={Theoretical Economics},
  volume={18},  
pages={1225-1269},
  year={2023},
}

@article{schwartz1982pricing,
  title={The pricing of commodity-linked bonds},
  author={Schwartz, Eduardo S},
  journal={The journal of Finance},
  volume={37},
  number={2},
  pages={525--539},
  year={1982},
  publisher={JSTOR}
}

@Article{lessard1977commodity,
  author  = {Lessard, Donald},
  journal = {Working Paper},
  title   = {Commodity-Linked Bonds from Less-Developed Countries: An Investment Opportunity?},
  year    = {1977},
}

@Unpublished{albul2015contingent,
  author = {Albul, Boris and Jaffee, Dwight M and Tchistyi, Alexei},
  note   = {Working Paper},
  title  = {Contingent Convertible Bonds and Capital Structure Decisions},
  year   = {2015},
}

@article{friedman2020optimal,
  title={Optimal Gerrymandering in a Competitive Environment},
  author={Friedman, John N and Holden, Richard},
  journal={Economic Theory Bulletin},
  volume={8},
  number={2},
  pages={347--367},
  year={2020},
  publisher={Springer}
}

@article{W88,
  title={Extreme Points of Moment Sets},
  author={Gerhard Winkler},
  journal={Mathematics of Operations Research},
  volume={13},
  number={4},
  pages={581--587},
  year={1988}
}

@article{Y87,
 author = {Menahem E. Yaari},
 journal = {Econometrica},
 number = {1},
 pages = {95--115},
 title = {The Dual Theory of Choice under Risk},
 volume = {55},
 year = {1987}
}

@article{DD99,
  title={A Liquidity-Based Model of Security Design},
  author={Peter DeMarzo and Darrell Duffie},
  journal={Econometrica},
  volume={67},
  number={1},
  pages={65--99},
  year={1999},
}

@article{Aetal22,
  author = "S. Nageeb Ali and Nima Haghpanah and Xiao Lin and Ron Siegel",
  year ="2022",
  title = "How to Sell Hard Information",
  journal = "Quarterly Journal of Economics",
  volume = "137",
  number = "1",
  pages = "619-678"
}

@book{B14,
  title={An Introduction to the Theory of Mechanism Design},
  author={B{\"{o}}rgers, Tilman},
  year={2015},
  publisher={Oxford University Press}
}

@article{B53,
  author = "David Blackwell",
  title = "Equivalent Comparisons of Experiments",
  journal = "Annals of Mathematical Statistics",
  volume = "24 ",
  number= "2",
  pages ="265-272",
  year ="1953",
}

@book{cinlarprobability2010,
  title={Probability and Stochastics},
  author={{\c{C}}inlar, Erhan},
  year={2010},
  publisher={Springer}
}

@book{hart1995firms,
  title={Firms, Contracts, and Financial Structure},
  author={Hart, Oliver},
  year={1995},
  publisher={Clarendon Press}
}

@book{shiller1994macro,
  title={Macro Markets: Creating Institutions for Managing Society's Largest Economic Risks},
  author={Shiller, Robert J},
  year={1994},
  publisher={OUP Oxford}
}

@book{lessard1987capital,
  title={Capital Flight and the Third World Debt},
  author={Lessard, Donald R and Williamson, John},
  year={1987},
  publisher={Institute for International Economics}
}

@article{borensztein2004case,
  title={The Case for GDP-Indexed Bonds},
  author={Borensztein, Eduardo and Mauro, Paolo},
  journal={Economic Policy},
  volume={19},
  number={38},
  pages={166--216},
  year={2004},
  publisher={Oxford University Press}
}

@article{nachman1994optimal,
  title={Optimal Design of Securities under Asymmetric Information},
  author={Nachman, David C and Noe, Thomas H},
  journal={Review of Financial Studies},
  volume={7},
  number={1},
  pages={1--44},
  year={1994},
  publisher={Oxford University Press}
}

@article{casamatta2003financing,
  title={Financing and Advising: Optimal Financial Contracts with Venture Capitalists},
  author={Casamatta, Catherine},
  journal={Journal of Finance},
  volume={58},
  number={5},
  pages={2059--2085},
  year={2003},
  publisher={Wiley Online Library}
}

@article{schmidt1997managerial,
  title={Managerial Incentives and Product Market Competition},
  author={Schmidt, Klaus M},
  journal={Review of Economic Studies},
  volume={64},
  number={2},
  pages={191--213},
  year={1997},
  publisher={Wiley-Blackwell}
}

@article{eisfeldt2004endogenous,
  title={Endogenous Liquidity in Asset Markets},
  author={Eisfeldt, Andrea L},
  journal={Journal of Finance},
  volume={59},
  number={1},
  pages={1--30},
  year={2004},
  publisher={Wiley Online Library}
}

@article{allen1988optimal,
  title={Optimal Security Design},
  author={Allen, Franklin and Gale, Douglas},
  journal={Review of Financial Studies},
  volume={1},
  number={3},
  pages={229--263},
  year={1988},
  publisher={Oxford University Press}
}

@article{buchler2005competition,
  title={Competition, Representation and Redistricting: The Case Against Competitive Congressional Districts},
  author={Buchler, Justin},
  journal={Journal of Theoretical Politics},
  volume={17},
  number={4},
  pages={431--463},
  year={2005},
  publisher={Sage Publications Sage CA: Thousand Oaks, CA}
}

@article{Getal23,
  title={Optimal Insurance: Dual Utility, Random Losses, and Adverse Selection},
  author={Gershkov, Alex and Moldovanu, Benny and Strack, Philipp and Zhang, Mengxi},
  journal={American Economic Review},
  volume={113},
  number={10},
  pages={2581-2614},
  year={2023},
}

@Unpublished{gershkov2023optimal,
  author = {Gershkov, Alex and Moldovanu, Benny and Strack, Philipp and Zhang, Mengxi},
  note   = {Working Paper},
  title  = {Optimal Security Design for Risk-Averse Investors},
  year   = {2023},
}

@Unpublished{malamud2010optimal,
  author = {Malamud, Semyon and Rui, Huaxia and Whinston, Andrew B},
  note   = {Working Paper},
  title  = {Optimal Securitization with Heterogeneous Investors},
  year   = {2010},
}

@book{krehbiel2010pivotal,
  title={Pivotal Politics},
  author={Krehbiel, Keith},
  year={2010},
  publisher={University of Chicago Press}
}

@book{AN87,
  title={Linear Programming in Infinite-Dimensional Space},
  author={Edward J. Anderson and Peter Nash},
  year={1987},
  publisher={Wiley}
}

@article{HLP29,
  title={Some Simple Inequalities Satisfied by Convex Functions},
  author={Godfrey Hardy and John E. Littlewood and George P{\'o}lya },
  journal={Messenger Math},
  volume={58},
  number={4},
  pages={145--152},
  year={1929},
}

@article{S65,
  title={The Existence of Probability Measures with Given Marginals},
  author={Strassen, Volker},
  journal={Annuals of Mathematical Statistics},
  volume={36},
  pages={423-439},
  year={1965}
}

@article{R67,
  author = "John V. Ryff",
  year ="1967",
  title = "Extreme Points of Some Convex Subsets of $L^1(0,1)$",
  journal = "Proceedings of the American Mathematical Society",
  volume = "18",
  number = "6",
  pages = "1026-1034"
}

@article{KMS21,
  author = "Andreas Kleiner and Benny Moldovanu and Philipp Strack",
  year ="2021",
  title = "Extreme Points and Majorization: Economic Applications",
  journal = "Econometrica",
  volume = "89",
  number = "4",
  pages = "1557-1593"
}

@article{LM17,
  author = "Elliot Lipnowski and Laurent Mathevet",
  year ="2018",
  title = "Disclosure to a Psychological Audience",
  journal = "American Economic Journal: Microeconomics",
  volume = "10",
  number = "4",
  pages = "67-93"
}

@article{ABSY22,
  author = "Itai Arieli and Yakov Babichenko and Rann Smorodinsky and Takuro Yamashita",
  year ="2023",
  title = "Optimal Persuasion via Bi-Pooling",
  journal = "Theoretical Economics",
  volume = "18",
  number = "1", 
  pages = "15-36",
}

@article{cho2009bargaining,
  title={Bargaining Foundations of the Median Voter Theorem},
  author={Cho, Seok-ju and Duggan, John},
  journal={Journal of Economic Theory},
  volume={144},
  number={2},
  pages={851--868},
  year={2009},
  publisher={Elsevier}
}

@article{gilligan2006public,
  title={Public Choice Principles of Redistricting},
  author={Gilligan, Thomas W and Matsusaka, John G},
  journal={Public Choice},
  volume={129},
  number={3},
  pages={381--398},
  year={2006},
  publisher={Springer}
}

@Unpublished{ely2019cake,
  author  = {Ely, Jeffrey C},
  note    = {Working Paper},
  title   = {A Cake-Cutting Solution to Gerrymandering},
  year    = {2019},
}

@Unpublished{pegdenetal2017,
  author  = {Pegden, Wesley and Procaccia, Ariel D. and Yu, Dingli},
  note    = {Working Paper},
  title   = {A Partisan Districting Protocol with Provably Nonpartisan Outcomes},
  year    = {2017},
  journal = {Carnegie Mellon University},
}

@Unpublished{Yang2020generating,
  author  = {Yang, Kai Hao},
  note    = {Working Paper},
  title   = {A Note on Generating Arbitrary Joint Distributions Using Partitions},
  year    = {2020},
}

@Unpublished{N23,
  author  = {Afshin Nikzad},
  note    = {Working Paper},
  title   = {Constrained Majorization: Applications in Mechanism Design},
  year    = {2023},
  journal = {University of Southern California},
}

@article{shotts2002gerrymandering,
  title={Gerrymandering, Legislative Composition, and National Policy Outcomes},
  author={Shotts, Kenneth W},
  journal={American Journal of Political Science},
  pages={398--414},
  year={2002},
  publisher={JSTOR}
}

@article{biais2005strategic,
  title={Strategic Liquidity Supply and Security Design},
  author={Biais, Bruno and Mariotti, Thomas},
  journal={Review of Economic Studies},
  volume={72},
  number={3},
  pages={615--649},
  year={2005},
  publisher={Wiley-Blackwell}
}

@Unpublished{allen2022security,
  author  = {Allen, Franklin and Barbalau, Adelina},
  note = {Working Paper},
  title   = {Security Design: A Review},
  year    = {2022},
}

@article{mccarty2001hunt,
  title={The Hunt for Party Discipline in Congress},
  author={McCarty, Nolan and Poole, Keith T and Rosenthal, Howard},
  journal={American Political Science Review},
  volume={95},
  number={3},
  pages={673--687},
  year={2001},
  publisher={Cambridge University Press}
}

@article{bradbury2005legislative,
  title={Legislative District Configurations and Fiscal Policy in American States},
  author={Bradbury, John Charles and Crain, W Mark},
  journal={Public Choice},
  volume={125},
  number={3},
  pages={385--407},
  year={2005},
  publisher={Springer}
}

@article{BBM15,
  author = "Dirk Bergemann and Benjamin Brooks and Stephen Morris",
  title = "The Limits of Price Discrimination",
  journal = "American Economic Review",
  volume = "105",
  number = "3",
  pages ="921-957",
  year ="2015",
}

@article{RS17,
  author = "Roesler, Anne-Katrin and Bal\'{a}zs Szentes",
  year ="2017",
  title = "Buyer-Optimal Learning and Monopoly Pricing",
  journal = "American Economic Review",
  volume = "107",
  number = "7",
  pages = "2072-2080"
}

@article{KG11,
  author = "Emir Kamenica and Matthew Gentzkow",
   title = "Bayesian Persuasion",
 journal = "American Economic Review",
    year = "2011",
    pages = "2560-2615",
    volume = "101",
    number = "6"
    }

@article{GK16,
  author = "Matthew Gentzkow and Emir Kamenica",
   title = "A Rothschild-Stiglitz Approach to Bayesian Persuasion",
 journal = "American Economic Review: Papers and Proceedings",
    year = "2016",
    pages= "597-601",
    volume = "106",
    number = "5",
    
    }

@article{K19,
  author = "Emir Kamenica",
   title = "Bayesian Persuasion and Information Design",
 journal = "Annual Review of Economics",
    year = "2019",
    volume = "11",
    pages= "249-272"}

@article{Ketal17,
  author = "Kolotilin, Anton and Ming Li and Tymofiy Mylovanov and Andriy Zapechelnyuk",
  year ="2017",
  title = "Persuasion of a Privately Informed Receiver",
  journal = "Econometrica",
  volume = "85",
  number = "6",
  pages = "1949-1964"
}

@article{M86,
  author = "Paul Milgrom",
  year ="1981",
  title = "Good News and Bad News: Representation Theorems and Applications",
  journal = "Bell Journal of Economics",
  volume = "12",
  number = "2",
  pages = "380-391"
}

@article{S06,
  author = "Vasiliki Skreta",
  year ="2006",
  title = "Sequentially Optimal Mechanisms",
  journal = "Review of Economic Studies",
  volume = "73",
  number = "4",
  pages = "1085-1111"
}

@article{DM19,
  author = "Piotr Dworczak and Giorgio Martini",
  title = "The Simple Economics of Optimal Persuasion",
  journal = "Journal of Political Economy",
  volume = "127",
  year = "2019",
  pages = "1993-2048"
}

@article{KMZ22,
  author = "Anton Kolotilin and  Tymofiy Mylovanov and Andriy Zapechelnyuk",
  year ="2022",  
  title = "Censorship as Optimal Persuasion",
  journal = "Theoretical Economics",
  volume = "17",
  number = "2",
  pages = "561-585"
}

@article{gomberg2021electoral,
  author = "Andrei Gomberg and Romans Pancs and Tridib Sharma",
    volume = 64,
    number = 3,
  year =2023,
  title = "Electoral Maldistricting",
  journal = "International Economic Review",
 }

@article{rostek2010quantile,
  title={Quantile Maximization in Decision Theory},
  author={Rostek, Marzena},
  journal={Review of Economic Studies},
  volume={77},
  number={1},
  pages={339--371},
  year={2010},
  publisher={Wiley-Blackwell}
}

@article{chambers2007ordinal,
  title={Ordinal Aggregation and Quantiles},
  author={Chambers, Christopher P},
  journal={Journal of Economic Theory},
  volume={137},
  number={1},
  pages={416--431},
  year={2007},
  publisher={Elsevier}
}

@article{manski1988ordinal,
  title={Ordinal Utility Models of Decision Making under Uncertainty},
  author={Manski, Charles F},
  journal={Theory and Decision},
  volume={25},
  number={1},
  pages={79--104},
  year={1988},
  publisher={Springer}
}

@article{de2021static,
  title={Static and Dynamic Quantile Preferences},
  author={de Castro, Luciano and Galvao, Antonio F},
  journal={Economic Theory},
  pages={1--33},
  year={2021},
  publisher={Springer}
}

@article{R86,
  title={On Extremal Points of the Unit Ball in the Banach Space of Lipschitz Continuous Functions},
  author={Stefan Rolewicz},
  journal={Journal of the Australian Mathematical Society},
  volume={41},
  number={1},
  pages={95-98},
  year={1986}
}

@incollection{R84,
  author      = "Stefan Rolewicz",
  title       = "On Optimal Observability of Lipschitz Systems.",
  editor      = " G. Hammer and Diethard Pallaschke",
  booktitle   = "Selected Topics in Operations Research and Mathematical Economics: Proceedings of the 8th Symposium on Operations Research",
  publisher   = "Springer",
  address     = "Berlin",
  year        = 1984,
  pages       = "152-158"
}

@article{F94,
  title={Extreme Points of the Unit Ball of the Space of Lipschitz Functions},
  author={Jeff D. Farmer},
  journal={Proceedings of the American Mathematical Society},
  volume={121},
  number={3},
  pages={807-813},
  year={1994},
  publisher={JSTOR}
}

@article{S97,
  title={Extreme Points of Unit Balls in Lipschitz Function Spaces},
  author={Ryszard Smarzewski},
  journal={Proceedings of the American Mathematical Society},
  volume={125},
  number={5},
  pages={1391-1397},
  year={1997},
  publisher={JSTOR}
}

@article{K73,
  title={Integraldarstellung von Dilationen},
  author={Hans G. Kellerer},
  journal={in Transactions of the Sixth Prague Conference
on Information Theory, Statistical Decision Functions, Random Processes },
  pages={341-374},
  year={1973},
}

@article{L75,
  title={On an Ordering for Probability Measures and a Corresponding Integral Representation},
  author={Hans Joachim Lakeit},
  journal={Mathematische Annalen},
  volume={217},
  pages={229-240},
  year={1975},
}

@Unpublished{KW24,
  author = {Anton Kolotilin and Alexander Wolitzky},
  note   = {Working Paper},
  title  = {Distributions of Posterior Quantiles via Matching},
  year   = {2024},
}

@article{innes1990limited,
  title={Limited Liability and Incentive Contracting with Ex-ante Action Choices},
  author={Innes, Robert D},
  journal={Journal of Economic Theory},
  volume={52},
  number={1},
  pages={45--67},
  year={1990},
  publisher={Elsevier}
}

\end{spacing}

\clearpage{}

\newgeometry{left=1.8cm, bottom=2.2cm, right=1.8cm, top=2.2cm}
\appendix
\noindent\textbf{\LARGE{\labeltext{Appendix}{appendix}}}
\small
\renewcommand{\theequation}{A.\arabic{equation}}
\renewcommand{\thesubsection}{A.\arabic{subsection}}

\subsection{Proof of Theorem \ref{thm1}}\label{thm1p}
Consider any $\overline{F},\lb{F},H \in \mathcal{F}$ such that $\lb{F}(x)\leq H(x) \leq \ub{F}(x)$ for all $x \in \mathbb{R}$. We first show that if $H$ satisfies 1 and 2 for a countable collection of intervals $\{[\underline{x}_n,\overline{x}_n)\}_{n=1}^\infty$, then $H$ must be an extreme point of $\mathcal{I}(\underline{F},\overline{F})$. To this end, first note that $\mathcal{I}(\underline{F},\overline{F}) \subseteq \mathcal{F}$ is a convex subset of the collection of Borel-measurable functions on $\mathbb{R}$. Since the collection of Borel-measurable functions on $\mathbb{R}$ is a real vector space, it suffices to show that for any Borel-measurable $\widehat{H}$ with $\widehat{H} \neq 0$, either $H+\widehat{H} \notin \mathcal{I}(\underline{F},\overline{F})$ or $H-\widehat{H} \notin \mathcal{I}(\underline{F},\overline{F})$. Clearly, if either $H+\widehat{H} \notin \mathcal{F}$ or $H-\widehat{H} \notin \mathcal{F}$, then it must be that either $H+\widehat{H} \notin \mathcal{I}(\underline{F},\overline{F})$ or $H-\widehat{H} \notin \mathcal{I}(\underline{F},\overline{F})$. Thus, we may suppose that both $H+\widehat{H}$ and $H-\widehat{H}$ are in $\mathcal{F}$. Now notice that since $\widehat{H} \neq 0$, there exists $x_0 \in \mathbb{R}$ such that $\widehat{H}(x_0) \neq 0$. If $x_0 \notin \cup_{n=1}^\infty [\underline{x}_n,\overline{x}_n)$, then $H(x_0) \in \{\lb{F}(x_0),\ub{F}(x_0)\}$ and hence either $H(x_0)+|\widehat{H}(x_0)|>\ub{F}(x_0)$ or $H(x_0)-|\widehat{H}(x_0)|<\lb{F}(x_0)$. Thus, it must be that either $H+\widehat{H} \notin \mathcal{I}(\underline{F},\overline{F})$ or $H-\widehat{H} \notin \mathcal{I}(\underline{F},\overline{F})$. Meanwhile, if $x_0 \in [\underline{x}_n,\overline{x}_n)$ for some $n \in \mathbb{N}$, then $\widehat{H}$ must be constant on $[\underline{x}_n,\overline{x}_n)$, as $H$ is constant on $[\underline{x}_n,\overline{x}_n)$ and both $H+\widehat{H}$ and $H-\widehat{H}$ are nondecreasing. Thus, either $H(\underline{x}_n)+|\widehat{H}(\underline{x}_n)|=\ub{F}(\underline{x}_n)+|\widehat{H}(x_0)|>\ub{F}(\underline{x}_n)$, or $H(\overline{x}_n^-)-|\widehat{H}(
\overline{x}_n^-)|=\lb{F}(\overline{x}_n^-)-|\widehat{H}(x_0)|<\lb{F}(\overline{x}_n^-)$, and hence either $H+\widehat{H} \notin \mathcal{I}(\underline{F},\overline{F})$ or $H-\widehat{H} \notin \mathcal{I}(\underline{F},\overline{F})$, as desired. 

Conversely, suppose that $H$ is an extreme point of $\mathcal{I}(\underline{F},\overline{F})$. To show that $H$ must satisfy 1 and 2 for some countable collection of intervals $\{[\underline{x}_n,\overline{x}_n)\}_{n=1}^\infty$, we first claim that if $\lb{F}(x_0^-)<H(x_0):=\eta<\ub{F}(x_0)$ for some $x_0 \in \mathbb{R}$, then it must be that either $H(x)=H(x_0)$ for all $x \in [\ub{F}^{-1}(\eta^+),x_0]$ or $H(x)=H(x_0)$ for all $x \in [x_0,\lb{F}^{-1}(\eta))$. Indeed, suppose the contrary, so that there exists $\underline{x} \in [\ub{F}^{-1}(\eta^+),x_0)$ and $\overline{x} \in (x_0,\lb{F}^{-1}(\eta))$ such that $H(\underline{x})<H(x_0)<H(\overline{x}^-)$. Then, since $H$ is right-continuous, and since $H(\underline{x})<H(x_0)<H(\overline{x}^-)$, it must be that $H^{-1}(\eta)>\ub{F}^{-1}(\eta^+)$ and $H^{-1}(\eta^+)<\lb{F}^{-1}(\eta)$. Moreover, since $x \mapsto F(x^-)$ is left-continuous, $H^{-1}(\eta)>\underline{x} \geq \ub{F}^{-1}(\eta^+)$ implies $ \ub{F}(H^{-1}(\eta)^-)>\eta$. Likewise, $H^{-1}(\eta^+)<\overline{x}<\lb{F}^{-1}(\eta)$ implies that $\lb{F}(H^{-1}(\eta^+))<\eta$. 
Now define a function $\Phi:[0,1]^2 \to \mathbb{R}^2$ as 
\[
\Phi(\varepsilon_1,\varepsilon_2):=\begin{pmatrix}\eta-\varepsilon_2-\lb{F}(H^{-1}((\eta+\varepsilon_1)^+))\\\ub{F}(H^{-1}(\eta-\varepsilon_2)^-)-\eta-\varepsilon_1\end{pmatrix},
\]
for all $(\varepsilon_1,\varepsilon_2) \in [0,1]^2$. Then $\Phi$ is continuous at $(0,0)$ and $\Phi(0,0) \in \mathbb{R}_{++}^2$. Therefore, there exists $(\hat{\varepsilon}_1,\hat{\varepsilon}_2) \in [0,1]^2\backslash \{(0,0)\}$ such that $\Phi(\hat{\varepsilon_1},\hat{\varepsilon}_2)\in \mathbb{R}_{++}^2$. Let $\underline{\eta}:=\eta-\hat{\varepsilon}_2$ and $\overline{\eta}:=\eta+\hat{\varepsilon}_1$, it then follows that 
\begin{equation}\label{bound}
\lb{F}(H^{-1}(\underline{\eta}^+)^-) \leq \lb{F}(H^{-1}(\overline{\eta}^+))<\underline{\eta}<\eta<\overline{\eta}<\ub{F}(H^{-1}(\underline{\eta})^-) \leq \ub{F}(H^{-1}(\underline{\eta})).
\end{equation}
Now consider the function $h:[H^{-1}(\underline{\eta}),H^{-1}(\overline{\eta}^+)] \to [\underline{\eta},\overline{\eta}]$, defined as $h(x):=H(x)$, for all $x \in [H^{-1}(\underline{\eta}),H^{-1}(\overline{\eta}^+)]$. Clearly, $h$ is nondecreasing. As a result, since the extreme points of the collection of uniformly bounded monotone functions are step functions with at most one jump (see, for instances, \citealp{S06} and \citealp{B14}), $\underline{\eta}<h(x_0)=H(x_0)=\eta<\overline{\eta}$ implies that there exists distinct nondecreasing, right-continuous functions $h_1,h_2$ that map from $[H^{-1}(\underline{\eta}),H^{-1}(\overline{\eta}^+)]$ to $[\underline{\eta},\overline{\eta}]$, as well as a constant $\lambda \in (0,1)$ such that $h(x)=\lambda h_1(x) +(1-\lambda) h_2(x)$, for all $x \in [H^{-1}(\underline{\eta}),H^{-1}(\overline{\eta}^+)]$. Now define $\widehat{H}_1, \widehat{H}_2$ as 
\[
\widehat{H}_1(x):=\left\{
\begin{array}{cc}
H(x),&\mbox{if } x \notin [H^{-1}(\underline{\eta}),H^{-1}(\overline{\eta}^+)]\\
h_1(x),&\mbox{if } x \in [H^{-1}(\underline{\eta}),H^{-1}(\overline{\eta}^+)]
\end{array}
\right.;
\]
and
\[
\widehat{H}_2(x):=\left\{
\begin{array}{cc}
H(x),&\mbox{if } x \notin [H^{-1}(\underline{\eta}),H^{-1}(\overline{\eta}^+)]\\
h_2(x),&\mbox{if } x \in [H^{-1}(\underline{\eta}),H^{-1}(\overline{\eta}^+)]
\end{array}
\right.
\]
Clearly, $\lambda \widehat{H}_1+(1-\lambda) \widehat{H}_2=H$. 

It now remains to show that $\widehat{H}_1,\widehat{H}_2 \in \mathcal{I}(\underline{F},\overline{F})$. Indeed, for any $i \in \{1,2\}$ and for any $x,y \in \mathbb{R}$ with $x<y$, if $x,y \notin [H^{-1}(\underline{\eta}),H^{-1}(\overline{\eta}^+)]$, then $\widehat{H}_i(x)=H(x) \leq H(y) =\widehat{H}_i(y)$, since $H$ is nondecreasing. Meanwhile, if $x,y \in [H^{-1}(\underline{\eta}),H^{-1}(\overline{\eta}^+)]$, then $\widehat{H}_i(x)=h_i(x) \leq h_i(y)=\widehat{H}_i(y)$. If $x<H^{-1}(\underline{\eta})$ and $y \in [H^{-1}(\underline{\eta}),H^{-1}(\overline{\eta}^+)]$, then $\widehat{H}_i(x)=H(x) \leq \underline{\eta} \leq h_i(y)=\widehat{H}_i(y)$. Likewise, if $y>H^{-1}(\overline{\eta}^+)$ and $x \in [H^{-1}(\underline{\eta}),H^{-1}(\overline{\eta}^+)]$, then $\widehat{H}_i(x)=h_i(x) \leq \overline{\eta} \leq H(y)=\widehat{H}_i(y)$. Together, $\widehat{H}_i$ must be nondecreasing, and hence $\widehat{H}_i \in \mathcal{F}$ for all $i \in \{1,2\}$. Moreover, for any $i \in \{1,2\}$ and for all $x \in [H^{-1}(\underline{\eta}),H^{-1}(\overline{\eta}^+)]$, from \eqref{bound}, we have 
\[
\lb{F}(x) \leq \lb{F}(H^{-1}(\overline{\eta}^+))<\underline{\eta} \leq h_i(x)\leq \overline{\eta}<\ub{F}(H^{-1}(\eta)^-) \leq \ub{F}(x).
\]
Together with $H \in \mathcal{I}(\underline{F},\overline{F})$, it then follows that $\lb{F}(x) \leq \widehat{H}_i(x) \leq \ub{F}(x)$ for all $x \in \mathbb{R}$, and hence $\widehat{H}_i \in \mathcal{I}(\underline{F},\overline{F})$ for all $i \in \{1,2\}$. Consequently, there exists distinct $\widehat{H}_1,\widehat{H}_2 \in \mathcal{I}(\underline{F},\overline{F})$ and $\lambda \in (0,1)$ such that $H=\lambda\widehat{H}_1+(1-\lambda)\widehat{H}_2$. Thus, $H$ is not an extreme point of $\mathcal{I}(\underline{F},\overline{F})$, as desired. 

As a result, for any extreme point $H$ of $\mathcal{I}(\underline{F},\overline{F})$, the set $\{x \in \mathbb{R}|\lb{F}(x)<H(x)<\ub{F}(x)\}$ can be partitioned into three classes of open intervals: ${I}^{\ub{F}}$, ${I}^{\lb{F}}$, and ${I}^{\ub{F},\lb{F}}$ such that for any open interval $(\underline{x},\overline{x}) \in {I}^{\ub{F}}$, $H$ is a constant on $[\underline{x},\overline{x})$ and $H(\underline{x})=\ub{F}(\underline{x})$; for any open interval $(\underline{x},\overline{x}) \in {I}^{\lb{F}}$, $H$ is a constant on $[\underline{x},\overline{x})$ and $H(\overline{x}^-)=\lb{F}(\overline{x}^-)$; and for any open interval $(\underline{x},\overline{x}) \in {I}^{\ub{F},\lb{F}}$, $H$ is a constant on $[\underline{x},\overline{x})$ and $\ub{F
}(\underline{x})=H(\underline{x})=H(\overline{x}^-)=\lb{F}(\overline{x}^-)$. Note that since $\ub{F},\lb{F},H$ are nondecreasing and since $H \in \mathcal{I}(\underline{F},\overline{F})$, every interval in $I^{\ub{{F}}}$ and $I^{\lb{F}}$ must have at least one of its end points being a discontinuity point of $H$. Since $H$ has at most countably many discontinuity points, $I^{\ub{F}}$ and $I^{\lb{F}}$ must be countable. Meanwhile, any distinct intervals $(\underline{x}_1,\overline{x}_1),(\underline{x}_2,\overline{x}_2) \in I^{\ub{F},\lb{F}}$  must be disjoint. Moreover, for any pair of these intervals with $\overline{x}_1<\underline{x}_2$, there must exist some $x_0 \in (\overline{x}_1,\underline{x}_2)$ at which $H$ is discontinuous. Therefore, since $H$ has at most countably many discontinuity points, $I^{\ub{F},\lb{F}}$ must be countable as well. 

Together, for any extreme point $H$ of $\mathcal{I}(\underline{F},\overline{F})$, there exists countably many intervals $\{[\underline{x}_n,\overline{x}_n)\}_{n=1}^\infty:=I^{\ub{F}} \cup I^{\lb{F}} \cup I^{\ub{F},\lb{F}}$ such that $H$ satisfies 1 and 2. This completes the proof. \hfill $\blacksquare$

\subsection{Proof of Theorem \ref{thm2}}\label{thm2p}
To show that $\mathcal{H}_\tau \subseteq \mathcal{I}(F_R^\tau,F_L^\tau)$, consider any $H \in \mathcal{H}_\tau$. Let $\mu \in \mathcal{M}$ and any $r \in \mathcal{R}_\tau$ be a signal and a selection rule, respectively, such that $H^\tau(\cdot|\mu,r)=H$. By the definition of $H^\tau(\cdot|\mu,r)$, it must be that, for all $\omega \in \mathbb{R}$,
\begin{align*}
H^\tau(\omega|\mu,r) \leq\mu(\{G \in \mathcal{F}_0|G^{-1}(\tau) \leq \omega\})=\mu(\{G \in \mathcal{F}_0|G(x)\geq \tau\}).
\end{align*}
Now consider any $\omega \in \mathbb{R}$. Clearly, $\mu(\{G \in \mathcal{F}_0|G(\omega)\geq \tau\}) \leq 1$, since $\mu$ is a probability measure. Moreover, let $M^+_\omega(q):=\mu(\left\{G \in \mathcal{F}_0|G(\omega) \geq q\right\})$ for all $q \in [0,1]$. From \eqref{bp}, it follows that the left-limit of $1-M^+_x$ is a CDF and a mean-preserving spread of a Dirac measure at $F(\omega)$. Therefore, whenever $\tau \geq F(\omega)$, then $M^+_\omega(\tau)$ can be at most $F(\omega)/\tau$ to have a mean of $F(\omega)$.\footnote{More specifically, to maximize the probability at $\tau$, a mean-preserving spread of $F(\omega)$ must assign probability $F(\omega)/\tau$ at $\tau$, and probability $1-F(x)/\tau$ at $0$.} Together, this implies that $\mu(\{G\in \mathcal{F}_0|G(x) \geq \tau\}) \leq F_L^\tau(\omega)$ for all $\omega \in \mathbb{R}$. 

At the same time, by the definition of $H^\tau(\cdot|\mu,r)$, it must be that, for all $\omega \in \mathbb{R}$,
\begin{align*}
H^\tau(\omega^-|\mu,r) \geq \mu(\{G \in \mathcal{F}_0|G^{-1}(\tau^+)<\omega\})=\mu(\{G \in \mathcal{F}_0|G(x)>\tau\}). 
\end{align*}
Consider any $\omega \in \mathbb{R}$. Since $\mu$ is a probability measure, it must be that $\mu(\{G \in \mathcal{F}_0|G(\omega)>\tau\}) \geq 0$. Furthermore, let $M^-_\omega(q):=\mu(\{G \in \mathcal{F}_0|G(\omega)>q\})$ for all $q \in [0,1]$. From \eqref{bp}, it follows that $1-M^-_x$ is a CDF and a mean-preserving spread of a Dirac measure at $F(\omega)$. Therefore, whenever $\tau \leq F(\omega)$, then $M^-_\omega(\tau)$ must be at least $(F(\omega)-\tau)/(1-\tau)$ to have a mean of $F(\omega)$.\footnote{More specifically, to minimize the probability at $\tau$, a mean-preserving spread of $F_0(x)$ must assign probability $(F(\omega)-\tau)/(1-\tau)$ at $1$, and probability $1-(F(\omega)-\tau)/(1-\tau)$ at $0$.} Together, this implies that $\mu(\{G \in \mathcal{F}_0|G(\omega)>\tau\}) \geq F_R^\tau$ for all $\omega \in \mathbb{R}$, which, in turn, implies that $F_R^\tau(\omega) \leq H^\tau(\omega^-|\mu,r) \leq H^\tau(\omega|\mu,r) \leq F_L^\tau(\omega)$ for all $\omega \in \mathbb{R}$, as desired.  

To prove that $\mathcal{I}(F_R^\tau,F_L^\tau) \subseteq \mathcal{H}_\tau$, we first show that for any extreme point $H$ of $\mathcal{I}(F_R^\tau,F_L^\tau)$, there exists a signal $\mu \in \mathcal{M}$ and a selection rule $r\in \mathcal{R}_\tau$ such that $H(\omega)=H^\tau(\omega|\mu,r)$ for all $\omega \in \mathbb{R}$. Consider any extreme point $H$ of $\mathcal{I}(F_R^\tau,F_L^\tau)$. By \bref{thm1}, there exists a countable collection of intervals $\{(\underline{x}_n,\overline{x}_n)\}_{n=1}^\infty$ such that $H$ satisfies 1 and 2. Since $(1-F_L^\tau(x))F_R^\tau(x)=0$ for all $x \notin [F^{-1}(\tau),F^{-1}(\tau^+)]$, there exists at most one $n \in \mathbb{N}$ such that $0<H(\underline{x}_n)=F_L^\tau(\underline{x}_n)=F_R^\tau(\overline{x}_n^-)=H(\overline{x}_n^-)<1$. Therefore, for $\underline{x}$ and $\overline{x}$ defined as
\[
\underline{x}:=\sup\{\underline{x}_n|n \in \mathbb{N}, \, H(\underline{x}_n)=F_L^\tau(\underline{x}_n)\} ,
\]
and
\[
\overline{x}:=\inf \{\overline{x}_n|n \in \mathbb{N}, \, H(\overline{x}_n^-)=F_R^\tau(\overline{x}_n^-)\}, 
\]
respectively, it must be that $\overline{x} \geq \underline{x}$, and that for all $n \in \mathbb{N}$, either $\overline{x}_n \leq \underline{x}$ and $H(\underline{x_n})=F_L^\tau(\underline{x}_n)$; or $\underline{x}_n \geq \overline{x}$ and $H(\overline{x}_n^-)=F_R^\tau(\overline{x}_n^-)$. Henceforth, let $\mathbb{N}_1$ be the collection of $n \in \mathbb{N}$ such that $\overline{x}_n \leq \overline{x}$ and $H(\underline{x}_n)=F_L^\tau(\underline{x}_n)$, and let $\mathbb{N}_2$ be the collection of $n \in \mathbb{N}$ such that $\underline{x}_n \geq \underline{x}$ and $H(\overline{x}_n^-)=F_R^\tau(\overline{x}_n^-)$. Note that $\mathbb{N}_1 \cup \mathbb{N}_2=\mathbb{N}$ and that $|\mathbb{N}_1 \cap \mathbb{N}_2| \leq 1$, with $\underline{x}_n=\underline{x}$ and $\overline{x}_n=\overline{x}$ whenever $n \in \mathbb{N}_1 \cap \mathbb{N}_2$. 

We now construct a signal $\mu \in \mathcal{M}$ and a selection rule $r \in \mathcal{R}_\tau$ such that $H^{\tau}(\cdot|\mu,r)=H$. To this end, let $\eta:=H(\overline{x}^-)-H(\underline{x})$ and let $\hat{x}:=\inf\{x \in [\underline{x},\overline{x}]|H(x)=H(\overline{x}^-)\}$. Note that by the definition of $\underline{x}$ and $\overline{x}$, if $\eta>0$, then $\hat{x} \in (\underline{x},\overline{x})$ and $H(x)=H(\underline{x})$ for all $x \in [\underline{x},\hat{x})$, while $H(x)=H(\overline{x}^-)$ for all $x \in [\hat{x},\overline{x})$. In particular, $F_L^\tau(\hat{x}) \geq H(\hat{x})=F_L^\tau(\underline{x})+\eta$, and hence $F(\hat{x})-\tau\eta \geq F(\underline{x})$. Likewise, $F(\hat{x})+(1-\tau)\eta \leq F(\overline{x}^-)$. Let
\[
\underline{y}:=F^{-1}([F(\hat{x})-\tau\eta]^+), \quad \mbox{ and } \quad \overline{y}:=F^{-1}(F(\hat{x})+(1-\tau)\eta)).
\]
It then follows that $\underline{x} \leq \underline{y} \leq \hat{x} \leq \overline{y} \leq \overline{x}$, with at least one inequality being strict if $\eta>0$. Next, define $\widehat{F}$ as follows: $\widehat{F} \equiv 0$ if $\eta=0$; and 
\[
\widehat{F}(x):=\left\{
\begin{array}{cc}
0,&\mbox{if } x<\underline{y}\\
\frac{F(x)-(F(\hat{x})-\tau\eta)}{\eta},&\mbox{if } x \in [\underline{y},\overline{y})\\
1,&\mbox{if } x\geq \overline{y}
\end{array}
\right.,
\]
if $\eta>0$. Clearly $\widehat{F} \in \mathcal{F}_0$ if $\eta>0$, and $\hat{x} \in [\widehat{F}^{-1}(\tau),\widehat{F}^{-1}(\tau^+)]$. Moreover, for all $x \in \mathbb{R}$, let 
\[
\widetilde{F}(x):=\frac{F(x)-\eta\widehat{F}(x)}{1-\eta}.
\]
By construction, $\eta \widehat{F}+(1-\eta)\widetilde{F}=F$. From the definition of $\underline{y}$ and $\overline{y}$, it can be shown that $\widetilde{F} \in \mathcal{F}_0$ as well. Furthermore, 
\[
\widetilde{F}(\overline{x}^-)-\widetilde{F}(\underline{x})=\frac{F(\overline{x}^-)-F(\underline{x})-\eta}{1-\eta}=\frac{1}{1-\eta}\left[\frac{\tau}{1-\tau}(1-F(\overline{x}^-))+\frac{1-\tau}{\tau}F(\underline{x})\right].  
\]
Next, define $\widetilde{F}_1$ and $\widetilde{F}_2$ as follows: 
\[
\widetilde{F}_1(x):=\left\{
\begin{array}{cc}
\frac{F(x)}{F(\underline{x})+\alpha(F(\overline{x}^-)-F(\underline{x})-\eta)},&\mbox{if } x<\underline{x}\\
\frac{F(\underline{x})\alpha(F(x)-F(\underline{x})-\eta)}{F(\underline{x})+\alpha(F(\overline{x}^-)-F(\underline{x})-\eta)},& \mbox{if } x \in [\underline{x},\overline{x})\\
1,&\mbox{if } x \geq \overline{x}
\end{array}
\right.; 
\]
and
\[
\widetilde{F}_2(x):=\left\{
\begin{array}{cc}
0,&\mbox{if } x<\underline{x}\\
\frac{(1-\alpha)(F(x)-F(\underline{x})-\eta)}{1-F(\overline{x}^-)+(1-\alpha)(F(\overline{x}^-)-F(\underline{x})-\eta)},&\mbox{if } x \in [\underline{x},\overline{x})\\
\frac{F(x)-F(\underline{x})+(1-\alpha)(F(\overline{x}^-)-F(\underline{x})-\eta)}{1-F(\overline{x}^-)+(1-\alpha)(\widetilde{F}(\overline{x}^-)-\widetilde{F}(\underline{x})-\eta)},& \mbox{if } x \geq \overline{x}
\end{array}
\right.,
\]
where 
\[
\alpha:=\frac{\frac{1-\tau}{\tau}F(\underline{x})}{\frac{\tau}{1-\tau}(1-F(\overline{x}^-))+\frac{1-\tau}{\tau}F(\underline{x})}.
\]
By construction, $\widetilde{\alpha}\widetilde{F}_1+(1-\widetilde{\alpha})\widetilde{F}_2=\widetilde{F}$, where $\widetilde{\alpha} \in (0,1)$ is given by $\widetilde{\alpha}:=[F(\underline{x})+\alpha (F(\overline{x}^-)-F(\underline{x})-\eta)]/(1-\eta)$. Moreover, $\widetilde{F}_1(\underline{x}) \geq \tau$, and $\widetilde{F}_2(\overline{x}^-) \leq \tau$.

Now define two classes of distributions, $\{\widetilde{F}_1^x\}_{x \leq \underline{x}}$ and $\{\widetilde{F}_2^x\}_{x \geq \overline{x}}$, as follows: 
\[
\widetilde{F}^x_1(z):=\left\{
\begin{array}{cc}
0,&\mbox{if }z < x\\
\widetilde{F}(\underline{x}),&\mbox{if } z \in [x,\underline{x})\\
\widetilde{F}(z),&\mbox{if } z \geq \underline{x}\\
\end{array}
\right.;\mbox{ and }
\widetilde{F}_2^x(z):=\left\{
\begin{array}{cc}
\widetilde{F}(z),&\mbox{if } z <\overline{x}\\
\widetilde{F}(\overline{x}^-),&\mbox{if } z \in [\overline{x},x)\\
1,&\mbox{if } z \geq x
\end{array}
\right..
\]
Note that since $\widetilde{F}_1(\underline{x}) \geq \tau$ and $\widetilde{F}_2(\overline{x}^-) \leq \tau$, $x \in [(\widetilde{F}_1^x)^{-1}(\tau),(\widetilde{F}_1^x)^{-1}(\tau^+)]$ for all $x \leq \underline{x}$ and $x \in [(\widetilde{F}_2^x)^{-1}(\tau),(\widetilde{F}_2^x)^{-1}(\tau^+)]$ for all $x \geq \overline{x}$. Moreover, for any $n \in \mathbb{N}_1$ and for any $m \in \mathbb{N}_2$, let 
\[
\widetilde{F}_1^n(z):=\frac{1}{\widetilde{F}(\overline{x}_n)-\widetilde{F}(\underline{x}_n)}\int_{\underline{x}_n}^{\overline{x}_n} \widetilde{F}_1^\omega(z) \widetilde{F}(\diff x), 
\]
and
\[
\widetilde{F}_2^{m}(z):=\frac{1}{\widetilde{F}(\overline{x}_m)-\widetilde{F}(\underline{x}_m)}\int_{\underline{x}_m}^{\overline{x}_m} \widetilde{F}_2^\omega(z)\diff \widetilde{F}(\diff \omega),
\]
for all $z \in \mathbb{R}$. By construction, $\widetilde{F}_1^{n}, \widetilde{F}_2^{m} \in \mathcal{F}_0$ and $\overline{x}_n \in [(\widetilde{F}_1^n)^{-1}(\tau),(\widetilde{F}_1^n)^{-1}(\tau^+)]$, $\underline{x}_m \in [(\widetilde{F}_2^m)^{-1}(\tau),(\widetilde{F}_2^m)^{-1}(\tau^+)]$ for all $n \in \mathbb{N}_1$ and $m \in \mathbb{N}_2$. 

Next, for any $\omega \in \mathbb{R}$, let $\widetilde{G}^x \in \mathcal{F}_0$ be defined as
\[
\widetilde{G}^\omega(z):=\left\{
\begin{array}{cc}
\widetilde{F}_1^\omega(z),& \mbox{if } \omega \in (-\infty,\overline{x}]\backslash \cup_{n \in \mathbb{N}_1} [\underline{x}_n,\overline{x}_n)\\
\widetilde{F}_1^n(z),& \mbox{if } \omega \in [\underline{x}_n,\overline{x}_n), \, n \in \mathbb{N}_1\\
\widetilde{F}_2^\omega(z),& \mbox{if } \omega \in [\overline{x},\infty) \backslash \cup_{m \in \mathbb{N}_2} [\underline{x}_m,\overline{x}_m)\\
\widetilde{F}_2^m(z),&\mbox{if } \omega \in [\underline{x}_m,\overline{x}_m), \, m \in \mathbb{N}_2
\end{array}
\right.,
\]
for all $z \in \mathbb{R}$. Let 
\[
\widetilde{H}(x):=\left\{
\begin{array}{cc}
\frac{H(x)}{1-\eta},&\mbox{if } x<\underline{x}\\
\frac{H(\underline{x})}{1-\eta},&\mbox{if } x \in [\underline{x},\overline{x})\\
\frac{H(x)-\eta}{1-\eta},&\mbox{if } x \geq \overline{x}
\end{array}
\right.,
\]
and define $\tilde{\mu}$ as 
\[
\tilde{\mu}(\{\widetilde{G}^x \in \mathcal{F}_0|\omega \leq z\}):=\widetilde{H}(z),
\]
for all $z \in \mathbb{R}$. Then, by construction, for any $z \in \mathbb{R}$,
\begin{equation}\label{avgmu}
\int_{\mathcal{F}} F(z)\tilde{\mu}(\diff F)=\int_\mathbb{R} \widetilde{G}^{\omega}(z)\widetilde{H}(\diff \omega)=\widetilde{F}(z).
\end{equation} 
Moreover, let $\tilde{r}:\mathcal{F}_0 \to \Delta(\mathbb{R})$ be defined as 
\[
\tilde{r}(G):=\left\{
\begin{array}{cc}
\delta_{\{G^{-1}(\tau^+)\}}, &\mbox{if } G=\widetilde{G}^x, \, \omega \geq \overline{x}\\  
\delta_{\{G^{-1}(\tau)\}}, &\mbox{otherwise} 
\end{array}
\right.,
\]
for all $G \in \mathcal{F}_0$. It then follows that $H^{\tau}(x|\tilde{\mu},\tilde{r})=\widetilde{H}(x)$ for all $x \in \mathbb{R}$. Next, let $\mu \in \Delta(\mathcal{F}_0),r \in \mathcal{R}_\tau$ together be defined as 
\[
\mu:=(1-\eta)\tilde{\mu}+\eta\delta_{\{\widehat{F}\}}, 
\]
and
\[
r(G):=\left\{
\begin{array}{cc}
\delta_{\{\hat{x}\}},& \mbox{if } G=\widehat{F}\\
\tilde{r}(G),&\mbox{otherwise}
\end{array}
\right.,
\]
for all $G \in \mathcal{F}_0$. Since $F=\eta \widehat{F}+(1-\eta)\widetilde{F}$, together with \eqref{avgmu}, we have $\mu \in \mathcal{M}$. Moreover, since $H^\tau(\cdot|\tilde{\mu},\tilde{r})=\widetilde{H}$, we have $H^\tau(x|\mu,r)=H(x)$ for all $x \in \mathbb{R}$.

Lastly, let $\Gamma$ be a collection of probability measures $\gamma \in \Delta(\mathbb{R} \times \mathcal{F}_0)$ such that $\gamma(\{(\omega,G) \in \mathbb{R}\times\mathcal{F}_0|\omega \in [G^{-1}(\tau),
G^{-1}(\tau^+)]\})=1$ and 
\[
\int_{\mathbb{R} \times \mathcal{F}_0} G(\omega) \gamma(\diff \omega,\diff G)=F(\omega),
\]
for all $\omega \in \mathbb{R}$. Define a linear functional $\Xi:\Gamma \to \mathcal{F}_0$ as 
\[
\Xi(\gamma)[\omega]:=\gamma((-\infty,\omega],\mathcal{F}_0),
\]
for all $\gamma \in \Gamma$ and for all $\omega \in \mathbb{R}$. Then, since for any $\widehat{H}$ in the set of extreme points $\mathrm{ext}(\mathcal{I}(F_R^\tau,F_L^\tau))$ of $\mathcal{I}(F_R^\tau,F_L^\tau)$, there exists $\hat{\mu} \in \mathcal{M}$ and $\hat{r} \in \mathcal{R}_\tau$ such that $H^{\tau}(\omega|\hat{\mu},\hat{r})=\widehat{H}(\omega)$ for all $\omega \in \mathbb{R}$, it must be that $\mathrm{ext}(\mathcal{I}(F_R^\tau,F_L^\tau)) \subseteq \Xi(\Gamma)$.

Now consider any $H \in\mathcal{I}(F_R^\tau,F_L^\tau)$. Since $\mathcal{I}(F_R^\tau,F_L^\tau)$ is a compact and convex set of a metrizable, locally convex topological space,\footnote{To see this, recall that for any sequence $\{H_n\} \subseteq \mathcal{I}(F_R^\tau,F_L^\tau)$, Helly's selection theorem implies that there exists a subsequence $\{H_{n_k}\} \subseteq \{H_n\}$ that converges pointwise (and hence, in weak-*) to some $H \in \mathcal{I}(F_R^\tau,F_L^\tau)$.} Choquet's theorem implies that there exists a probability measure $\Lambda_H \in \Delta(\mathcal{I}(F_R^\tau,F_L^\tau))$ with $\Lambda_H(\mathrm{ext}(\mathcal{I}(F_R^\tau,F_L^\tau)))=1$ such that 
\[
\int_{\mathcal{I}(F_R^\tau,F_L^\tau)}\widehat{H}(\omega)\Lambda_H(\diff \widehat{H})=H(\omega),
\]
for all $\omega \in \mathbb{R}$. Define a measure $\widetilde{\Lambda}_H$ by 
\[
\widetilde{\Lambda}_H(A):=\Lambda_H(\{\Xi(\gamma)|\gamma \in A\}),
\]
for all measurable $A \subseteq \Gamma$. Since $\Lambda_H(\mathrm{ext}(\mathcal{I}(F_R^\tau,F_L^\tau)))=1$ and $\mathrm{ext}(\mathcal{I}(F_R^\tau,F_L^\tau)) \subseteq \Xi(\Gamma)$, $\widetilde{\Lambda}_H$ is a probability measure on $\Gamma$. For any $\omega \in \mathbb{R}$ and for any measurable $A \subseteq \mathcal{F}_0$, let 
\[
\gamma((-\infty,\omega],A):=\int_{\Gamma} \tilde{\gamma}((-\infty,\omega],A) \widetilde{\Lambda}_H(\diff \tilde{\gamma}),
\]
and let $\mu(A):=\gamma(\mathbb{R},A)$. By construction, for all $\omega \in \mathbb{R}$,
\[
\int_{\mathcal{F}} G(\omega) \mu(\diff G)=\int_{\Gamma}\left(\int_{\mathbb{R}\times\mathcal{F}_0}G(\omega) \tilde{\gamma}(\diff \tilde{\omega},\diff G)\right)\widetilde{\Lambda}_H(\diff \tilde{\gamma})=F(\omega),
\]
and hence $\mu \in \mathcal{M}$. Furthermore, by the disintegration theorem (c.f., \citealp*{cinlarprobability2010}, theorem 2.18), there exists a transition probability $r:\mathcal{F}_0 \to \Delta(\mathbb{R})$ such that $\gamma(\diff \omega,\diff G)=r(\diff \omega|G)\mu(\diff G)$. Since $\widetilde{\Lambda}_H(\Gamma)=1$, and since $r$ is a transition probability, we have $r \in \mathcal{R}_\tau$. Finally, for any $\omega \in \mathbb{R}$, since $\Xi$ is affine, 
\begin{align*}
H^\tau(\omega|\mu,r)=\gamma((-\infty,\omega],\mathcal{F}_0)=&\Xi(\gamma)[\omega]\\
=& \int_{\Gamma}\Xi(\tilde{\gamma})[\omega]\widetilde{\Lambda}_H(\diff \tilde{\gamma})\\
=& \int_{\mathrm{ext}(\mathcal{I}(F_R^\tau,F_L^\tau) )}\widehat{H}(\omega)\Lambda_H(\diff \widehat{H})\\
=&H(\omega),
\end{align*}
as desired. This completes the proof. \hfill $\blacksquare$ 

\subsection{Proof of Theorem \ref{thm3}}
By \bref{thm2}, 
\[
\widetilde{\mathcal{H}}_\tau \subseteq \mathcal{H}_\tau = \mathcal{I}(F_R^\tau,F_L^\tau).
\]
It remains to show that 
\[
\bigcup_{\varepsilon>0} \mathcal{I}(F_R^{\tau,\varepsilon},F_L^{\tau,\varepsilon}) \subseteq \widetilde{\mathcal{H}}_\tau.
\]
To this end, let $\widetilde{\mathcal{M}}_\tau$ be the collection of $\mu \in \mathcal{M}$ such that $\mu(\{G \in \mathcal{F}_0|G^{-1}(\tau)<G^{-1}(\tau^+)\})=0$. Consider any $\varepsilon>0$ and any extreme point $H$ of  $\mathcal{I}(F_R^{\tau,\varepsilon},F_L^{\tau,\varepsilon})$. By \bref{thm1}, there exists a countable collection of intervals $\{(\underline{x}_n,\overline{x}_n)\}_{n=1}^\infty$ such that $H$ satisfies 1 and 2. Since $(1-F_R^{\tau,\varepsilon}(x))F_L^{\tau,\varepsilon}(x)=0$ for all $x \neq F_0^{-1}(\tau)$, there exists at most one $n \in \mathbb{N}$ such that $0<H(\underline{x}_n)=F_R^{\tau,\varepsilon}(\underline{x}_n)=F_L^{\tau,\varepsilon}(\overline{x}_n^-)=H(\overline{x}_n^-)<1$. Therefore, for $\underline{x}$ and $\overline{x}$ defined as
\[
\underline{x}:=\sup\{\underline{x}_n|n \in \mathbb{N}, \, H(\underline{x}_n)=F_R^{\tau,\varepsilon}(\underline{x}_n)\} \quad \mbox{ and } \quad \overline{x}:=\inf \{\overline{x}_n|n \in \mathbb{N}, \, H(\overline{x}_n^-)=F_L^{\tau,\varepsilon}(\overline{x}_n^-)\}, 
\]
respectively, it must be that $\overline{x} \geq \underline{x}$, and that, for all $n \in \mathbb{N}$, either $\overline{x}_n \leq \underline{x}$ and $H(\underline{x_n})=F_L^{\tau,\varepsilon}(\underline{x}_n)$, or $\underline{x}_n \geq \overline{x}$ and $H(\overline{x}_n^-)=F_R^{\tau,\varepsilon}(\overline{x}_n^-)$. Henceforth, let $\mathbb{N}_1$ be the collection of $n \in \mathbb{N}$ such that $\overline{x}_n \leq \overline{x}$ and $H(\underline{x}_n)=F_L^{\tau,\varepsilon}(\underline{x}_n)$, and let $\mathbb{N}_2$ be the collection of $n \in \mathbb{N}$ such that $\underline{x}_n \geq \underline{x}$ and $H(\overline{x}_n^-)=F_R^{\tau,\varepsilon}(\overline{x}_n^-)$. Note that $\mathbb{N}_1 \cup \mathbb{N}_2=\mathbb{N}$ and that $|\mathbb{N}_1 \cap \mathbb{N}_2| \leq 1$, with $\underline{x}_n=\underline{x}$ and $\overline{x}_n=\overline{x}$ whenever $n \in \mathbb{N}_1 \cap \mathbb{N}_2$. 

We now construct a signal $\mu \in \widetilde{\mathcal{M}}_\tau$ such that $H^{\tau}(\cdot|\mu)=H$. First, let $\eta:=H(\overline{x}^-)-H(\underline{x})$ and let $\hat{x}:=\inf\{x \in [\underline{x},\overline{x}]|H(x)=H(\overline{x}^-)\}$. Note that, by the definition of $\underline{x}$ and $\overline{x}$, if $\eta>0$, then $\hat{x} \in (\underline{x},\overline{x})$ and $H(x)=H(\underline{x})$ for all $x \in [\underline{x},\hat{x})$, while $H(x)=H(\overline{x}^-)$ for all $x \in [\hat{x},\overline{x})$. In particular, $F_L^{\tau,\varepsilon}(\hat{x}) \geq H(\hat{x})=F_L^{\tau,\varepsilon}(\underline{x})+\eta$, and hence $F(\hat{x})-(\tau+\varepsilon)\eta \geq F(\underline{x})$. Likewise, $F(\hat{x})+(1-\tau+\varepsilon)\eta \leq F(\overline{x}^-)$. Now let
\[
\underline{y}:=F^{-1}(F(\hat{x})-(\tau+\varepsilon)\eta), \quad \mbox{ and } \quad \overline{y}:=F^{-1}(F(\hat{x})+(1-\tau+\varepsilon)\eta).
\]
It then follows that $\underline{x} \leq \underline{y} \leq \hat{x} \leq \overline{y} \leq \overline{x}$, with at least one inequality being strict if $\eta>0$. Next, define $\widehat{F}$ as follows: $\widehat{F} \equiv 0$ if $\eta=0$; and 
\[
\widehat{F}(x):=\left\{
\begin{array}{cc}
0,&\mbox{if } x<\underline{y}\\
\frac{F(x)-(F(\hat{x})-(\tau+\varepsilon)\eta)}{\eta},&\mbox{if } x \in [\underline{y},\overline{y})\\
1,&\mbox{if } x\geq \overline{y}
\end{array}
\right.,
\]
if $\eta>0$. Clearly $\widehat{F} \in \mathcal{F}_0$ if $\eta>0$, and $\hat{x} =\widehat{F}^{-1}(\tau)$. Moreover, for all $x \in \mathbb{R}$, let 
\[
\widetilde{F}(x):=\frac{F(x)-\eta\widehat{F}(x)}{1-\eta}.
\]
By construction, $\eta \widehat{F}+(1-\eta)\widetilde{F}=F$. From the definition of $\underline{y}$ and $\overline{y}$, it can be shown that $\widetilde{F} \in \mathcal{F}_0$ as well. Furthermore, 
\[
\widetilde{F}(\overline{x}^-)-\widetilde{F}(\underline{x})=\frac{F(\overline{x}^-)-F(\underline{x})-\eta}{1-\eta}=\frac{1}{1-\eta}\left[\frac{\tau-\varepsilon}{1-(\tau-\varepsilon)}(1-F(\overline{x}^-))+\frac{1-(\tau+\varepsilon)}{\tau+\varepsilon}F(\underline{x})\right].  
\]
Next, define $\widetilde{F}_1$ and $\widetilde{F}_2$ as follows: 
\[
\widetilde{F}_1(x):=\left\{
\begin{array}{cc}
\frac{F(x)}{F(\underline{x})+\alpha(F(\overline{x}^-)-F(\underline{x})-\eta)},&\mbox{if } x<\underline{x}\\
\frac{F(\underline{x})\alpha(F(x)-F(\underline{x})-\eta)}{F(\underline{x})+\alpha(F(\overline{x}^-)-F(\underline{x})-\eta)},& \mbox{if } x \in [\underline{x},\overline{x})\\
1,&\mbox{if } x \geq \overline{x}
\end{array}
\right.; 
\]
and
\[
\widetilde{F}_2(x):=\left\{
\begin{array}{cc}
0,&\mbox{if } x<\underline{x}\\
\frac{(1-\alpha)(F(x)-F(\underline{x})-\eta)}{1-F(\overline{x}^-)+(1-\alpha)(F(\overline{x}^-)-F(\underline{x})-\eta)},&\mbox{if } x \in [\underline{x},\overline{x})\\
\frac{F(x)-F(\underline{x})+(1-\alpha)(F(\overline{x}^-)-F(\underline{x})-\eta)}{1-F(\overline{x}^-)+(1-\alpha)(\widetilde{F}(\overline{x}^-)-\widetilde{F}(\underline{x})-\eta)},& \mbox{if } x \geq \overline{x}
\end{array}
\right.,
\]
where 
\[
\alpha:=\frac{\frac{1-(\tau+\varepsilon)}{\tau+\varepsilon}F(\underline{x})}{\frac{\tau-\varepsilon}{1-(\tau-\varepsilon)}(1-F(\overline{x}^-))+\frac{1-(\tau+\varepsilon)}{\tau+\varepsilon}F(\underline{x})}.
\]
By construction, $\widetilde{\alpha}\widetilde{F}_1+(1-\widetilde{\alpha})\widetilde{F}_2=\widetilde{F}$, where $\widetilde{\alpha} \in (0,1)$ is given by $\widetilde{\alpha}:=[F(\underline{x})+\alpha (F(\overline{x}^-)-F(\underline{x})-\eta)]/(1-\eta)$. Moreover, $\widetilde{F}_1(\underline{x}) = \tau+\varepsilon>\tau$, and $\widetilde{F}_2(\overline{x}^-) = \tau-\varepsilon<\tau$.

Now define two classes of distributions, $\{\widetilde{F}_1^x\}_{x \leq \underline{x}}$ and $\{\widetilde{F}_2^x\}_{x \geq \overline{x}}$, as follows: 
\[
\widetilde{F}^x_1(z):=\left\{
\begin{array}{cc}
0,&\mbox{if }z < x\\
\widetilde{F}(\underline{x}),&\mbox{if } z \in [x,\underline{x})\\
\widetilde{F}(z),&\mbox{if } z \geq \underline{x}\\
\end{array}
\right.;\mbox{ and }
\widetilde{F}_2^x(z):=\left\{
\begin{array}{cc}
\widetilde{F}(z),&\mbox{if } z <\overline{x}\\
\widetilde{F}(\overline{x}^-),&\mbox{if } z \in [\overline{x},x)\\
1,&\mbox{if } z \geq x
\end{array}
\right..
\]
Note that since $\widetilde{F}_1(\underline{x}) >\tau$ and $\widetilde{F}_2(\overline{x}^-) < \tau$, $x =(\widetilde{F}_1^x)^{-1}(\tau)=(\widetilde{F}_1^x)^{-1}(\tau^+)$ for all $x \leq \underline{x}$ and $x =(\widetilde{F}_2^x)^{-1}(\tau)=(\widetilde{F}_2^x)^{-1}(\tau^+)$ for all $x \geq \overline{x}$. Moreover, for any $n \in \mathbb{N}_1$ and for any $m \in \mathbb{N}_2$, let 
\[
\widetilde{F}_1^n(z):=\frac{1}{\widetilde{F}(\overline{x}_n)-\widetilde{F}(\underline{x}_n)}\int_{\underline{x}_n}^{\overline{x}_n} \widetilde{F}_1^\omega(z) \widetilde{F}(\diff x), 
\]
and
\[
\widetilde{F}_2^{m}(z):=\frac{1}{\widetilde{F}(\overline{x}_m)-\widetilde{F}(\underline{x}_m)}\int_{\underline{x}_m}^{\overline{x}_m} \widetilde{F}_2^\omega(z) \widetilde{F}(\diff \omega),
\]
for all $z \in \mathbb{R}$. By construction, $\widetilde{F}_1^{n}, \widetilde{F}_2^{m} \in \mathcal{F}_0$ and $\overline{x}_n=(\widetilde{F}_1^n)^{-1}(\tau)=(\widetilde{F}_1^n)^{-1}(\tau^+)$, $\underline{x}_m =(\widetilde{F}_2^m)^{-1}(\tau)=(\widetilde{F}_2^m)^{-1}(\tau^+)$ for all $n \in \mathbb{N}_1$ and $m \in \mathbb{N}_2$. Next, for any $\omega \in \mathbb{R}$, let $\widetilde{G}^x \in \mathcal{F}_0$ be defined as
\[
\widetilde{G}^\omega(z):=\left\{
\begin{array}{cc}
\widetilde{F}_1^\omega(z),& \mbox{if } \omega \in (-\infty,\overline{x}]\backslash \cup_{n \in \mathbb{N}_1} [\underline{x}_n,\overline{x}_n)\\
\widetilde{F}_1^n(z),& \mbox{if } \omega \in [\underline{x}_n,\overline{x}_n), \, n \in \mathbb{N}_1\\
\widetilde{F}_2^\omega(z),& \mbox{if } \omega \in [\overline{x},\infty) \backslash \cup_{m \in \mathbb{N}_2} [\underline{x}_m,\overline{x}_m)\\
\widetilde{F}_2^m(z),&\mbox{if } \omega \in [\underline{x}_m,\overline{x}_m), \, m \in \mathbb{N}_2
\end{array}
\right.,
\]
for all $z \in \mathbb{R}$. Let 
\[
\widetilde{H}(x):=\left\{
\begin{array}{cc}
\frac{H(x)}{1-\eta},&\mbox{if } x<\underline{x}\\
\frac{H(\underline{x})}{1-\eta},&\mbox{if } x \in [\underline{x},\overline{x})\\
\frac{H(x)-\eta}{1-\eta},&\mbox{if } x \geq \overline{x}
\end{array}
\right.,
\]
and define $\tilde{\mu}$ as 
\[
\tilde{\mu}(\{\widetilde{G}^x \in \mathcal{F}_0|\omega \leq z\}):=\widetilde{H}(z),
\]
for all $z \in \mathbb{R}$. Then, by construction, for any $z \in \mathbb{R}$,
\begin{equation}\label{avgmu2}
\int_{\mathcal{F}_0} G(z)\tilde{\mu}(\diff G)=\int_\mathbb{R} \widetilde{G}^{\omega}(z)\widetilde{H}(\diff \omega)=\widetilde{F}(z).
\end{equation} 
Furthermore, $H^\tau(x|\tilde{\mu})=\widetilde{H}(x)$ for all $x \in \mathbb{R}$. As a result, from \eqref{avgmu2}, for $\mu \in \Delta(\mathcal{F}_0)$ defined as 
\[
\mu:=(1-\eta)\tilde{\mu}+\eta\delta_{\{\widehat{F}\}},
\]
since $F=\eta \widehat{F}+(1-\eta)\widetilde{F}$, it must be that $\mu \in \widetilde{\mathcal{M}}_\tau$. Moreover, since $H^\tau(\cdot|\tilde{\mu})=\widetilde{H}$, we have $H^\tau(x|\mu)=H(x)$ for all $x \in \mathbb{R}$. 

Lastly, consider any $H \in \mathcal{I}(F_R^{\tau,\varepsilon},F_L^{\tau,\varepsilon})$. Since $\mathcal{I}(F_R^{\tau,\varepsilon},F_L^{\tau,\varepsilon})$ is a convex and compact set in a metrizable space, Choquet's theorem implies that there exists a probability measure $\Lambda_H \in \Delta(\mathcal{I}(F_R^{\tau,\varepsilon},F_L^{\tau,\varepsilon}))$ that assigns probability 1 to $\mathrm{ext}(\mathcal{I}(F_R^{\tau,\varepsilon},F_L^{\tau,\varepsilon}))$ such that 
\[
H(\omega)=\int_{\mathcal{I}(F_R^{\tau,\varepsilon},F_L^{\tau,\varepsilon})} \widetilde{H}(\omega)\Lambda_H(\diff \widetilde{H}).
\]

Meanwhile, define the linear functional $\Xi:\widetilde{\mathcal{M}}_\tau \to \mathcal{F}_0$ as 
\[
\Xi(\tilde{\mu})[\omega]:=\tilde{\mu}(\{G \in \mathcal{F}_0|G^{-1}(\tau) \leq \omega\}),
\]
for all $\tilde{\mu} \in \widetilde{\mathcal{M}}_\tau$ and for all $\omega \in \mathbb{R}$. Now, define a probability measure $\widetilde{\Lambda}$ on $\widetilde{\mathcal{M}}_\tau$ by 
\[
\widetilde{\Lambda}_H(A):=\Lambda_H(\{\Xi(\tilde{\mu})|\tilde{\mu} \in A\}),
\]
for all $A \subseteq \widetilde{\mathcal{M}}_\tau$. Then, since $\Lambda_H(\mathrm{ext}(\mathcal{I}(F_R^{\tau,\varepsilon},F_L^{\tau,\varepsilon})))=1$ and since, for any $\widetilde{H} \in \mathrm{ext}(\mathcal{I}(F_R^{\tau,\varepsilon},F_L^{\tau,\varepsilon}))$, there exists $\tilde{\mu}\in \widetilde{\mathcal{M}}_\tau$ such that $H(\omega)=H^\tau(\omega|\tilde{\mu})$, it must be that $\widetilde{\Lambda}_H(\widetilde{\mathcal{M}}_\tau)=1$, and hence $\widetilde{\Lambda}_H$ is a probability measure on $\widetilde{\mathcal{M}}_\tau$. Let $\tilde{\mu} \in \widetilde{\mathcal{M}}_\tau$ be defined as 
\[
\tilde{\mu}(A):=\int_{\widetilde{\mathcal{M}}_\tau}\mu(A)\widetilde{\Lambda}_H(\diff \mu),
\]
for all measurable $A \subseteq \mathcal{F}_0$. Then, since $\Xi$ is linear, it follows that
\begin{align*}
H(\omega)=\int_{\mathcal{I}(F_R^{\tau,\varepsilon},F_L^{\tau,\varepsilon})}\widetilde{H}(\omega)\Lambda_H(\diff \widetilde{H})=&\int_{\widetilde{\mathcal{M}}_\tau}\Xi(\mu)[\omega]\widetilde{\Lambda}_H(\diff \mu)\\
=&\Xi(\tilde{\mu})[\omega]\\
=&H^\tau(\omega|\tilde{\mu}),
\end{align*}
and therefore, $H \in \widetilde{\mathcal{H}}_\tau$. Together, for any $\varepsilon>0$, any $H \in \mathcal{I}(F_R^{\tau,\varepsilon},F_L^{\tau,\varepsilon})$ must be in $\widetilde{\mathcal{H}}_\tau$. In other words, 
\[
\bigcup_{\varepsilon>0} \mathcal{I}(F_R^{\tau,\varepsilon},F_L^{\tau,\varepsilon}) \subseteq \widetilde{\mathcal{H}}_\tau.
\]
This completes the proof. \hfill $\blacksquare$

\subsection{Proof of Corollary \ref{lit}}
For 1, consider any $H \in \mathcal{H}_{q}$. By \bref{thm2}, $H \in \mathcal{I}(F_R^q,F_L^q)$. Thus, 
$
(F_L^q)^{-1}(\tau) \leq H^{-1}(\tau) \leq H^{-1}(\tau^+) \leq  (F_R^q)^{-1}(\tau^+),
$
and therefore $[H^{-1}(\tau),H^{-1}(\tau^+)] \subseteq [(F_L^q)^{-1}(\tau),(F_R^q)^{-1}(\tau^+)]$. Conversely, consider any interval $Q=[\underline{\omega},\overline{\omega}] \subseteq [(F_L^q)^{-1}(\tau),(F_R^q)^{-1}(\tau^+)]$. Then, let $\widehat{H}$ be defined as 
\[
\widehat{H}(\omega):=\left\{
\begin{array}{cc}
0,&\mbox{if } \omega<\underline{\omega}\\
\tau,&\mbox{if } \omega \in [\underline{\omega},\overline{\omega})\\
1,&\mbox{if } \omega\geq \overline{\omega}
\end{array}
\right.,
\]
for all $\omega \in \mathbb{R}$. Then $\widehat{H} \in \mathcal{I}(F_L^q,F_R^q)$ and $Q=[H^{-1}(\tau),H^{-1}(\tau^+)]$. Moreover, by \bref{thm2}, $\widehat{H} \in \mathcal{H}_{q}$, as desired. 

For 2, consider any $H \in \widetilde{\mathcal{H}}_{q}$. By \bref{thm2}, $H \in \mathcal{I}(F_R^q,F_L^q)$. Thus, it must be that $[H^{-1}(\tau),H^{-1}(\tau^+)] \subseteq [(F_L^q)^{-1}(\tau), (F_R^q)^{-1}(\tau)]$. Conversely, for any $\hat{x} \in ((F_L^q)^{-1}(\tau),(F_R^q)^{-1}(\tau^+))$, note that since $\hat{x}>(F_L^q)^{-1}(\tau)$ and since $F$ is continuous, we have $F(\hat{x})/\tau>q$. Similarly, we also have $(F(\hat{x})-\tau)/(1-\tau)<q$. Let $\varepsilon:=\min\{F(\hat{x})/{\tau}-q, q-(F(\hat{x})-\tau)/(1-\tau)\}$. Then, either $\hat{x}=(F_L^{q,\varepsilon})^{-1}(\tau)$ or $\hat{x}=(F_R^{q,\varepsilon})^{-1}(\tau)$. Since both $F_L^{q,\varepsilon}$ and $F_R^{q,\varepsilon}$ are in $\mathcal{I}(F_R^{q,\varepsilon},F_L^{q,\varepsilon})$, \bref{thm3} implies that $\hat{x}=H^{-1}(\tau)$ for some $H \in \widetilde{H}_q$. Lastly, note that under a signal $\mu \in \mathcal{M}$ such that $\mu$ assigns probability $\tau$ to $F_L^\tau$ and probability $1-\tau$ to $F_R^\tau$, we have $\mu \in \widetilde{M}_q$ and $H^{q}(x|\mu)=\tau$ for all $x \in [(F_L^q)^{-1}(\tau),(F_R^q)^{-1}(\tau)]$. Hence, $[(F_L^q)^{-1}(\tau),(F_R^q)^{-1}(\tau)] \subseteq [H^{-1}(\tau),H^{-1}(\tau^+)]$ for some $H \in \widetilde{H}_q$, as desired. \hfill $\blacksquare$

\subsection{Proof of Corollary \ref{optqp}}
$(i)$ and $(ii)$ follow immediately from the fact that any $H \in \mathcal{I}(F_R^\tau,F_L^\tau)$ is dominated by $F_R^\tau$ and dominates $F_L^\tau$, and that $v_S(x)$ is increasing in $x$ for all $x \leq a$ and is decreasing in $x$ for all $x>a$. 

For $(iii)$, suppose that for any $\underline{a} \leq \overline{a}$, $H^{C}_{\underline{a},\overline{a}}$ is not optimal. Then, since at least one extreme point of $\mathcal{I}(F_R^\tau,F_L^\tau)$ must be the solution of \eqref{eq:simplified_bayesian_persuasion},  consider any such extreme point and denote it by $H$. By \bref{thm1}, there exists a countable collection of intervals $\{[\underline{x}_n,\overline{x}_n)\}_{n=1}^\infty$ such that conditions 1 and 2 of \bref{thm1} hold. Since $H^{C}_{\underline{a},\overline{a}}$ is not optimal for any $\underline{a} \leq \overline{a}$, $H \neq H^{C}_{\underline{a},\overline{a}}$ for all $\underline{a} \leq \overline{a}$. In particular, there must exist $n \in \mathbb{N}$ such that $\underline{x}_n<\overline{x}_n$ and either $H(\overline{x_n}^-)>F_R^\tau(\overline{x}_n)$ or $H(\underline{x}_n)<F_L^\tau(\underline{x}_n)$. Let $a$ be the minimizer of $v_S$ and suppose that $a \leq F^{-1}(\tau)$. Suppose that $H(\overline{x}_n^-)>F_R^\tau(\overline{x}_n)$. Then it must be that $H(\underline{x}_n)=F_L^\tau(\underline{x}_n)$. Moreover, since $H(\overline{x}_n^-)>F_R^\tau(\overline{x}_n)$, $H(\overline{x}_n)>F_R^\tau(\overline{x}_n)$ as well. If $\overline{x}_n \leq a$, then by replacing $H(x)$ with $\min\{F_L^\tau(x),H(\overline{x}_n)\}$ for all $x \in [\underline{x}_n,\overline{x}_n)$ and otherwise leaving $H$ unchanged, the resulting distribution $\widehat{H}$ must still be in $\mathcal{I}(F_R^\tau,F_L^\tau)$. Since $v_S$ is strictly decreasing on $[\underline{x}_n,\overline{x}_n)$, $\widehat{H}$ must give a higher value, a contradiction. If, on the other hand, $\overline{x}_n>a$, then since $H(\overline{x}_n)>F_R^\tau(\overline{x}_n)$ and since $F$ is continuous, there exists $y>\overline{x}_n$ such that $H(\overline{x}_n^-)>F_R^\tau(y)$. Moreover, since $H$ satisfies conditions 1 and 2, $H(x)>H(\overline{x}_n^-)$ for all $x \in [\overline{x}_n,y)$. Therefore, by replacing $H(x)$ with $H(\overline{x}_n^-)$ for all $x \in [\overline{x},y)$ and leaving $H$ unchanged otherwise, the resulting $\widehat{H}$ must still be in $\mathcal{I}(F_R^\tau,F_L^\tau)$. Since $v_S$ is strictly increasing on $[\overline{x}_n,y)$, $\widehat{H}$ must give a higher value, a contradiction. Analogous arguments also lead to a contradiction for the case of $H(\underline{x}_n)<F_L^\tau(\underline{x}_n)$, as well as $a>F^{-1}(\tau)$. Therefore, $H^{C}_{\underline{a}.\overline{a}}$ must be optimal for some $\underline{a} \leq \overline{a}$. 

For $(iv)$, note that $F$ is not an extreme point of $\mathcal{I}(F_R^\tau,F_L^\tau)$ according to \bref{thm1}. Therefore, it is never the unique solution of \eqref{eq:simplified_bayesian_persuasion}. \hfill $\blacksquare$

\subsection{Proof of Proposition \ref{qp}}
Let $\bar{v}(G):=\sup_{x \in [G^{-1}(\tau),G^{-1}(\tau^+)]}v_S(x)$ for all $G \in \mathcal{F}_0$. Then, by \bref{thm2},
\[
\mathrm{cav}(\hat{v})[F] \leq \mathrm{cav}(\bar{v})[F] = \sup_{H \in \mathcal{I}(F_R^\tau, F_L^\tau)} \int_\mathbb{R}v_S(\omega)H(\diff \omega).
\]
Meanwhile, by \bref{thm3}, 
\[
\sup_{H \in \cup_{\varepsilon>0}\mathcal{I}(F_R^{\tau,\varepsilon},F_L^{\tau,\varepsilon})} \int_\mathbb{R}v_S(\omega)H(\diff \omega) \leq \mathrm{cav}(\hat{v})[F].
\]
Together, since $\mathrm{cl}(\{\mathcal{I}(F_R^{\tau,\varepsilon},F_L^{\tau,\varepsilon})\})=\mathcal{I}(F_R^\tau, F_L^\tau)$, \eqref{eq:simplified_bayesian_persuasion} then follows. \hfill $\blacksquare$

\subsection{Proof of Corollary \ref{cor2}}\label{BDp}
For necessity, consider any $H \in \widetilde{\mathcal{H}}_\tau$ such that $H(z_k^-)-H(z_{k-1}^-)=\theta_k$ for all $k \in \{1,\ldots,K\}$. Then for any $k \in \{1,\ldots,K\}$,  $\sum_{i=1}^k \theta_i=H(z_k^-)$. Since $H \in \widetilde{\mathcal{H}}_\tau$, there exists a signal $\mu \in \mathcal{M}$ for which $\mu$-almost all posteriors have a unique $\tau$-quantile and $H(z_k^-)=\mu(\{G \in \mathcal{F}_0|G^{-1}(\tau)<z_k\})=\mu(\{G \in \mathcal{F}_0|\tau<G(z_k)\})$. Since $\mu \in \mathcal{M}$, $G(z_k)$ is a mean-preserving spread of $F(z_k)$ when $G \sim \mu$. Thus, $\mu(\{G \in \mathcal{F}_0|\tau<G(z_k)\})<F(z_k)/\tau$, and hence \eqref{lb} holds. Analogous arguments can be applied to show that \eqref{ub} holds as well. 

For sufficiency, consider any prediction dataset $(\theta_k)_{k=1}^K$ such that \eqref{lb} and \eqref{ub} hold. Let $H$ be the distribution that assigns probability $\theta_k$ at $(z_k+z_{k-1})/2$. Then, there exists $\varepsilon>0$ such that $H \in \mathcal{I}(F_R^{\tau,\varepsilon},F_L^{\tau,\varepsilon})$. By \bref{thm3}, there exists a signal $\mu$ with $\mu(\{G \in \mathcal{F}_0|G^{-1}(\tau)<G^{-1}(\tau^+)\})=0$ such that $H(x)=H^\tau(x|\mu)$ for all $x \in \mathbb{R}$, which in turn implies that $\mu$ $\tau$-quantile-rationalizes $(\theta_k)_{k=1}^K$, as desired. \hfill $\blacksquare$

\subsection{Proofs of Proposition \ref{moral} and Proposition \ref{prop2}}\label{security}
We prove the following result that leads to \bref{moral} and \bref{prop2} immediately.\footnote{\citet{R84} characterizes the extreme points of bounded Lipschitz functions defined on the unit interval that vanish at zero, and he shows that a function is an extreme point of the unit ball of this set if and only if the absolute value of its derivative equals 1 almost everywhere (see also \citealp{R86,F94,S97}). The convex set of interest here is different. First, functions in $\mathcal{I}(\lb{F},\ub{F})$ are subject to an additional monotonicity constraint. Second, these functions are bounded by $\lb{F}$ and $\ub{F}$ under the pointwise dominance order, rather than the Lipschitz (semi) norm. In particular, functions in $\mathcal{I}(\lb{F},\ub{F})$ may have unbounded derivatives, whenever well-defined. Lastly, \bref{thma1} below characterizes the extreme points of this set subject to finitely many other linear constraints, which are not present in the characterization of \citet{R84}.} 

\begin{taggedtheorem}{A.1}\label{thma1}
Let $\ub{F}(x):=x$ and $\lb{F}(x):=0$ for all $x \in [0,1]$. For any $J \in \mathbb{N}$, for any collection of bounded linear functionals $\{\Gamma_j\}_{j=1}^J$ on $L^1([0,1])$ and for any collection $\{\gamma_j\}_{j=1}^J \subseteq \mathbb{R}$, let $\mathcal{I}^c$ be a convex subset of $\mathcal{I}(\underline{F},\overline{F})$ defined as 
\[
\mathcal{I}^c:=\left\{H \in \mathcal{I}(\underline{F},\overline{F})| \Gamma_j(H) \geq \gamma_j,\, \forall j \in \{1,\ldots,J\}\right\}. 
\]
Suppose that $H \in \mathcal{I}^c$ is an extreme point of $\mathcal{I}^c$. Then there exists countably many intervals $\{[\underline{x}_n,\overline{x}_n)\}_{n=1}^\infty$ such that:
\begin{enumerate}
\item $H(x)=x$ for all $x \notin \cup_{n=1}^\infty [\underline{x}_n,\overline{x}_n)$. 
\item For all $n,m \in \mathbb{N}$, with $n\neq m$, $H$ is constant on $[\underline{x}_n,\overline{x}_n)$ and $H(\underline{x}_n)\neq H(\underline{x}_m)$.
\item For all but at most $J$ many $n \in \mathbb{N}$, $H(\underline{x}_n)=\underline{x}_n$. 
\end{enumerate}
\end{taggedtheorem}

\begin{proof}
Consider any extreme point $H$ of $\mathcal{I}^c$. We first claim that for any $x \in (0,1)$, it must be either $H(x)=x$ or $H(y)=H(x)$ for all $y \in (x,x+\delta)$ for some $\delta>0$. To see this, note that since $\mathcal{I}^c$ is a subset of $\mathcal{I}(\underline{F},\overline{F})$ defined by $J$ linear constraints, Proposition 2.1 of \citet{W88} implies that there exists $\{H_j\}_{j=1}^{J+1} \subseteq \mathrm{ext}(\mathcal{I}(\underline{F},\overline{F}))$ and $\{\lambda_j\}_{j=1}^{J+1} \subseteq [0,1]$ such that $H(x)=\sum_{j=1}^{J+1} \lambda_jH_j(x)$ for all $x \in [0,1]$ and $\sum_{j=1}^{J+1} \lambda_j=1$. Now suppose that $H(x)<x$ for some $x \in (0,1)$. Then there must exist a nonempty subset $\mathcal{J} \subseteq \{1,\ldots,J+1\}$ such that $H_j(x)<x$ for all $j \in \mathcal{J}$ and that $H_j(x)=x$ for all $j \in \{1,\ldots, J+1\} \backslash \mathcal{J}$. Since $H_j$ is an extreme point of $\mathcal{I}(\underline{F},\overline{F})$ for all $j \in \mathcal{J}$, \bref{thm1} implies that for each $j \in \mathcal{J}$, there exists an interval $[\underline{x}^j,\overline{x}^j)$ containing $x$ on which $H_j$ is constant. Let $(\underline{x},\overline{x})$ be the interior of the intersection of $\{[\underline{x}^j,\overline{x}^j)\}_{j \in \mathcal{J}}$. Then it must be that 
\[
H(y)=\alpha y+(1-\alpha) \eta
\]
for all $y \in (\underline{x},\overline{x})$, for some $\eta<x$, and $\alpha \in (0,1)$. Now suppose that for any $\delta>0$, there exists $y \in (x,x+\delta)$ such that $H(x)<H(y)$. Take any $\hat{\delta} \in (0,\min\{(1-\alpha)(x-\eta)/(1+\alpha),x-\underline{x},\overline{x}-x\})$ and let $x_*:=x-\hat{\delta}$ and $x^*:=x+\hat{\delta}$. Then it must be that $H(y)<x$ for any $y \in [x_*,x^*]$ and that $H(x^*)<x_*$. Moreover, the function $h:[x_*,x^*] \to [H(x_*),H(x^*)]$ defined as $h(y):=H(y)$ for all $y \in [x_*,x^*]$ must not be a step function, since otherwise, as $h$ is right-continuous on $(x_*,x^*)$, there must be some $\delta>0$ such that $H(y)=h(y)=h(x)=H(x)$ for all $y \in [x,x+\delta)$, a contradiction. Meanwhile, since each functional $\Gamma_j:L^1([0,1]) \to \mathbb{R}$ is bounded, Riesz's representation implies that there must exist $\Phi_j \in L^\infty([0,1])$ such that 
\[
\Gamma_j(\widetilde{H})=\int_0^1 \widetilde{H}(x) \Phi_j(x) \diff x,
\]
for all $\widetilde{H} \in \mathcal{I}(\underline{F},\overline{F})$. Therefore, since any extreme point of the collection of nondecreasing, right-continuous functions $\tilde{h}$ from $[x_*,x^*]$ to $[H(x_*),H(x^*)]$ such that 
\[
\int_{x_*}^{x^*} \tilde{h}(x) \Phi_j(x)\diff x \geq \gamma_j
\]
for all $j \in \{1,\ldots,J\}$ is a step function with at most $J+1$ steps, as implied by Proposition 2.1 of \citet{W88}, the function $h$ is not an extreme point of this collection. Thus, there exists two distinct functions $h_1,h_2:[x_*,x^*] \to [H(x_*),H(x^*)]$ and $\lambda \in (0,1)$ such that $h(y)=\lambda h_1(y)+(1-\lambda)h_2(y)$ for all $y \in [x_*,x^*]$ and that 
\begin{equation}\label{xstar}
\int_{x_*}^{x^*} h_l(x) \Phi_j(x) \diff x=\int_{x_*}^{x^*} H(x) \Phi_j(x) \diff x,
\end{equation}
for all $j \in \{1,\ldots,J\}$ and for all $l \in \{1,2\}$. Now let $H_1$, $H_2$ be defined as 
\[
H_1(y):=\left\{
\begin{array}{cc}
H(y),&\mbox{if } y \notin [x_*,x^*]\\
h_1(y),&\mbox{if } y \in [x_*,x^*]
\end{array}
\right.; \quad
H_2(y):=\left\{
\begin{array}{cc}
H(y),&\mbox{if } y \notin [x_*,x^*]\\
h_2(y),&\mbox{if } y \in [x_*,x^*]
\end{array}
\right..
\]
Then, $H=\lambda H_1+(1-\lambda)H_2$ and $H_1 \neq H_2$. Moreover, since $h_1(y),h_2(y) \leq H(x^*) <x_*$ for all $y \in [x_*,x^*]$, and since $H \in \mathcal{I}(\underline{F},\overline{F})$, it must be that both $H_1$ and $H_2$ are in $\mathcal{I}(\underline{F},\overline{F})$. Furthermore, by \eqref{xstar}, it must be that 
\begin{align*}
\Gamma_j(H_l)=\int_0^1 H_l(x) \Phi_j(x) \diff x=&\int_{[0,1]\backslash [x_*,x^*]} H(x) \Phi_j(x) \diff x+\int_{x_*}^{x^*} h_l(x)\Phi_j(x) \diff x\\
=&\int_{[0,1]\backslash [x_*,x^*]} H(x) \Phi_j(x) \diff x+\int_{x_*}^{x^*} H(x)\Phi_j(x) \diff x\\
=&\int_0^1 H(x) \Phi_j(x) \diff x \\
\geq& \gamma_j,
\end{align*}
for all $j \in \{1,\ldots,J\}$ and for all $l \in \{1,2\}$. Thus, $H_1,H_2 \in \mathcal{I}^c$, a contradiction. Together, for any $x \in (0,1)$, it must be either $H(x)=x$ or $H(y)=H(x)$ for all $y \in (x,x+\delta)$ for some $\delta>0$.

Let $X \subseteq [0,1]$ be the collection of $x \in [0,1]$ such that $H(x)=x$. For any $x \notin X$, let $\overline{\delta}_x:=\sup\{y \in [0,1]|H(y)=H(x)\}$ and $\underline{\delta}_x:=\inf\{y \in [0,1]|H(y)=H(x)\}$. Then it must be $\underline{\delta}_x<\overline{\delta}_x$ for all $x \notin X$. Moreover, for any $x,y \in [0,1]\backslash X$ with $x<y$, $H(x)<H(y)$ if and only if $\overline{\delta}_x<\underline{\delta}_y$. Therefore, $[0,1]\backslash X$ can be expressed as a union of a collection $I$ of disjoint intervals. Since $I$ is a collection of disjoint intervals on $[0,1]$, each element of $I$ must uniquely contain at least one rational number. Thus, there exists an injective map from the collection $I$ to a subset of rational numbers in $[0,1]$, and therefore the collection $I$ must be countable. 

Enumerate $I$ as $\{[\underline{x}_n,\overline{x}_n)\}_{n=1}^\infty$ and suppose that there is a subset $\mathcal{N}$ of these intervals, with $|\mathcal{N}| > J$, such that $H(\underline{x}_n)<\underline{x_n}$. For each $n \in \mathcal{N}$, since $H(\underline{x}_n)<\underline{x}_n$ and since $H(x)=x$ for all $x \notin \cup_{n=1}^\infty [\underline{x}_n,\overline{x}_n)$, $H$ must be discontinuous at $\underline{x}_n$. Let $\eta_n:=H(\underline{x}_n)-H(\underline{x}_n^-)$ for all $n \in \mathcal{N}$, and let $\eta:=\min\{\eta_n\}_{n \in \mathcal{N}}$. Furthermore, let $\phi_j^n \in \mathbb{R}$ be defined as 
\[
\phi_{j}^n:= \int_{\underline{x}_n}^{\overline{x}_n} \Phi_j(x) \diff x,
\]
for all $n \in \mathcal{N}$ and for all $j \in \{1,\ldots,J\}$. Then the $|\mathcal{N}| \times J$ matrix $\Phi:=(\phi_j^n)_{j \in \{1,\ldots,J\}}^{n \in \mathcal{N}}$ is a linear map from $\mathbb{R}^{|\mathcal{N}|}$ to $\mathbb{R}^J$. Since $|\mathcal{N}|>J$, $\mathrm{dim}(\mathrm{null}(\Phi)) \geq 1$, and thus there must exist a nonzero vector $\{\hat{h}_n\}_{n \in \mathcal{N}}$ such that 
\begin{equation}\label{null}
\sum_{n \in \mathcal{N}} \phi_j^n \hat{h}_n=0,
\end{equation}
for all $j \in \{1,\ldots,J\}$. Let $\varepsilon:=\min\{\eta/4|\hat{h}_n|,(\underline{x}_n-H(\underline{x}_n))/4|\hat{h}_n|\}_{n \in \mathcal{N}}$, and let $\widehat{H}$ be defined as 
\[
\widehat{H}(x):=\left\{
\begin{array}{cc}
0,&\mbox{if } x \notin \cup_{n \in \mathcal{N}} [\underline{x}_n,\overline{x}_n)\\
\varepsilon\hat{h}_n,&\mbox{if } x \in [\underline{x}_n,\overline{x}_n), \, n \in \mathcal{N}
\end{array}
\right..
\]
Then, since $\{\hat{h}_n\}_{n \in \mathcal{N}}$ is a nonzero vector in $\mathbb{R}^{|\mathcal{N}|}$ and since $\varepsilon>0$, $\hat{H} \neq 0$. Moreover, since $\varepsilon<\eta/4|\hat{h}_n|$ for all $n \in \mathcal{N}$, $H(x)-|\widehat{H}(x)|=H(\underline{x}_n)-\varepsilon|\hat{h}_n|>H(\underline{x}_n)-\eta/2>H(\underline{x}_n^-)+\eta/4>H(x)+|\widehat{H}(x)|$ for all $x <\underline{x}_n$ and for all $n \in \mathcal{N}$. Therefore, both $H+\widehat{H}$ and $H-\widehat{H}$ are nondecreasing. Meanwhile, since for any $n \in \mathcal{N}$ and for any $x \in [\underline{x}_n,\overline{x}_n)$, $H(x)+|\widehat{H}(x)|=H(\underline{x}_n)+\varepsilon|\hat{h}_n|<\underline{x}_n$, both $H+\widehat{H}$ and $H-\widehat{H}$ are in $\mathcal{I}(\underline{F},\overline{F})$. In addition, by \eqref{null}, for any $j \in \{1,\ldots,J\}$, 
\begin{align*}
\int_0^1 (H(x)+\widehat{H}(x))\Phi_j(x)\diff x=&\int_{[0,1]\backslash \cup_{n \in \mathcal{N}}[\underline{x}_n,\overline{x}_n)} H(x)\Phi_j(x)\diff x+\int_{\cup_{n \in \mathcal{N}}[\underline{x}_n,\overline{x}_n)}H(x)\Phi_j(x)\diff x+\varepsilon\sum_{n \in \mathcal{N}}\hat{h}_n\phi_j^n\\
=& \int_0^1 H(x)\Phi_j(x)\diff x \\
\geq& \gamma_j,
\end{align*}
and 
\begin{align*}
\int_0^1 (H(x)-\widehat{H}(x))\Phi_j(x)\diff x=&\int_{[0,1]\backslash \cup_{n \in \mathcal{N}}[\underline{x}_n,\overline{x}_n)} H(x)\Phi_j(x)\diff x+\int_{\cup_{n \in \mathcal{N}}[\underline{x}_n,\overline{x}_n)}H(x)\Phi_j(x)\diff x-\varepsilon\sum_{n \in \mathcal{N}}\hat{h}_n\phi_j^n\\
=& \int_0^1 H(x)\Phi_j(x)\diff x \\
\geq& \gamma_j.
\end{align*}
Together, both $H+\widehat{H}$ and $H-\widehat{H}$ are in $\mathcal{I}^c$, and hence, $H$ is not an extreme point, a contradiction. This completes the proof. 
\end{proof}

\renewcommand{\proofname}{Proofs of Proposition 3 and Proposition 5}

\begin{proof}
Note that since $|\phi_e(x|e)|$ is dominated by an integrable function on $[0,1]$, one can apply the dominated convergence theorem to show that the objective function of both \eqref{ent} and \eqref{iss} are continuous in $(H,e)$ and $(H,\underline{z})$, respectively. Similarly, the constraint set can be shown to be closed. Therefore, both \eqref{ent} and \eqref{iss} admit a solution. 

Consequently, since for any fixed $e$ and $\underline{z}$, the objective is continuous in $H$ and the feasible set is compact and convex in \eqref{ent} and \eqref{iss}, respectively, \bref{moral} and the first part of \bref{prop2} follow immediately from \bref{thma1}, with $J=2$ and $J=1$, respectively. This is because any $H$ satisfying conditions 1 through 3 correspond to a contingent debt contract with at most $J$ non-defaultable face values. The uniqueness part of \bref{prop2} further follows from the fact that the objective of \eqref{iss} is strictly convex in $H$ when $\Phi(\cdot|s)$ has full support for all $s \in S$, and hence, every solution must be an extreme point of the feasible set. 
\end{proof}

\subsection{Proof of Proposition \ref{finite}}\label{afinite}
Let $\Pi^*(e)$ be the value of the entrepreneur's problem $\eqref{ent}$ for a fixed $e \in [0,\bar{e}]$. We first show that there exists Lagrange multipliers $\lambda_1^* \neq 0$ and $\lambda_2^* \geq 0$ such that 
\begin{align}\label{entir}
\Pi^*(e)=\sup_{H \in \mathcal{I}(\underline{F},\overline{F})} \bigg[\int_0^1 (x-H(x))\phi(x|e)\diff x+&\lambda_1^*\left(\int_0^1(x-H(x))\phi_e(x|e)\diff x-C'(e)\right)\notag\\
+&\lambda_2^*\left(\int_0^1 H(x)\phi(x|e)\diff x-(1+r)I\right)\bigg].
\end{align}
To this end, we adopt a similar argument as \citet{N23}. For any fixed $e \in [0,\bar{e}]$ and for any $\gamma \in \mathbb{R}$, let $M_e(\gamma)$ be the value of 
\begin{align}\label{entg}
\sup_{H \in \mathcal{I}(\underline{F},\overline{F})} &\left[\int_0^{1} [x-H(x)]\phi( x | e)\diff x-C(e)\right] \notag\\
\mbox{s.t. }& \int_0^{1} [x-H(x)]\phi_e(x|e)\diff x=C'(e)\\
&\int_0^{1} H(x)\phi( x|e)\diff x \geq \gamma \notag.
\end{align}
Note that 
\begin{equation}\label{sol}
M_e((1+r)I)=\Pi^*(e)=\int_0^1 (x-H^*(x))\phi(x|e)\diff x-C(e),
\end{equation}
where $H^*$ is a solution of \eqref{ent} with a fixed $e$. Moreover, $M_e$ is nonincreasing and concave in $\gamma$. Indeed, monotonicity follows from the ordered structure of the feasible set as $\gamma$ increases. For concavity, consider any $\gamma_1,\gamma_2 \in \mathbb{R}$ and let $\gamma^\lambda:=\lambda\gamma_1+(1-\lambda)\gamma_2$ for any $\lambda \in (0,1)$. Since \eqref{entg} admits a solution, there exists $H_1, H_2 \in \mathcal{I}(\underline{F},\overline{F})$ such that 
\[
\int_0^1 (x-H_1(x))\phi(x|e)\diff x-C(e)=M(\gamma_1); \quad \int_0^1 (x-H_2(x))\phi(x|e) \diff x-C(e)=M(\gamma_2). 
\]
Furthermore, 
\begin{align*}
&\int_0^{1} (x-H_i(x))\phi_e(x|e)\diff x=C'(e)\\
&\int_0^{1} H_i(x)\phi( x|e)\diff x \geq \gamma_i
\end{align*}
for $i \in \{1,2\}$. Let $H^\lambda:=\lambda H_1+(1-\lambda) H_2$, we must have $H^\lambda \in \mathcal{I}(\underline{F},\overline{F})$ and 
\begin{align*}
&\int_0^{1} (x-H^\lambda(x))\phi_e(x|e)\diff x=C'(e)\\
&\int_0^{1} H^\lambda(x)\phi( x|e)\diff x \geq \gamma^\lambda.
\end{align*}
Thus, 
\begin{align*}
M_e(\gamma^\lambda)\geq &\int_0^1 (x-H^\lambda(x))\phi(x|e)\diff x-C(e)\\
=&\lambda \int_0^1 (x-H_1(x))\phi(x|e)\diff x+ (1-\lambda) \int_0^1 (x-H_2(x))\phi(x|e)\diff x\\
=& \lambda M_e(\gamma_1)+(1-\lambda)M_e(\gamma_2). 
\end{align*}
Since $M_e$ is nonincreasing and concave, and since $(1+r)I$ is an interior of the set 
\[
\left\{\int_0^1 H(x)\phi(x|e)\diff x\bigg| H \in \mathcal{I}(\underline{F},\overline{F}), \, \int_0^1 (x-H(x))\phi_e(x|e)\diff x=C'(e)\right\},
\]
there exists $\lambda_2^* \geq 0$ such that 
\[
M_e(\gamma)\leq M_e((1+r)I)-\lambda_2^*(\gamma-(1+r)I)
\]
for all $\gamma \in \mathbb{R}$. Meanwhile, for any $H \in \mathcal{I}(\underline{F},\overline{F})$ such that 
\begin{equation}\label{ic}
\int_0^1 (x-H(x))\phi_e(x|e)\diff x=C'(e),
\end{equation}
it must be that 
\[
M_e\left(\int_0^1 H(x)\phi(x|e)\diff x\right) \geq \int_0^1 (x-H(x))\phi(x|e)\diff x-C(e),
\]
by the definition of $M_e$. Together with \eqref{sol}, we have 
\begin{align}\label{dual}
M_e((1+r)I)=&\int_0^1 (x-H^*(x))\phi(x|e)\diff x-C(e)\notag\\ 
\geq& \int_0^1 (x-H(x))\phi(x|e)\diff x-C(e)+\lambda_2^*\left(\int_0^1 H(x)\phi(x|e)\diff x-(1+r)I\right),
\end{align}
for all $H \in \mathcal{I}(\underline{F},\overline{F})$ such that \eqref{ic} holds. Since $H^*$ is feasible for \eqref{ent} with the fixed $e$, \eqref{dual} implies 
\begin{align}\label{dual2}
&\int_0^1 (x-H^*(x))\phi(x|e)\diff x+\lambda_2^*\left(\int_0^1 H^*(x)\phi(x|e)\diff x-(1+r)I\right)\notag \\
\geq& \int_0^1 (x-H(x))\phi(x|e)\diff x+\lambda_2^*\left(\int_0^1 H(x)\phi(x|e)\diff x-(1+r)I\right),
\end{align}
for all $H \in \mathcal{I}(\underline{F},\overline{F})$ satisfying \eqref{ic}. Now let 
\[
\mathcal{L}_e(H;\lambda):=\int_0^1(x-H(x))\phi(x|e)\diff x-C(e)+\lambda\left(\int_0^1 H(x)\phi(x|e)\diff x-(1+r)I\right),
\]
and let $\mathcal{L}_e(\lambda)$ be the value of 
\begin{align}\label{prime2}
\sup_{H \in \mathcal{I}(\underline{F},\overline{F})} &\mathcal{L}_e(H;\lambda)\\
\mbox{s.t. }& \int_0^1 (x-H(x))\phi_e(x|e)\diff x=C'(e). \notag
\end{align}
Then, \eqref{dual2} implies that $H^*$ solves \eqref{prime2} with $\lambda=\lambda_2^*$ and 
\[
\mathcal{L}_e(\lambda_2^*)=\int_0^1 (x-H^*(x))\phi(x|e)\diff x-C(e).
\]
Meanwhile, by the definition of $\mathcal{L}_e(\lambda)$,
\[
\mathcal{L}_e(\lambda) \geq \int_0^1 (x-H(x))\phi(x|e)\diff x-C(e)
\]
for all feasible $H$ of \eqref{ent} with fixed $e$. Finally, since the constraint in \eqref{prime2} is an equality, standard results (see, e.g., Theorem 3.12 of \citealp{AN87}) implies that there exits $\lambda_1 \neq 0$ such that \eqref{entir} holds.

For any fixed $e \in [0,\bar{e}]$, since the primal problem \eqref{ent} is convex for any fixed $e \in [0,\bar{e}]$, there exists an extreme point $H^*$ of the feasible set that attains $\Pi^*(e)$. By \bref{thma1}, there exists a countable collection of intervals $\{[\underline{x}_n,\overline{x}_n)\}_{n=1}^\infty$ such that $H^*$ satisfies conditions 1 through 3 for $J=2$. Meanwhile, as established above, $H^*$ must also solve the dual problem \eqref{entir} of \eqref{ent} for this fixed $e$. Note that the dual problem can be written as 
\[
\sup_{H \in \mathcal{I}(\underline{F},\overline{F})}\left[\int_0^1 H(x)[(1+\lambda_2^*)\phi(x|e)-\lambda_1^*\phi_e(x|e)]\diff x+\kappa\right],
\]
with $\kappa \in \mathbb{R}$ being a constant that does not depend on $H$. Moreover, 
\[
(1+\lambda_2^*)\phi(x|e)-\lambda_1^*\phi_e(x|e) \geq 0 \iff \frac{\phi_e(x|e)}{\phi(x|e)} \leq \frac{1+\lambda_2^*}{\lambda_1^*}. 
\]
Since $\phi_e(\cdot|e)/\phi(\cdot|e)$ is at most $N$-peaked, there must be a finite interval partition $\{I_k\}_{k=1}^K$ of $[0,1]$ with $K \leq 2N$ such that $\phi_e(x|e)/\phi(x|e)-(1+\lambda_2^*)/\lambda_1^*$ takes the same sign for all $x \in I_k$. 

Therefore, if there are more than $N+1$ intervals on which $H^*$ is constant, then either there are at least two of them contained in 
a single interval $I_k$ with $\phi_e(x|e)/\phi(x|e) < (1+\lambda_2^*)/\lambda_1^*$ for all $x \in I_k$, or there is at least one of them contained in an interval $I_j$ with $\phi_e(x|e)/\phi(x|e)>(1+\lambda_2^*)/\lambda_1^*$ for all $x \in I_j$. If there are two intervals $[\underline{x}_n,\overline{x}_n)$, $[\underline{x}_m,\overline{x}_m)$, with $\overline{x}_n \leq \underline{x}_m$, that are contained in some $I_k$ with $\phi_e(x|e)/\phi(x|e) < (1+\lambda_2^*)/\lambda_1^*$ for all $x \in I_k$, then, since by condition 2 of \bref{thma1}, $H^*(\underline{x}_n)<H^*(\underline{x}_m)$,  for $H^{**}$ defined as
\[
H^{**}(x):=\left\{
\begin{array}{cc}
H^*(x),&\mbox{if } x \notin [\underline{x}_n,\overline{x}_m)\\
H^*(\underline{x}_n),&\mbox{if } x \in [\underline{x}_n,\overline{x}_m)
\end{array}
\right.,
\]
for all $x \in [0,1]$, $H^{**} \in \mathcal{I}(\underline{F},\overline{F})$ and yields a higher value to the objective of \eqref{entir} than $H^*$. Likewise, if there is at least one interval on which $H^*$ is constant that is contained in some $I_j$ such that $\phi_e(x|e)/\phi(x|e) < (1+\lambda_2^*)/\lambda_1^*$ for all $x \in I_j$, then, since $H^*(x)<x$ for all $x \in (\underline{x}_n,\overline{x}_n)$, for $H^{**}$ defined as 
\[
H^{**}(x):=\left\{
\begin{array}{cc}
H^*(x),&\mbox{if } x \notin I_j\\
\max\{x,H^*(\overline{x}_n)\},&\mbox{if } x \in I_j
\end{array}
\right.,
\]
for all $x \in [0,1]$, $H^{**} \in \mathcal{I}(\underline{F},\overline{F})$ and yields a higher value to the objective of $\eqref{entir}$ than $H^*$. Thus, $H^*$ cannot be a solution of the dual problem \eqref{entir} for this fixed $e$, a contradiction. Consequently, the solution $H^*$ to the primal problem \eqref{ent} for any fixed $e \in [0,\bar{e}]$ cannot admit more than $N+1$ intervals where $H^*$ is constant. As a result, $H^*$ is a contingent debt contract with at most $N+1$ face values. Since $e \in [0,\bar{e}]$ is arbitrary, this completes the proof. \hfill $\blacksquare$
\end{document}